\documentclass[acmtog,screen,nonacm]{acmart}

\AtBeginDocument{%
  }

\setcctype{by}
\acmJournal{TOG}
\acmYear{2025} \acmVolume{44} \acmNumber{4} \acmArticle{} \acmMonth{8} \acmPrice{}\acmDOI{10.1145/3731155}

\acmSubmissionID{332}

\usepackage{amsmath,mleftright,mathtools}
\usepackage{graphicx}
\usepackage{subcaption}
\usepackage{wrapfig}
\usepackage{pdfpages}

\begin{document}

%%
%% The "title" command has an optional parameter,
%% allowing the author to define a "short title" to be used in page headers.
\title{CK-MPM: A Compact-Kernel Material Point Method}

%%
%% The "author" command and its associated commands are used to define
%% the authors and their affiliations.
%% Of note is the shared affiliation of the first two authors, and the
%% "authornote" and "authornotemark" commands
%% used to denote shared contribution to the research.
% \author{Ben Trovato}
% \authornote{Both authors contributed equally to this research.}
% \email{trovato@corporation.com}
% \orcid{1234-5678-9012}
% \author{G.K.M. Tobin}
% \authornotemark[1]
% \email{webmaster@marysville-ohio.com}
% \affiliation{%
%   \institution{Institute for Clarity in Documentation}
%   \city{Dublin}
%   \state{Ohio}
%   \country{USA}
% }

\author{Michael Liu}
\affiliation{%
  \institution{Carnegie Mellon University}
  \country{USA}
}
\email{appledorem.g@gmail.com}

\author{Xinlei Wang}
\affiliation{%
  \institution{NetEase Games Messiah Engine}
  \country{China}
}
\email{wxlwxl1993@zju.edu.cn}

\author{Minchen Li}
\affiliation{%
  \institution{Carnegie Mellon University}
  \country{USA}
}
\email{minchernl@gmail.com}

%%
%% By default, the full list of authors will be used in the page
%% headers. Often, this list is too long, and will overlap
%% other information printed in the page headers. This command allows
%% the author to define a more concise list
%% of authors' names for this purpose.
% \renewcommand{\shortauthors}{Liu et al.}

%%
%% The abstract is a short summary of the work to be presented in the
%% article.
\begin{abstract}
The Material Point Method (MPM) has become a cornerstone of physics-based simulation, widely used in geomechanics and computer graphics for modeling phenomena such as granular flows, viscoelasticity, fracture mechanics, etc. Despite its versatility, the original MPM suffers from cell-crossing instabilities caused by discontinuities in particle-grid transfer kernels. Existing solutions mostly mitigate these issues by adopting smoother shape functions, but at the cost of increased numerical diffusion and computational overhead due to larger kernel support.
In this paper, we propose a novel $C^2$-continuous compact kernel for MPM that achieves a unique balance in terms of stability, accuracy, and computational efficiency. Our method integrates seamlessly with Affine Particle-In-Cell (APIC) and Moving Least Squares (MLS) MPM, while only doubling the number of grid nodes associated with each particle compared to linear kernels. At its core is an innovative dual-grid framework, which associates particles with grid nodes exclusively within the cells they occupy on two staggered grids, ensuring consistent and stable force computations.
We demonstrate that our method can be conveniently implemented using a domain-specific language, Taichi, or based on open-source GPU MPM frameworks, achieving faster runtime and less numerical diffusion compared to quadratic B-spline MPM. Comprehensive validation through unit tests, comparative studies, and stress tests demonstrates the efficacy of our approach in conserving both linear and angular momentum, handling stiff materials, and scaling efficiently for large-scale simulations.
Our results highlight the transformative potential of compact, high-order kernels in advancing MPM's capabilities for stable, accurate, and high-performance simulations.
\end{abstract}

%%
%% The code below is generated by the tool at http://dl.acm.org/ccs.cfm.
%% Please copy and paste the code instead of the example below.
%%


%%
%% Keywords. The author(s) should pick words that accurately describe
%% the work being presented. Separate the keywords with commas.
%% A "teaser" image appears between the author and affiliation
%% information and the body of the  document, and typically spans the
%% page.
\begin{teaserfigure}
    \centering
    % \begin{subfigure}[b]{0.43\linewidth}
    %     \includegraphics[height=5.5cm]{figures/teaser_full.png}
    %     % \caption{Front}
    % \end{subfigure}%
    \includegraphics[width=\linewidth]{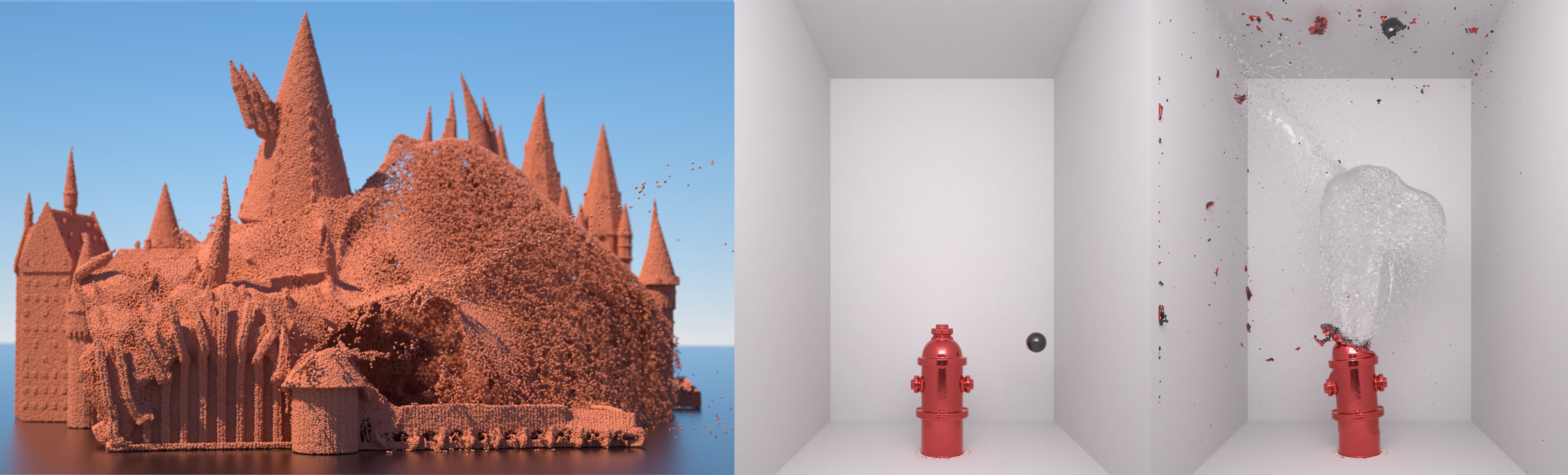}
  \caption{Large-scale simulations of a sandcastle (left) and a fire hydrant (right) destroyed by high-speed balls, both performed using our compact-kernel material point method (CK-MPM), exhibiting intricate and realistic dynamics.}
  \label{fig:teaser}\label{fig:castle-crasher-teaser}
\end{teaserfigure}

% \received{20 February 2007}
% \received[revised]{12 March 2009}
% \received[accepted]{5 June 2009}

%%
%% This command processes the author and affiliation and title
%% information and builds the first part of the formatted document.
\maketitle

\section{Introduction}

The Material Point Method (MPM), introduced by \citet{sulsky1995application} as an extension of the Particle-in-Cell (PIC) method \cite{harlow1962particle}, has found widespread applications in solid mechanics. It is widely used in fields such as geomechanics and computer graphics, including simulation of granular materials \cite{klar2016drucker,yue2018hybrid,chen2021hybrid,zhao2023coupled}, viscoelastic materials \cite{su2021unified,fang2019silly}, snow \cite{stomakhin2013material,gaume2018dynamic}, ductile fracture \cite{wolper2019cd,wolper2020anisompm}, solid-fluid interactions \cite{fang2020iq,fei2019multi,fei2018multi,fei2017multi}, frictional contact \cite{han2019hybrid,guo2018material,jiang2017anisotropic}, phase change effects \cite{stomakhin2014augmented,ding2019thermomechanical,su2021unified}, and even combustion \cite{kala2024thermomechanical} and explosions \cite{cao2022efficient}. For a complete survey, we refer the reader to \citet{de2020material}.

As a hybrid Lagrangian-Eulerian method, MPM employs Lagrangian particles to track the geometry of the simulation object while using a Eulerian background grid to compute forces and perform time integration. Alternatively, particles can be viewed as quadrature points, and the grid nodes represent the degrees of freedom (DOFs). In its original formulation, MPM adopted the piecewise linear particle-grid transfer kernel, same as the kernel used in PIC. However, since MPM requires the gradient of the kernel function to compute forces, linear kernels introduce discontinuities that cause numerical instability when particles move across grid cells -- a phenomenon known as cell-crossing instability. This issue could even lead to numerical explosions.

Most existing methods address cell-crossing instabilities by introducing smoother kernel functions, which often come at the cost of larger support regions, leading to more severe numerical diffusion. Numerical diffusion is a fundamental limitation of hybrid Lagrangian-Eulerian methods, arising from particle-grid transfers \cite{bridson2015fluid}. These transfers act as averaging operations that smooth the velocity field, often leading to the loss of sharp features. For instance, contact gaps may form between two colliding objects as the velocities of their boundary particles are averaged across the grid nodes between them, even before the objects physically interact. Similarly, high-frequency modes in vibrating elastic objects may dissipate because the solution space defined by kernels with large support may fail to capture these modes effectively.

Additionally, using kernels with larger support can lead to higher computational costs. For example, the widely used quadratic B-Spline kernel associates 27 grid nodes with each particle in 3D -- over three times the 8 grid nodes associated with linear kernels. This increases the complexity of the particle-to-grid transfer step, amplifying challenges such as write conflicts and memory bandwidth limitations. To address this, many works focus on high-performance computing solutions, leveraging specialized data structures and parallel algorithms on advanced computing hardware \cite{gao2018gpu,wang2020massively,fei2021principles,qiu2023sparse}. While these approaches achieve substantial speedups, their performance remains fundamentally constrained by the underlying discretization using quadratic B-Splines. 

In this paper, we propose a novel $C^2$-continuous compact kernel for particle-grid transfer that achieves a unique balance of stability, accuracy, and efficiency. Our approach, named compact-kernel (CK) MPM, is fully compatible with Affine Particle-In-Cell (APIC) \cite{jiang2015affine} and Moving Least Squares (MLS) MPM \cite{hu2018moving}, effectively avoiding cell-crossing instabilities while only doubling the number of grid nodes associated with each particle compared to linear kernels.
To achieve this, we introduce a dual grid framework, where each particle is associated exclusively with the nodes of the cell it resides in on both grids. Our method can be conveniently implemented using a domain-specific language, Taichi \cite{hu2019taichi}, or within state-of-the-art GPU MPM simulation frameworks, such as \citet{wang2020massively}, achieving faster performance and less numerical diffusion compared to quadratic B-spline MPM.
An extensive set of unit tests, comparative studies, and stress tests is performed to validate the efficacy of CK-MPM. These evaluations demonstrate its ability to conserve linear and angular momentum, handle stiff materials with robustness, reduce numerical diffusion, and scale efficiently for high-resolution simulations, highlighting the transformative potential of using more compact kernels in MPM. Our code is open-sourced at \url{https://github.com/Simulation-Intelligence/CK-MPM}.

\section{Related Work}

Our work focuses on the design, application, and analysis of compact kernels for the particle-grid transfer in MPM, with an emphasis on addressing cell-crossing instability. Accordingly, we primarily review relevant literature in this area, while briefly discussing related works on mitigating numerical diffusion.

\paragraph{Cell-Crossing Instability}
To address the cell-crossing instability, various methods have been proposed. \citet{bardenhagen2004generalized} introduced the Generalized Interpolation Material Point (GIMP) method, which treats particles as volumes during particle-grid transfer, resulting in effective shape functions derived from integrals of the linear kernel. \citet{sadeghirad2011convected} further extended GIMP with Convected Particle Domain Interpolation (CPDI), which accounts for particle volume changes during transfer and improves accuracy, particularly for cases involving large tensile deformations and rotations. Similarly, \citet{wilson2021distillation} proposed dynamically splitting particles near cell boundaries into subparticles for transfer, achieving higher efficiency while mitigating cell-crossing issues, albeit with some loss of accuracy compared to GIMP.
Another approach involves directly using smoother shape functions. For example, \citet{steffen2008analysis} proposed using quadratic and cubic B-spline shape functions, which produce smoother forces and effectively reduce cell crossing instabilities. B-spline MPM is widely adopted in the graphics community due to their simplicity and effectiveness \cite{jiang2016material}. Building on this idea, \citet{moutsanidis2020iga} introduced Isogeometric Analysis MPM (IGA-MPM), which uses non-uniform rational B-splines (NURBS) as shape functions. This approach provides additional flexibility, such as exact representations of conic sections and better preservation of symmetry in solutions.
An alternative strategy focuses on transferring stress from particles to the grid and then performing stress interpolation to calculate forces \cite{zhang2011material}. \citet{liang2019efficient} adapted this idea to a staggered grid, reducing the number of accumulation operations and improving computational efficiency. Our method introduces a smooth kernel function that is as compact as the linear kernel, applied within a staggered grid framework to achieve a unique balance of stability, accuracy, and efficiency.

\paragraph{Numerical Diffusion}  
Mitigating numerical diffusion is a widely studied topic in hybrid Lagrangian-Eulerian methods. Here, we focus on the context of MPM. As a variant of PIC, the Fluid-Implicit-Particle (FLIP) method \cite{brackbill1988flip} reduces numerical diffusion by transferring velocity difference instead of velocity. However, it is prone to numerical instability and is often blended with PIC for practical use.
To improve stability and preserve angular momentum, \citet{jiang2015affine,jiang2017angular} introduced the Affine Particle-In-Cell (APIC) method, which captures and transfers the local affine velocity field per particle. Extending this idea, \citet{fu2017polynomial} proposed the Polynomial Particle-In-Cell (PolyPIC) method, which uses higher-order polynomial functions to better preserve local velocity features. \citet{hu2018moving} showed that both APIC and PolyPIC can be interpreted as Galerkin-style MLS discretizations. For a comparative analysis of these methods, \citet{fei2021revisiting} evaluated their behaviors and introduced a new method combining FLIP and APIC to further reduce numerical dissipation.
Another strategy is to refine the grid in critical regions. \citet{zhao2024mapped} proposed Mapped MPM, which uses a nonuniform mapping to distort the grid, providing higher resolution in regions with sharp features. \citet{gao2017adaptive} extended GIMP to an adaptive Octree grid, dynamically allocating more degrees of freedom where needed.
While our method is not specifically designed to reduce numerical diffusion, we demonstrate that using compact kernels inherently helps mitigate diffusion.

\section{Background and Preliminaries}

We follow \citet{jiang2016material} to derive the weak form of the governing equations for continuum simulation and briefly introduce the traditional MPM pipeline in this section.
% \todo{introduce MPM starting from the strong and weak forms, all the way to algorithm steps including p2g, g2p, etc. can follow e.g. \url{https://dl.acm.org/doi/pdf/10.1145/3570160} sec3, \url{https://drive.google.com/file/d/1Lu9n_LXM5ztCNIrepHJj2wSa52OykhDA/preview} sec3, \url{https://arxiv.org/pdf/2312.10338} sec2, etc.}

\subsection{Spatial and Temporal Discretization}
We denote \(\Omega_0, \Omega_t \subset \mathbb{R}^3\) as the material space and world space, respectively, which are related through a deformation map \(\boldsymbol{\phi}: \Omega_0 \times [0, \infty) \rightarrow \mathbb{R}^3\), where \(\boldsymbol{\phi}(\mathbf{x}_{\Omega_0}, t) \in \Omega_t\) for \(\mathbf{x}_{\Omega_0} \in \Omega_0\). Subscripts are used to distinguish positions in material space (\(\mathbf{x}_{\Omega_0}\)) and world space (\(\mathbf{x}_{\Omega_t}\)) at time \(t\). 
Following a Lagrangian formulation, the dynamics of continua are described by a density field \(\rho_0: \Omega_0 \times [0, \infty) \rightarrow \mathbb{R}\), a velocity field \(\mathbf{v}_0: \Omega_0 \times [0, \infty) \rightarrow \mathbb{R}^3\), and the governing equations for the conservation of mass and momentum:
\begin{equation*}
     \begin{cases}
         \rho_0(\mathbf{x}_{\Omega_0}, t) J(\mathbf{x}_{\Omega_0}, t) = \rho_0(\mathbf{x}_{\Omega_0}, 0), \\
         \rho_0(\mathbf{x}_{\Omega_0}, 0) \frac{\partial \mathbf{v}_0}{\partial t}(\mathbf{x}_{\Omega_0}, t) = 
         \nabla_{\mathbf{x}_{\Omega_0}} \cdot \mathbf{P} + \rho_0(\mathbf{x}_{\Omega_0}, 0) \mathbf{g},
    \end{cases}
\end{equation*}
where \(J(\mathbf{x}_{\Omega_0}, t) = \det(\mathbf{F})\) measures the local volume change, \(\mathbf{F} = \frac{\partial \boldsymbol{\phi}}{\partial \mathbf{x}_{\Omega_0}}(\mathbf{x}_{\Omega_0}, t)\) is the deformation gradient, \(\mathbf{g}\) is the gravitational acceleration, and the first Piola-Kirchhoff stress tensor \(\mathbf{P}\) is a function of \(\mathbf{F}\). 
Finally, we define the Eulerian counterparts of the density and velocity fields as \(\rho\) and \(\mathbf{v}\), respectively.

We discretize time with an interval $\Delta t$, such that the equations are evaluated at time $t_n = n \Delta t$. The Lagrangian velocity $\mathbf{v}_0$ and acceleration $\mathbf{a}_0 = \partial \mathbf{v}_0 / \partial t$ at time $t_n$ can be approximated using finite difference methods via  
$
\mathbf{v}_0 \approx \frac{1}{\Delta t} (\mathbf{x}_{\Omega_{t_n}} - \mathbf{x}_{\Omega_{t_{n-1}}})$ and 
$\mathbf{a}_0 \approx \frac{1}{\Delta t} (\mathbf{v}_0(\cdot, t_{n+1}) - \mathbf{v}_0(\cdot, t_n))
$ when using Symplectic Euler. Assuming zero gravity and zero traction boundary conditions, and letting $\mathbf{q}_0 \ : \ \Omega_0 \rightarrow \mathbb{R}^3$ be an arbitrary test function in $\Omega_0$, the weak form in $\Omega_{t_n}$ can be obtained by pushing forward the one in $\Omega_0$:
% \begin{equation}
% \begin{aligned}
%     & \int_{\Omega_0} \rho_0(\cdot, 0) \frac{1}{\Delta t} (\mathbf{v}_0(\cdot, t_{n+1}) - \mathbf{v}_0(\cdot, t_n)) \cdot \mathbf{q}_0 (\cdot) \, d\mathbf{x}_{\Omega_0} \\
%     = & -\int_{\Omega_0} \mathrm{tr}(\mathbf{P}^T \nabla_{\mathbf{x}_{\Omega_0}} \mathbf{q}_0) \, d\mathbf{x}_{\Omega_0}.
% \end{aligned}
% \end{equation}
% By applying a push-forward operation, we transform the weak form to the deformed configuration $\Omega_{t_n}$, yielding:
\begin{equation}
\begin{aligned}
    & \int_{\Omega_{t_n}} \rho(\cdot, t_n) \frac{1}{\Delta t} (\mathbf{v}(\cdot, t_{n+1}) - \mathbf{v}(\cdot, t_n)) \cdot \mathbf{q}(\cdot) \, d\mathbf{x}_{\Omega_{t_n}} \\
    = & -\int_{\Omega_{t_n}} \mathrm{tr}((\frac{1}{J} \mathbf{P} \mathbf{F}^T)^T \nabla_{\mathbf{x}_{\Omega_{t_n}}} \mathbf{q}) \, d\mathbf{x}_{\Omega_{t_n}}.
\end{aligned}
\end{equation}
For numerical integration, we use the positions $\mathbf{x}_p$ of MPM particles as quadrature points and employ kernels $N$, e.g., quadratic B-spline functions, as weights $w_{i, p, t_n} = N(\mathbf{x}_{p, t_n} - \mathbf{x}_i)$, where $\mathbf{x}_i$ is the position of background grid nodes. By selecting appropriate test functions and applying mass lumping, the momentum update formula for MPM can be derived as:
\begin{equation}
    m_{i, t_n} (\mathbf{v}_{i, t_{n+1}} - \mathbf{v}_{i, t_n}) = -\Delta t \sum_p V_{p, 0} \mathbf{P} \mathbf{F}^T \nabla w_{i, p, t_n}, \label{eq:traditional-grid-update}
\end{equation}
where $m_{i, t_n}$ and $\mathbf{v}_{i, t_n}$ represent the mass and velocity of grid node $i$ at time $t_n$, and $V_{p, 0}$ is the volume of particle $p$ in the material space.

\subsection{MPM Pipeline Overview}

We provide an overview of the general pipeline for explicit MPM using a symplectic Euler time integrator. Each time step involves transferring quantities between particles and the background grid, as well as performing computations on the grid to update velocities:
\begin{enumerate}
    \item \textbf{Particle to Grid (P2G):} Particle masses are transferred to the grid using B-spline kernels, while momentum is transferred using either the PIC or APIC scheme.
    \item \textbf{Grid Update:} The grid momentum is updated (\autoref{eq:traditional-grid-update}) with grid-level boundary conditions applied as needed.
    \item \textbf{Grid to Particle (G2P):} Particle velocities are interpolated from the grid using B-spline kernels. For APIC, an additional matrix is computed to capture the local affine velocity field.
    \item \textbf{Update Deformation Gradient:} The deformation gradient $\mathbf{F}_{p, t_{n+1}}$ is updated using:
    \[
    \mathbf{F}_{p, t_{n+1}} = (\mathbf{I} + \Delta t \sum_i \mathbf{v}_{i, t_{n+1}} (\nabla w_{i, p, t_n})^T) \mathbf{F}_{p, t_n},
    \]
    where \(\mathbf{I}\) is the identity matrix. For elastoplastic materials, return mapping \cite{klar2016drucker,yue2015continuum} is applied to ensure the stress remains within the feasible region.
    \item \textbf{Particle Advection:} Particle positions are updated by integrating the interpolated velocities.
\end{enumerate}
Among these steps, the P2G operation is often the computational bottleneck.
% , accounting for more than 90\%
% \todo{confirm this is roughly correct}
% of the total computation time. 
This is primarily due to the scattering nature of the operation, where the scattering range depends on the kernel function. 
% Efficient parallelization of this step is challenging and often constrained by memory bandwidth and access patterns, making it a critical focus for optimization \cite{gao2018gpu,qiu2023sparse,fei2021principles}.

\section{Compact-Kernel MPM}

In this section, we introduce our novel compact kernel for particle-grid transfer, designed with a kernel radius of 2 and $C^2$-continuity. Our kernel satisfies all critical properties of kernel functions (\autoref{sec:smoothing_linear_kernel}). To effectively apply this kernel in discrete settings with 1st-order accuracy, we propose a staggered dual-grid discretization (\autoref{sec:discrete_CK}), which forms the foundation of our compact-kernel (CK) MPM (\autoref{sec:ck-mpm}). 
We further establish the theoretical foundation of our CK-MPM when combined with APIC (\autoref{sec:APIC-compatibility}) and MLS-MPM (\autoref{sec:MLS-compatibility}), proving its conservation properties for linear and angular momentum in the supplemental document. The symbols used in our derivation are explained in \autoref{table:notation}.

\begin{table}[ht]
    \centering
    \caption{Notation.}
    \label{table:notation}
    \begin{tabular}{|c|p{5cm}|}
        \hline
        \textbf{Symbol} & \textbf{Description} \\ \hline
        $\Delta x$ & Uniform, scalar distance between adjacent grid nodes  \\ \hline
        $\Delta t$ & Time step size  \\ \hline
        $\mathbf{x}^{\alpha_1 \cdots \alpha_p}_{\beta_1 \cdots \beta_q}$ & A $p$-times contravariant, $q$-times co-variant tensor. The Greek letters are used for indexing the components of the tensor. \\ \hline
        $\mathbf{x}_{a, b, \mathcal{G}}$ & The non-Greek-letter subscripts indicates additional information such as particle index and grid index. \\ \hline
        $\boldsymbol{\delta}$ & Kronecker delta tensor \\ \hline
        $\boldsymbol{\varepsilon}$ & Levi-Civita symbol for cross product. \\ \hline
        $\mathrm{sgn}(x)$ & The sign function. \\ \hline
        $\mathbf{x}^\alpha$ & Column vector using $\alpha$ as indices. \\ \hline
        $\mathbf{A}^\alpha_\beta$ & Matrix with row index $\alpha$ and column index $\beta$. \\ \hline
        $\mathbf{x}_\alpha \mathbf{y}^\alpha$ & Einstein sum on index $\alpha$. \\ \hline
        $(\mathbf{x}^T)_\alpha$ & Vector transpose, and thus a row vector with index $\alpha$. \\ \hline
    \end{tabular}
\end{table}

\subsection{Smoothing Linear B-Spline Kernel}\label{sec:smoothing_linear_kernel}

Although it is possible to directly construct a kernel with a radius smaller than $2$, we do not consider such an option, since it associates a particle with one single grid node, which is not practical.
Thus, in designing a new kernel function $\mathcal{K}(x)$ with a radius of 2 for MPM, our main intuition is to modify the linear B-spline kernel by adding a smoothing function $\mathcal{S}(x): [-1,1]\rightarrow \mathbb{R}$, that is, 
$$ 
\mathcal{K} (x) = 1 - |x| + \mathcal{S}(x),
$$
The primary motivation behind introducing the smoothing function $\mathcal{S}(x)$ is to address the non-differentiability of the linear B-spline kernel at x = 0 while preserving all other desirable properties. The derivation below serves as an intuitive guideline, outlining a general approach to systematically identify potential functional forms of $\mathcal{S}(x)$ that meet all requirements. For computational efficiency, we will choose a simplified form later. However, this general procedure is still meaningful, as it provides a theoretical foundation for further research. For the smoothed kernel function, we want the following properties to be satisfied:
\begin{enumerate}
    \item \textbf{Normalization:} $\int_\mathbb{R} \mathcal{K}(x) dx = 1$.
    \item \textbf{Monotonicity:} $\begin{cases} \mathcal{K}(a) > \mathcal{K}(b), &\text{if } 0 \leq a < b \\ \mathcal{K}(a) > \mathcal{K}(b), &\text{if } 0 \geq a > b  \end{cases}$.
    \item \textbf{Non-Negativity:} $\forall x \in \mathbb{R}, \mathcal{K}(x) \geq 0$.
    \item \textbf{Compactness:} $|x| \geq 1 \implies \mathcal{K}(x) = 0$.
    \item \textbf{Convergence to Dirac Delta:} $\lim_{h\rightarrow0} \mathcal{K}(\frac{x}{h}) = \delta(\frac{x}{h})$. \label{convergence-to-dirac-delta} 
    \item \textbf{Smoothness:} $\mathcal{K}(x)$ is  $C^2$-Continuous. \label{differentibility}
    \item \textbf{Partition of Unity:} $\mathcal{K}(x) + \mathcal{K}(1 - x) = 1$.
\end{enumerate}

We observe that the original linear B-spline kernel already satisfies all above properties except \autoref{differentibility}. To retain the satisfied properties, our smoothing function should at least satisfy
$$
\int_0^1 \mathcal{S}(x) dx = 0, \quad \text{and} \quad 
\mathcal{S}(0) = \mathcal{S}(1) = \mathcal{S}(-1) = 0.
$$
Note that the condition $\mathcal{S}(0) = 0$ is an artificial choice to resemble the behavior of linear kernels, and the condition $\mathcal{S}(-1) = \mathcal{S}(1) = 0$ ensures continuity. From the property on the right above, it is natural to consider periodic functions for potential candidates of $\mathcal{S}(x)$. Moreover, since $\{\sin(n\pi x), \cos(n \pi x) \ | \ \forall n \in \mathbb{N} \}$ forms a basis of $\mathcal{L}^2[-1, 1]$ (the Lebesgue space of square integrable functions defined on the $[-1, 1]$), we can fourier transform the function $\mathcal{S}(x) \in \mathcal{L}^2[-1, 1]$ into such basis and obtain:
$$
\mathcal{S}(x) = \sum_{n = 0}^\infty  a_n \sin(n \pi x) + \sum_{m = 0}^\infty b_m \cos(m \pi x),
$$
where $a_n, b_m \in \mathbb{R}, \forall n, m \in \mathbb{N}$. 

To ensure $\mathcal{S}(0) = 0$, we can set $\forall m \in \mathbb{N}, b_m = 0$, and we have
$$
\mathcal{S}(x) = \sum_{n = 0}^\infty a_n \sin(n\pi x).
$$

To smooth out the non-differentiability of the linear B-spline kernel at $x = 0$, we first observe that the derivative of the linear B-spline kernel when $x\neq0$ could be written as $-\mathrm{sgn}(x)$, where $\mathrm{sgn}$ denotes the sign function. Since $\mathcal{K}(x)$ should reach the maximum value of 1 at $x = 0$, it should have a derivative of 0 at $x = 0$, which requires $\lim_{x \rightarrow 0^-}\frac{d}{dx} \mathcal{S}(x) = -1$ and $\lim_{x \rightarrow 0^+}\frac{d}{dx} \mathcal{S}(x) = 1$. With a slight abuse of notation, we replace the $x$ in the decomposition of $\mathcal{S}(x)$ with $|x|$, leading us to observe that:
$$\frac{d}{dx}\mathcal{S}(x) = \frac{d}{dx}\bigg(\sum_{n = 0}^\infty a_n \sin(n\pi |x|)\bigg)  =   \pi \mathrm{sgn}(x) \sum_{n = 0}^\infty a_n n \cos(n\pi |x|).$$
For the above function, if we restrict $\pi (\sum_{n = 0}^\infty a_n n) = 1$, we have:
\begin{align*}
    \lim_{x \rightarrow 0^-} \frac{d}{dx} \mathcal{S}(x)
    =& \lim_{x \rightarrow 0^-} \pi \mathrm{sgn}(x) \sum_{n = 0}^\infty a_n \cos(n\pi x) n\\
    =& (\pi \sum_{n = 0}^\infty a_n n) \lim_{x \rightarrow 0^-} \mathrm{sgn}(x)\\
    =& -1,
\end{align*}
where it can be similarly shown that $\lim_{x \rightarrow 0^+} \frac{d}{dx} \mathcal{S}(x) = 1$, and they now nicely cancel out the unequal derivatives of the original linear B-spline kernel at $x=0$.

To ensure $C^2$-continuity of $\mathcal{K}(x)$, we also need its gradient to be smooth at $|x| = 1$. For $x = 1$, we observe that $\frac{d}{dx}\mathcal{S}(x) |_{x = 1} =  \pi \sum_{n = 0}^\infty(-1)^n a_n n $ is a non-convergent sequence. We therefore pick $a_n = 0$ for all odd $n$ so that it reduces to $\pi \sum_{n = 0}^\infty 2a_{2n}n$, which converges under the restriction $\pi (\sum_{n = 0}^\infty a_n n) = 1$, and so can be used to cancel out the derivative of linear B-spline functions, similar to when $x=0$ discussed above. The case for $x = -1$ is similar, and we reach a reduced form of the kernel function $\mathcal{S}(x)$, satisfying all required properties:
\begin{equation}
    \mathcal{S}(x) = \sum_{n = 0}^\infty 2a_{2n} n \sin(2n \pi |x|).
\end{equation}

\begin{figure}
    \centering
    \includegraphics[width=0.9\linewidth]{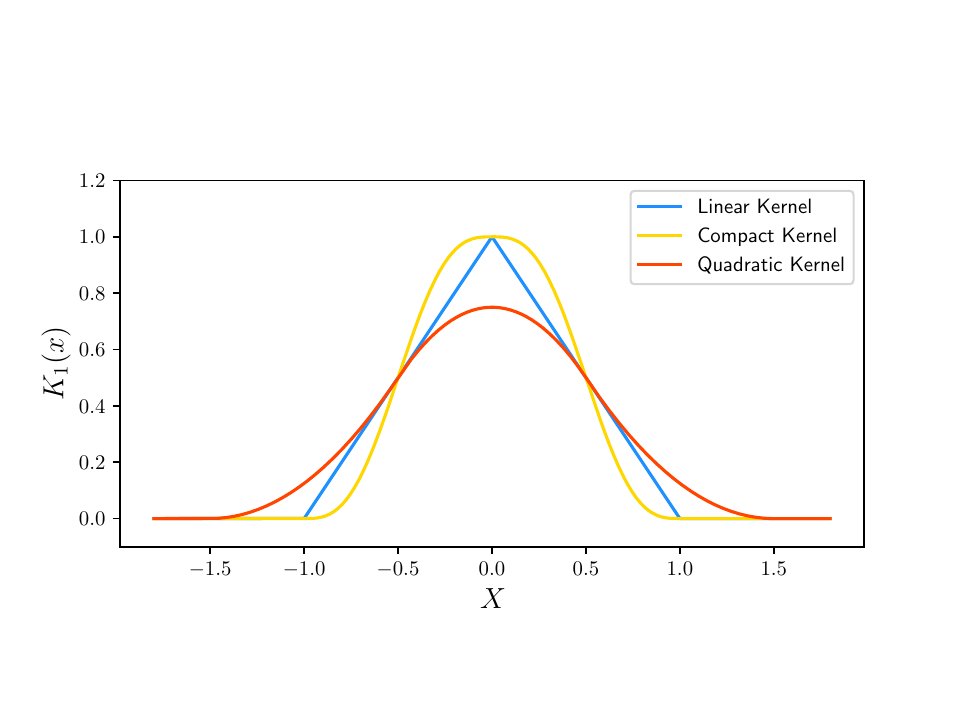}
    \caption{Plot of linear (blue), quadratic B-spline (red), and our compact (yellow) kernel functions.}
    \label{fig:kernel-functino-plot}
\end{figure}

For computational efficiency, we simply choose $a_n = 0$ for all $n$ except $n = 2$. Hence, our final kernel function in the 1-D setting is (\autoref{fig:kernel-functino-plot}):
\begin{equation}
    \mathcal{K}_1(x) = 1 - |x| + \frac{1}{2\pi} \sin(2\pi |x|).
\end{equation}

\subsection{Discrete Compact Kernel} \label{sec:discrete_CK}

\begin{figure}
    \begin{subfigure}[b]{0.32\linewidth}
        \centering
        \includegraphics[width=\linewidth]{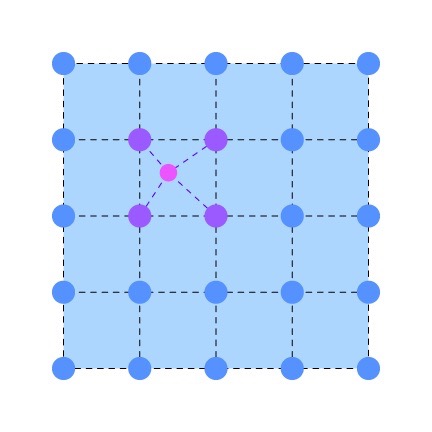}
        \caption{Linear kernel}
    \end{subfigure}
       \begin{subfigure}[b]{0.32\linewidth}
        \centering
        \includegraphics[width=\linewidth]{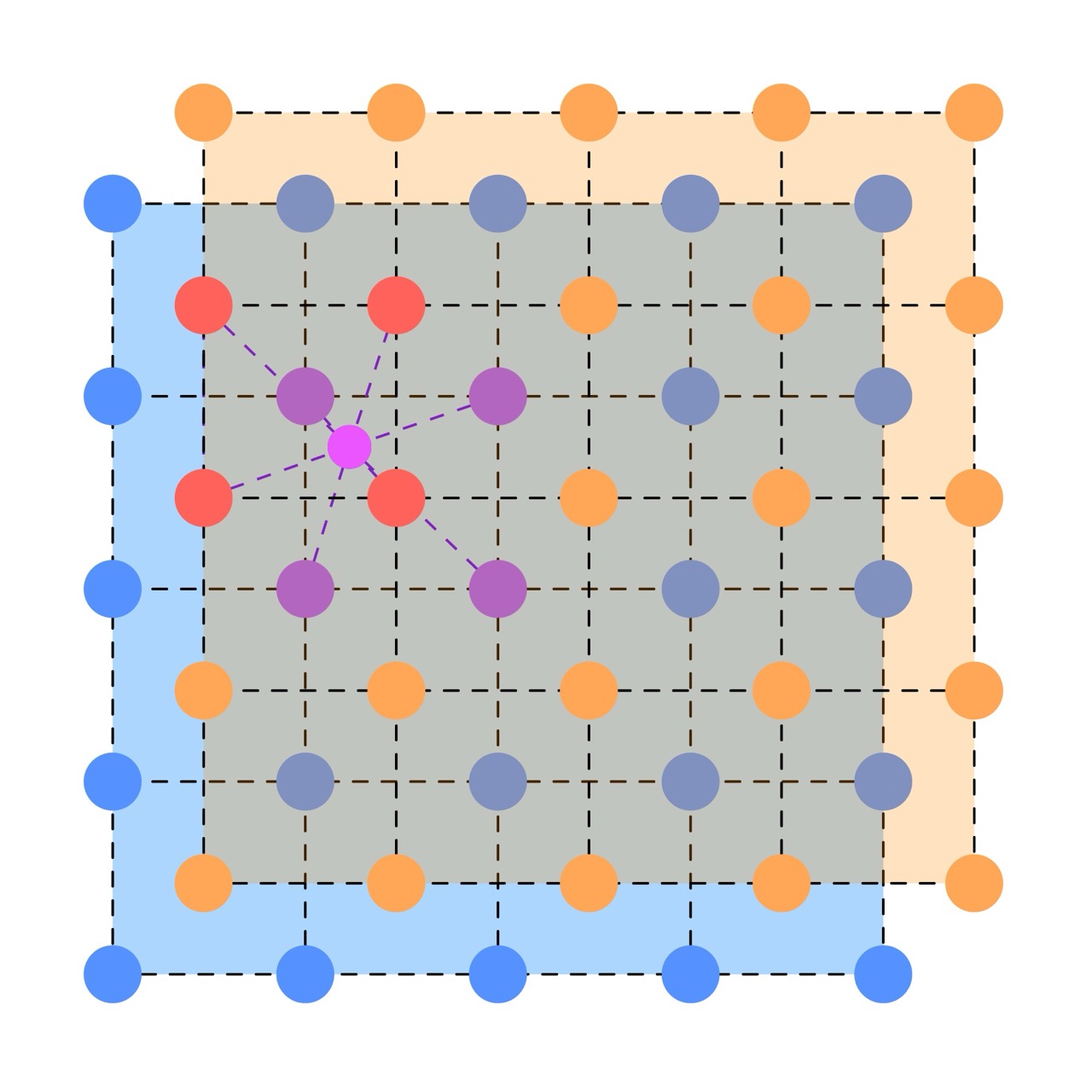}
        \caption{Our compact kernel}
    \end{subfigure}
    \begin{subfigure}[b]{0.32\linewidth}
        \centering
        \includegraphics[width=\linewidth]{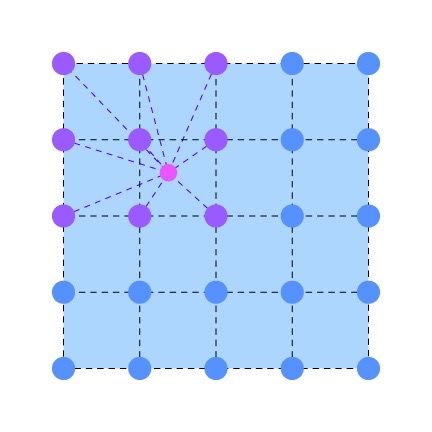}
        \caption{Quadratic kernel}
    \end{subfigure}
    \caption{Comparison of interpolation patterns of different kernels in 2D.}
    \label{fig:kernels}
\end{figure}

We now examine the property of our compact kernel in discrete settings. We define $\boldsymbol{\mathcal{K}}: \mathbb{R}^3 \rightarrow \mathbb{R}$ by:
\begin{equation}
    \boldsymbol{\mathcal{K}}(\mathbf{x}) = \prod_{\beta = 0}^2 \mathcal{K}_1(\mathbf{x}^\beta).
\end{equation}
Note that $\mathcal{K}_1$ is a scalar-valued function, whereas $\mathcal{K}$ operates on vectors in $\mathbb{R}^3$. For $\boldsymbol{\mathcal{K}}(x)$ to be an interpolation kernel, it must satisfy the following two properties to ensure 1st-order accuracy:
\begin{align}
    &\sum_i \boldsymbol{\mathcal{K}}\bigg(\frac{\mathbf{x}_p - \mathbf{x}_i}{\Delta x}\bigg) = 1, \label{interpolation-kernel-sum-to-1}\\
    &\sum_i \mathbf{x}_i^\alpha \boldsymbol{\mathcal{K}}\bigg(\frac{\mathbf{x}_p - \mathbf{x}_i}{\Delta x} \bigg) = \mathbf{x}_p^\alpha, \label{reconstruction-of-position}
\end{align}
where $\mathbf{x}_p$ and $\mathbf{x}_i$ represents the position of a particle $p$ and a grid node $i$, respectively. 

To show that $\boldsymbol{\mathcal{K}}(\mathbf{x})$ satisfies \autoref{interpolation-kernel-sum-to-1}, we consider the grid nodes that are associated with $\mathbf{x}_p^\alpha$. Since the kernel has a radius of $2$, there are eight grid nodes in total that are associated with $\mathbf{x}_p^\alpha$. For $0 \leq s, t, u \leq 1$, we denote $\mathbf{x}_{B_p(s, t, u)}^\alpha$ to be the grid nodes, which contain $\mathbf{x}_p^\alpha$, where $s, t, u$ denotes the nodal offset in the $x, y, z$-axis direction from the bottom left grid node $\mathbf{x}_{B_p(0, 0, 0)}^\alpha$. We observe that:
\begin{align*}
    &\sum_i \boldsymbol{\mathcal{K}}\bigg(\frac{\mathbf{x}_p - \mathbf{x}_i}{\Delta x}\bigg) \\
    =& \sum_{s = 0}^1 \sum_{t = 0}^1 \sum_{u = 0}^1 \boldsymbol{\mathcal{K}}\bigg(\frac{\mathbf{x}_p - \mathbf{x}_{B_p(s, t, u)}}{\Delta x}\bigg) \\
    =& \sum_{s = 0}^1 \sum_{t = 0}^1 \sum_{u = 0}^1 \prod_{\alpha = 0}^2 \mathcal{K}_1\bigg(\frac{\mathbf{x}_p^\alpha - \mathbf{x}_{B_p(s, t, u)}^\alpha}{\Delta x}\bigg). \\
    \end{align*}
We note that $\mathbf{x}_{B_p(s, t, u)}^0$ and $\mathbf{x}_{B_p(s, t, u)}^1$ are not affected by $u$. And, similarly,  $\mathbf{x}_{B_p(s, t, u)}^0$ is not affected by $t$.
    \begin{align*}
    =&\sum_{s = 0}^1 \sum_{t = 0}^1 \bigg( \sum_{u = 0}^1 \mathcal{K}_1\bigg(\frac{\mathbf{x}_p^2 - \mathbf{x}_{B_p(s, t, u)}^2}{\Delta x}\bigg) \prod_{\alpha = 0}^1 \mathcal{K}_1\bigg(\frac{\mathbf{x}_p^\alpha - \mathbf{x}_{B_p(s, t, 0)}^\alpha}{\Delta x}\bigg) \bigg)\\
    =& \sum_{s = 0}^1 \sum_{t = 0}^1 \bigg( \bigg( \mathcal{K}_1\bigg(\frac{\mathbf{x}_p^2 - \mathbf{x}_{B_p(s, t, 0)}^2}{\Delta x}\bigg) + \mathcal{K}_1 \bigg(\frac{\mathbf{x}_p^2 - \mathbf{x}_{B_p(s, t, 1)}^2}{\Delta x}\bigg)\bigg) \\
    &\prod_{\alpha = 0}^1 \mathcal{K}_1\bigg(\frac{\mathbf{x}_p^\alpha - \mathbf{x}_{B_p(s, t, 0)}^\alpha}{\Delta x}\bigg) \bigg),\\
    \end{align*}
 Now, recall our previous definition of $\mathcal{K}_1$, we see that the following equation holds:
\begin{equation}
    \mathcal{K}_1(x) + \mathcal{K}_1(1 - x) = 1, \label{resemble-linear}
\end{equation}
which resembles the behavior of linear kernels. Hence, we have: 
    \begin{align*}
    =& \sum_{s = 0}^1 \sum_{t = 0}^1 \prod_{\alpha = 0}^1 \mathcal{K}_1\bigg(\frac{\mathbf{x}_p^\alpha - \mathbf{x}_{B_p(s, t, 0)}^\alpha}{\Delta x}\bigg)\\
    =& \sum_{s = 0}^1 \mathcal{K}_1\bigg(\frac{\mathbf{x}_p^0 - \mathbf{x}_{B_p(s, 0, 0)}^0}{\Delta x}\bigg) \bigg(\sum_{t = 0}^1 \mathcal{K}_1\bigg(\frac{\mathbf{x}_p^1 - \mathbf{x}_{B_p(s, t, 0)}^1 }{\Delta x}\bigg)\bigg)\\
    =& \sum_{s = 0}^1 \mathcal{K}_1\bigg(\frac{\mathbf{x}_p^0 - \mathbf{x}_{B_p(s, 0, 0)}^0}{\Delta x}\bigg)\\
    =& 1.
\end{align*}

The property in \autoref{reconstruction-of-position} does not hold in general.
% \todo{it'd be great to briefly show that it is impossible to hold if the kernel radius is 2} 
To achieve this property in the discrete setting, we introduce a dual-grid system. Consider three grids $\{\mathcal{G}_0, \mathcal{G}_{-}, \mathcal{G}_{+}\}$ where $\mathcal{G}_0$ denotes a conceptual initial grid (\textbf{i.e. we will not store this grid}) and $\mathcal{G}_{-}, \mathcal{G}_+$ denote grids with an offset of $+\frac{1}{4}\Delta, -\frac{1}{4}\Delta x$ in all axes to $\mathcal{G}_0$ respectively (note that we may use $\mathcal{G}_{\pm1} \cong \mathcal{G}_{\pm}$ interchangeably). We will show that in the dual grid system with $\mathcal{G}_{-}$ and $\mathcal{G}_+$, the property in \autoref{reconstruction-of-position} can be achieved.

We first restate the two properties (\autoref{interpolation-kernel-sum-to-1} and \autoref{reconstruction-of-position}) in the new dual grid setting as:
\begin{equation}
    \frac{1}{2}\sum_{k \in \{\pm1\}} \sum_i \boldsymbol{\mathcal{K}}\bigg(\frac{\mathbf{x}_{p, \mathcal{G}_0} - \mathbf{x}_{i, \mathcal{G}_k}}{\Delta x}\bigg) = 1, \label{new-interpolation-sum-to-1}
\end{equation}
\begin{equation}
    \frac{1}{2} \sum_{k \in \{\pm1\}} \sum_i \mathbf{x}_{i, \mathcal{G}_k}^\alpha \boldsymbol{\mathcal{K}}\bigg(\frac{\mathbf{x}_{\mathcal{G}_0} - \mathbf{x}_{i, \mathcal{G}_k}}{\Delta x}\bigg) = \mathbf{x}^\alpha_{\mathcal{G}_0} , \label{new-reconstruction-of-position}
\end{equation}
where $\mathbf{x}_{\cdots, \mathcal{G}_k}^\nu$ indicates it is a position in grid $\mathcal{G}_k$. We can compute the position in grid $\mathcal{G}_k$ through the canonical transformation function:  
\begin{equation}
    \mathbf{x}^\nu_{\mathcal{G}_k} = \mathbf{x}^\nu_{\mathcal{G}_0} - k \frac{1}{4} \Delta x \mathbf{e}^\nu,
\end{equation}
where $\mathbf{e}^\nu$ denotes the vector of $1$ in all dimensions. It is trivial to see that \autoref{new-interpolation-sum-to-1} holds since it follows directly from \autoref{interpolation-kernel-sum-to-1}. To prove \autoref{new-reconstruction-of-position}, we repeatedly apply the partition of unity property of $\mathcal{K}_1(x)$. See details in the supplemental document.

\subsection{Compact-Kernel MPM} \label{sec:ck-mpm}

With the introduction of the dual-grid system, we now present a variation of the traditional MPM pipeline. In this subsection, we introduce a modified version of the PIC scheme and demonstrate its linear momentum conservation property. In \autoref{sec:APIC-compatibility} and \autoref{sec:MLS-compatibility}, we further extend the PIC pipeline to support APIC and MLS, respectively, demonstrating angular momentum conservation.

In all the following schemes, particle positions are stored in grid $\mathcal{G}_0$, while the other two grids are treated as offset grids. For clarity and brevity, we define the notation $w_{i, p, k, t_n} \coloneqq \boldsymbol{\mathcal{K}}\bigg(\frac{\mathbf{x}_{i, \mathcal{G}_k, t_n} - \mathbf{x}_{p, \mathcal{G}_0, t_n}}{\Delta x}\bigg)$ to represent the kernel function.

\subsubsection{Transfer to grid}
For mass and momentum, the new particle-to-grid transfer equations in the dual grid system are:
\begin{equation}
    m_{i, \mathcal{G}_k, t_n} = \sum_p w_{i, p, \mathcal{G}_k, t_n} m_p,
\end{equation}
\begin{equation}
    m_{i, \mathcal{G}_k, t_n} \mathbf{v}_{i, \mathcal{G}_k, t_n}^\alpha = \sum_p w_{i, p, \mathcal{G}_k, t_n} m_p \mathbf{v}_{p, t_n}^\alpha,\label{eq:pic-momentum-transfer-to-grid}
\end{equation}
for $k=\pm1$. We are thus distributing the particle information onto both grids. An analogy to the classical MPM on one grid could be made by considering
\begin{equation*}
    m_{i, t_n} \mathbf{v}_{i, t_n}^\alpha = \sum_p w_{i, p, t_n} m_p \mathbf{v}_{p, t_n}^\alpha.
\end{equation*}

\subsubsection{Compute force}
We first obtain the first Piola-Kirchoff stress on grid $\mathcal{G}_0$:
\begin{equation}
    (\mathbf{P}_{p, \mathcal{G}_0, t_n})^\alpha_\beta = \frac{\partial \Psi}{\partial (\mathbf{F})^\beta_\alpha}(\mathbf{F}_{p, \mathcal{G}_0, t_n}).
\end{equation}
Observe that this step matches the classical MPM pipeline, since it is performed on the background grid $\mathcal{G}_0$. Then, we can compute the force on each grid $\mathcal{G}_k$ by:
\begin{equation}
    \mathbf{f}^\alpha_{i, \mathcal{G}_k, t_n} = \sum_p V_p (\mathbf{P}_{p, \mathcal{G}_0, t_n})^\alpha_\beta (\mathbf{F}_{p, \mathcal{G}_0, t_n})^\beta_\nu (\nabla w_{i, p, \mathcal{G}_k, t_n})^\nu.
\end{equation}
We again note that this expression resembles the classical MPM pipeline, differing only in that it is evaluated on grid $\mathcal{G}_k$.

\subsubsection{Grid update}

We then update two grids independently:
\begin{equation}
    m_{i, \mathcal{G}_k, t_n} \tilde{\mathbf{v}}^\alpha_{i, \mathcal{G}_k, t^{n + 1}} =  m_{i, \mathcal{G}_k, t_n} \mathbf{v}^\alpha_{i, \mathcal{G}_k, t_n} + \Delta t \mathbf{f}^\alpha_{i, \mathcal{G}_k, t_n},
\end{equation}
for each $k \in \{\pm1\}$. 

\subsubsection{Transfer to particles}

When transferring to particles, we gather information from both grids as implied by the proof of \autoref{new-reconstruction-of-position}:
\begin{equation}
    \mathbf{v}^\alpha_{p, t^{n + 1}} = \frac{1}{2m_p} \sum_{k \in \{\pm1\}} \sum_i w_{i, p, \mathcal{G}_k, t_n} m_{i, \mathcal{G}_k, t_n} \tilde{\mathbf{v}}^\alpha_{i,\mathcal{G}_k, t^{n + 1}}. \label{eq:velocity-grid-to-particle}
\end{equation}

\subsubsection{Update deformation gradient}
To compute forces in PIC scheme, we calculate the covariant derivative of velocity from both grids:

\begin{equation}
    \frac{\partial \mathbf{v}^\alpha_{p, \mathcal{G}_0, t^{n + 1}}}{\partial \mathbf{x}^\beta_{p, \mathcal{G}_0, t^{n + 1}}} = \frac{1}{2} \sum_{k \in \{\pm 1\}} \sum_i \tilde{\mathbf{v}}^\alpha_{i, \mathcal{G}_k, t^{n  +1}} ((\nabla w_{i, p, k, t_n})^T)_\beta.
\end{equation}
And, we update deformation gradient as in the traditional MPM pipeline:
\begin{equation}
    (\mathbf{F}_{p, \mathcal{G}_0, t^{n + 1}})^\alpha_\beta = \bigg(\boldsymbol{\delta}^\alpha_\nu + \Delta t \frac{\partial \mathbf{v}^\alpha_{p, \mathcal{G}_0, t^{n + 1}}}{\partial \mathbf{x}^\nu_{p, \mathcal{G}_0, t^{n + 1}}} \bigg) (\mathbf{F}_{p, \mathcal{G}_0, t^{n + 1}})^\nu_\beta.
\end{equation}

In our supplemental document, we prove that our CK-MPM with the PIC particle-grid transfer scheme conserves linear momentum.

\subsection{Compatibility with APIC} \label{sec:APIC-compatibility}  
The Affine Particle-In-Cell (APIC) \cite{jiang2015affine}
method is widely recognized for its ability to preserve angular momentum by incorporating affine matrices to capture additional particle information. In this subsection, we present the adapted formulation of APIC within our dual-grid system.

\subsubsection{Particle-to-Grid Transfer}  
The additional matrix $\mathcal{D}_{p, \mathcal{G}_0, t_n}$ plays an essential role in APIC for angular momentum conservation by capturing the affine motion of particles. In our CK-MPM, the computation of $\mathcal{D}_{p, \mathcal{G}_0, t_n}$ is adapted as follows:
\begin{align*}
    (\mathbf{D}_{p, \mathcal{G}_0, t_n})^\alpha_\beta = \frac{1}{2} \sum_{k \in \{\pm1\}} \sum_i &w_{i, p, \mathcal{G}_k, t_n} (\mathbf{x}^\alpha_{i, \mathcal{G}_k, t_n} - \mathbf{x}^\alpha_{p, \mathcal{G}_0, t_n}) \\&(\mathbf{x}^\beta_{i, \mathcal{G}_k, t_n} - \mathbf{x}^\beta_{p, \mathcal{G}_0, t_n})^T.
\end{align*}
Then, the particle-to-grid transfer in the dual grid system is similar to the PIC setting:
\begin{equation}
\begin{aligned}
    &m_{i, \mathcal{G}_k, t_n} \mathbf{v}^\alpha_{i, \mathcal{G}_k, t_n} =\\ &\sum_p w_{i, p, \mathcal{G}_k, t_n} m_p (\mathbf{v}^\alpha_{p, t_n} + (\mathbf{B}_{p, t_n})^\alpha_\nu ((\mathbf{D}_{p, t_n})^{-1})^\nu_\beta(\mathbf{x}^\beta_{i, \mathcal{G}_k, t_n} - \mathbf{x}^\beta_{p, t_n}).
\end{aligned}
\label{eq:apic-momentum-transfer-to-grid}
\end{equation}

\subsubsection{Grid-to-particle Transfer}
Another core component of the APIC scheme is the computation of the matrix $\mathbf{B}_{p, \mathcal{G}_0, t^{n + 1}}$. In our dual-grid system, we define the computation as:
\begin{equation}
\begin{aligned}
&(\mathbf{B}_{p, \mathcal{G}_0, t^{n + 1}})^\alpha_\beta =\\ &\frac{1}{2} \sum_{k \in \{\pm1\}} \sum_i w_{i, p, \mathcal{G}_k, t_n}  \tilde{\mathbf{v}}^\alpha_{i, \mathcal{G}_k, t^{n + 1}} ((\mathbf{x}_{i, \mathcal{G}_k, t_n} - \mathbf{x}_{p, \mathcal{G}_0, t_n})^T)_\beta.
\end{aligned}
\end{equation}

The rest of the pipeline and formulation will follow the same form as shown above in the PIC scheme. We prove that our CK-MPM will also conserve total linear and angular momentum with the adopted APIC scheme in our supplemental document.

\subsection{Compatibility with MLS-MPM} \label{sec:MLS-compatibility}
The Moving Least Squares (MLS) MPM method provides an accurate and efficient approximation of the traditional MPM algorithm and is fully compatible with APIC. Here, we demonstrate that our CK-MPM algorithm is also compatible with the MLS scheme, allowing it to benefit from the performance acceleration offered by MLS.

Using the dual-grid notation introduced earlier, let us consider two sets of samples of a scalar function $u: \mathbb{R}^3 \rightarrow \mathbb{R}$, taken at locations $\mathbf{x}_{i, \mathcal{G}_-}$ and $\mathbf{x}_{i, \mathcal{G}_+}$. Our goal is to approximate $u$ in a local neighborhood around a fixed point $\mathbf{x}$. This is achieved by performing a polynomial least-squares fit.

Let $\mathbf{P}: \mathbb{R}^3 \rightarrow \mathbb{R}^d$ denote the vector of polynomial basis functions, we aim to approximate the value of $u$ at a query point $\mathbf{z}_{\mathcal{G}_0}$ near $\mathbf{x}_{\mathcal{G}_0}$ using the following formulation:
\begin{equation}
    u(\mathbf{z}_{\mathcal{G}_0}) = (\mathbf{P}^T(\mathbf{z}_{\mathcal{G}_0} - \mathbf{x}_{\mathcal{G}_0}))_\beta \mathbf{c}^\beta(\mathbf{x}),\label{eq:mls-u-approx}
\end{equation}
where $\mathbf{c}^\beta(\mathbf{x})$ are the coefficients obtained from the least-squares fit. 

In our supplemental document, we show that we may approximate $u(\mathbf{z})$ with:
\begin{equation}
\begin{aligned}
    &u(\mathbf{z}_{\mathcal{G}_0}) =\\& \frac{1}{2} \sum_{k \in \{ \pm 1 \}} \sum_i \boldsymbol{\mathcal{K}}(\mathbf{x}_{i, \mathcal{G}_k} - \mathbf{x}_{\mathcal{G}_0}) \mathbf{P}^T(\mathbf{z}_{\mathcal{G}_0} - \mathbf{x}_{\mathcal{G}_0}) \mathbf{M}^{-1} \mathbf{P}(\mathbf{x}_{i, \mathcal{G}_k} - \mathbf{x}_{\mathcal{G}_0}) u_i \label{eq:mls-full-u-approx}.
\end{aligned}
\end{equation}
With \autoref{eq:mls-full-u-approx} above, we can construct a nodal shape function for $\mathbf{x}_{i, \mathcal{G}_k}$ as:
\begin{equation}
    \Phi_{i, \mathcal{G}_k}(\mathbf{z}_{\mathcal{G}_0}) = \boldsymbol{\mathcal{K}}(\mathbf{x}_{i, \mathcal{G}_k} - \mathbf{x}_{\mathcal{G}_0}) \mathbf{P}^T(\mathbf{z}_{\mathcal{G}_0} - \mathbf{x}_{\mathcal{G}_0}) \mathbf{M}^{-1}(\mathbf{x}_{\mathcal{G}_0}) \mathbf{P}(\mathbf{x}_{i, \mathcal{G}_k} - \mathbf{x}_{\mathcal{G}_0}).
\end{equation}
Thus, the approximation of $u$ at $\mathbf{z}_{\mathcal{G}_0}$ can be expressed as:  
\begin{equation}
    u(\mathbf{z}_{\mathcal{G}_0}) = \frac{1}{2} \sum_{k \in \{\pm1\}} \sum_i \Phi_{i, \mathcal{G}_k}(\mathbf{z}_{\mathcal{G}_0}) u_i,
\end{equation}  
where $\Phi_{i, \mathcal{G}_k}$ are the shape functions. If we choose the polynomial basis to consist of monomials in 3D, this formulation aligns with the original MLS-MPM approximation \cite{hu2018moving}. However, note that our momentum matrix $\mathbf{M}(\mathbf{x}_{\mathcal{G}_0})$ cannot be simplified into a closed-form expression due to the dual-grid system. Consequently, we compute the momentum matrix for each particle independently at every timestep.

\section{Implementation}\label{sec:implementation}  
% In this section, we describe the implementation details of our CK-MPM scheme and outline optimizations to the computational pipeline.

\subsection{CUDA}  
We adopt the state-of-the-art open-source GPU MPM framework proposed by \citet{wang2020massively} as the foundation for our implementation. Specifically, we utilize the Grid-to-Particle-to-Grid (G2P2G) algorithm and the Array-of-Structures-of-Arrays (AoSoA) data structure described in their work, incorporating our modified compact kernel and dual-grid system. 

%One limitation of their implementation is that test-case parameters are hard-coded, making configuration adjustments labor-intensive. To address this, we follow the policy-based design pattern and extensively use C++ templates and `constexpr` to encode material parameters and grid settings separately in each test file, improving flexibility and ease of use while maintaining runtime efficiency.

\paragraph{Dual-Grid Storage}  
The original grid data structure in \citet{wang2020massively} stores information for each grid block (of size $4 \times 4 \times 4$ cells) in a contiguous memory segment. Within this block granularity, grid attributes are grouped in a \textit{Structure-of-Arrays} (SoA) layout, enabling efficient coalesced read/write access to GPU global memory. We extend this scheme by additionally grouping two blocks from different grids.

%We extend this scheme to a four-level index system, descending in the order of block index, grid index, attribute index, and cell index. In our dual-grid system, two grid blocks (from $\mathcal{G}_{+}$ and $\mathcal{G}_{-}$) share the same block index if the bottom-left cell of the block in $\mathcal{G}_{+}$ is offset by half of the cell width in all dimensions relative to the corresponding block in $\mathcal{G}_{-}$. By treating paired grid blocks as a single entity and introducing the grid index as an additional level, we efficiently manage data access within the dual-grid framework. 
% \todo{can provide a bit more details, is ours 3 by 3 by 3? why?}

\paragraph{Particle Block}  
In the original implementation by \citet{wang2020massively}, particle blocks are offset by two cells from their corresponding grid blocks to ensure that the particles stay in the same grid blocks after CFL-bounded advection in G2P2G. In our dual-grid scheme, we modify this relationship to account for the additional grid. Without loss of generality, we designate grid blocks from $\mathcal{G}_{-}$ as the reference blocks for defining the new particle blocks. This ensures compatibility with the dual-grid configuration. 
% \todo{can provide a bit more details, are we also offset by two cells? etc.}

% \paragraph{Kernel Optimization}  
% The G2P2G algorithm requires grid data to be loaded into shared memory. In our dual-grid scheme, this results in twice the shared memory usage compared to \citet{wang2020massively}, which can limit the number of parallel CUDA blocks running concurrently due to the finite shared memory available per Streaming Multiprocessor (SM). To address this, we optimize the memory usage by reducing the number of grid cells loaded into shared memory.

% As illustrated in \autoref{fig:kernels}, under the CFL restriction, particles in a particle block will never reach the top, right, or back layers of their corresponding grid blocks in $\mathcal{G}_{+}$. Exploiting this observation, we reduce the number of cells loaded from $2 \times 8^3 = 1024$ to $8^3 + 7^3 = 855$, achieving a $16.5\%$ reduction in shared memory usage per CUDA block. For each cell, 3 floating point numbers (floats) are required for velocity in the G2P process and 4 floats for mass and momentum in the P2G process. Consequently, the shared memory usage decreases from $1024 \cdot 7 \cdot 4 = 28,672$ bytes to $855 \cdot 7 \cdot 4 = 23,940$ bytes.
% On an NVIDIA RTX 3090 GPU with CUDA compute capability 8.6, the maximum available shared memory per SM is 100 KB. This reduction increases the maximum number of concurrent CUDA blocks per SM from 3 to 4, resulting in a $33.3\%$ improvement and effectively enhancing the performance of our dual-grid implementation.

\subsection{Taichi Implementation}
In addition to the CUDA implementation, we also provide a Taichi-based implementation of our compact kernel. This implementation includes the standard P2G, grid update, and G2P processes, in contrast to the fused G2P2G kernel. For comparison, we have also implemented the same procedure using a quadratic kernel. The entire implementation and comparison consist of fewer than 300 lines of Python code.

\section{Experiment}
% \subsection{Overview}
In the following subsections, we first validate our CK-MPM by demonstrating that it conserves linear and angular momentum (\autoref{sec:unit-test}). Next, we compare the performance and behavior of CK-MPM against the state-of-the-art open-source GPU MPM framework \cite{wang2020massively} (\autoref{sec:comparisons}). Finally, we showcase the robustness and versatility of our method through stress tests involving large-scale scenes and complex geometries (\autoref{sec:stress-tests}).

Except for stress tests, all experiments are conducted on a machine equipped with an Intel Core i9-12900KF CPU and an NVIDIA RTX 3090 GPU. The system runs CUDA 12.4 with NVIDIA driver version 550.54. For code compilation, we use gcc/g++ 12.3 with the C++20 standard enforced.

\subsection{Unit tests}\label{sec:unit-test}
In this subsection, we evaluate the conservation properties of CK-MPM using the Fixed Corotated hyperelasticity model \cite{stomakhin2012energetically} with $E = 10^6\,\text{Pa}$, $\nu = 0.4$, and $\rho = 10^3\,\text{kg/m}^3$ for both test cases. To reduce numerical errors, double-precision floating-point numbers are used in the simulation, and we assume unit particle mass when calculating the momentum.

\paragraph{Conservation of Linear Momentum}
\begin{figure}
    \centering
    %  \begin{subfigure}[b]{0.65\linewidth}
    %     \includegraphics[height=0.7\linewidth]{figures/conservation_linear_momentum.png}
    % \end{subfigure}%
    %  \begin{subfigure}[b]{0.4\linewidth}
    %     \includegraphics[height=0.75\linewidth]{figures/colliding_sphere_158_clipped.jpg}
    % \end{subfigure}%
    \includegraphics[width=\linewidth]{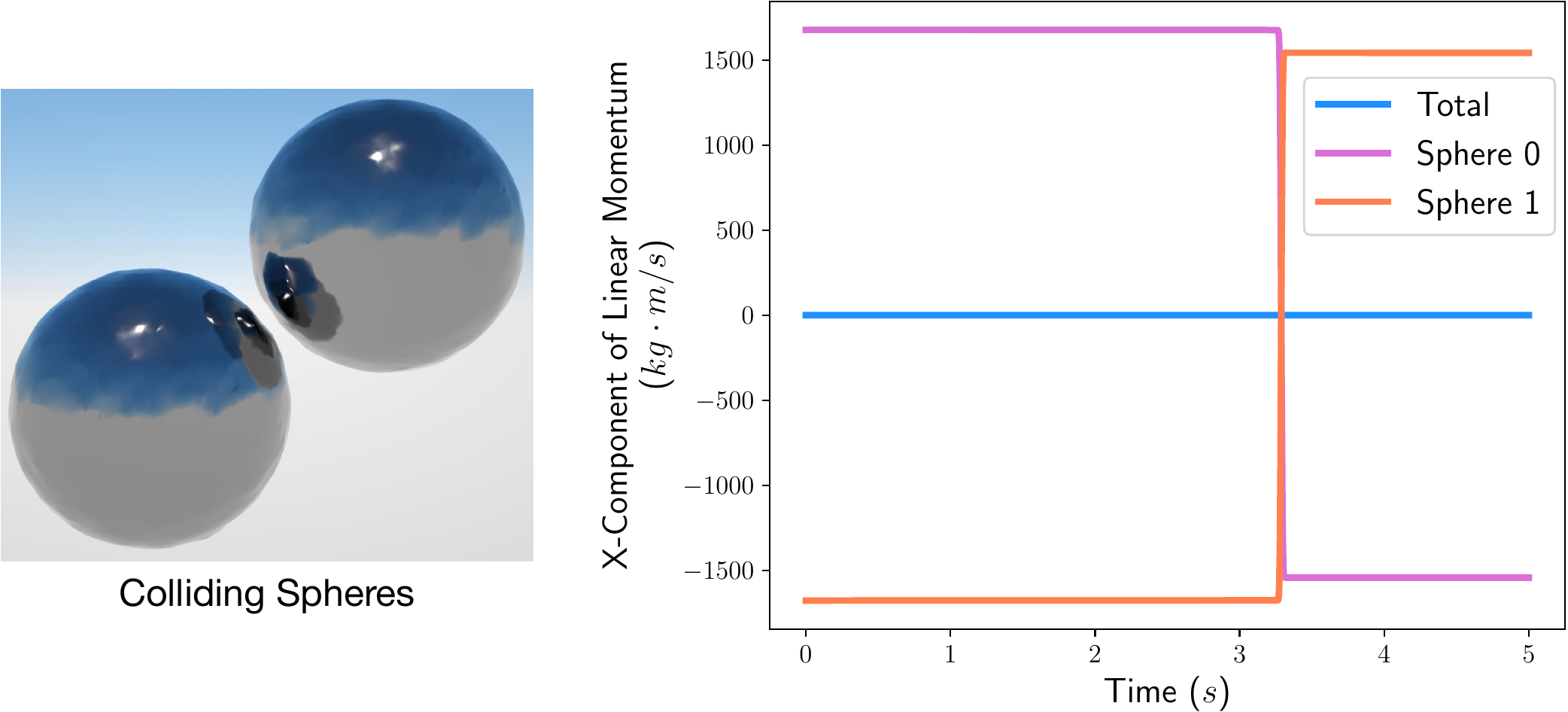}
    \caption{\textbf{Linear Momentum Conservation.} Plot of the conserved $x-$component total linear momentum (blue) and the linear momentum of two spheres colliding (green and orange).}
    \label{fig:unit-test-conservation-of-linear-momentum}
\end{figure}
To verify the conservation of linear momentum, we simulate a standard test case involving two colliding spheres. Each sphere has a radius of $\frac{10}{256}\,\text{m}$ and is discretized with 8 particles per cell, using a cell spacing of $dx = \frac{1}{256}\,\text{m}$. This results in $33,552$ particles per sphere. The spheres are initially located at $(\frac{32}{256}, \frac{32}{256}, \frac{32}{256})\,\text{m}$ and $(\frac{128}{256}, \frac{128}{256}, \frac{128}{256})\,\text{m}$, with initial velocities of $(0.05, 0.05, 0.05)\,\text{m/s}$ and $(-0.05, -0.05, -0.05)\,\text{m/s}$, respectively. Thus, the norm of the total linear momentum for each sphere is approximately $2905.69\,\text{kg} \cdot \text{m/s}$, while the total initial linear momentum of the system is zero.
We simulate the system for 5 seconds and measure the ratio of the norm of the total linear momentum of the system to the initial linear momentum of one sphere. As shown in \autoref{fig:unit-test-conservation-of-linear-momentum}, the $x-$component total linear momentum remains nearly zero (the maximum value is $0.0309\,\text{kg} \cdot \text{m/s}$, achieving an $L_\infty$-error rate of $\frac{0.0309\,\text{kg} \cdot \text{m/s}}{2905.69\,\text{kg} \cdot \text{m/s}} \approx 1.063 \cdot 10^{-5}$), while the momentum of two spheres interchanges after collision, demonstrating the strong capability of our method to conserve total linear momentum.

\paragraph{Conservation of Angular Momentum}
\begin{figure}
    % \begin{subfigure}[b]{0.65\linewidth}
    %     \includegraphics[height=0.7\linewidth]{figures/conservation_angular_momentum.png}
    % \end{subfigure}%
    % \begin{subfigure}[b]{0.35\linewidth}
    %     \includegraphics[height=\linewidth]{figures/rotating_rod_0_clipped.jpg}
    % \end{subfigure}
    \includegraphics[width=\linewidth]{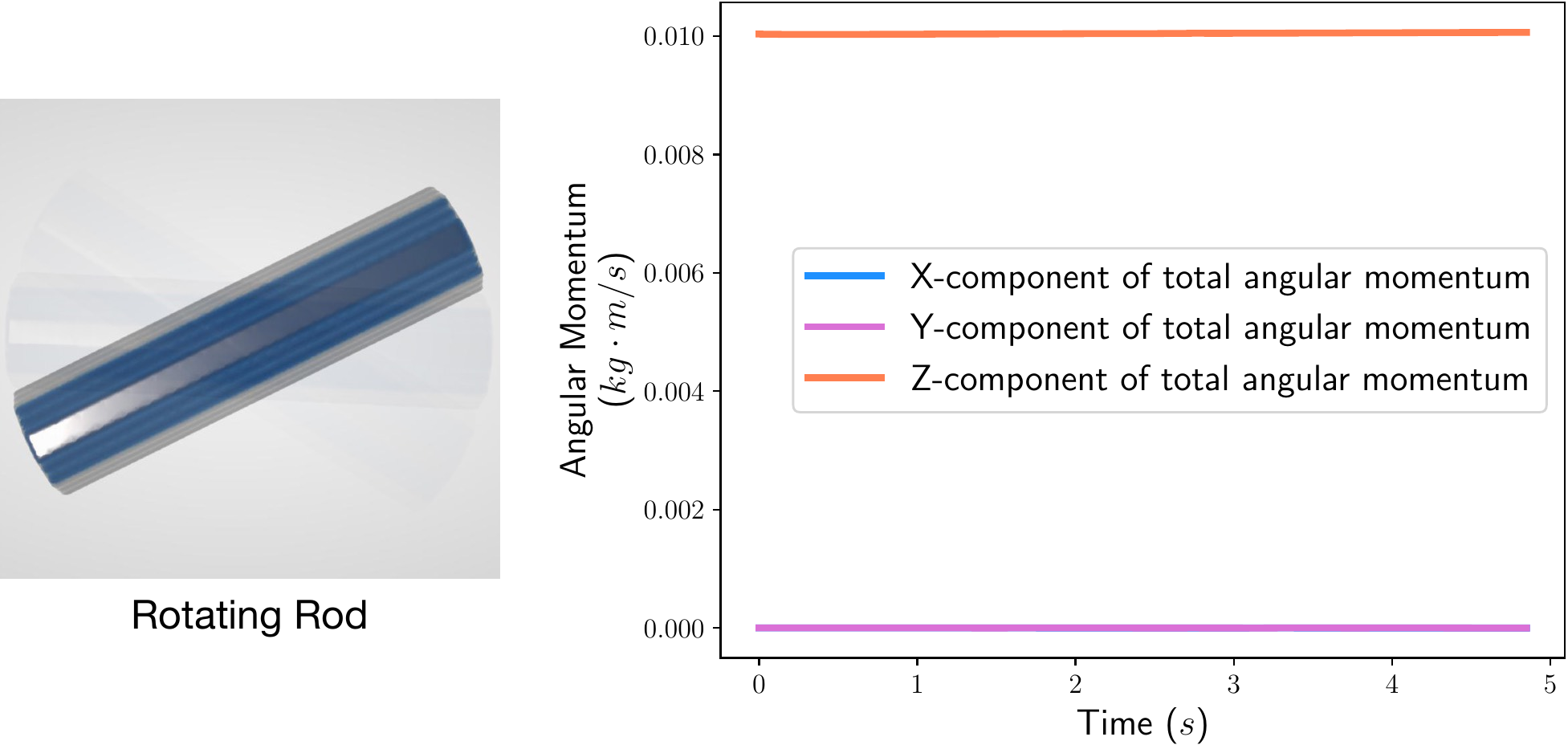}
    \caption{\textbf{Angular Momentum Conservation.} Evolution of x-,y-,z-component of the total angular momentum of a rotating bar simulation; all components are accurately preserved over time. }
    \label{fig:unit-test-conservation-of-angular-momentum}
\end{figure}
Next, we evaluate the conservation of angular momentum using a rotating rod test case. The rod is initialized as a cylinder with a radius of $\frac{5}{256}\,\text{m}$ and a length of $\frac{40}{256}\,\text{m}$, with its central axis aligned with the $y$-axis. The center of the cylinder is positioned at $(\frac{128}{256}, \frac{128}{256}, \frac{128}{256})\,\text{m}$. For each particle at a $y$-direction distance $\Delta r_p$ from the center of the cylinder, we assign velocities of $(\pm\frac{256 \Delta r_p}{20}, 0, 0)\,\text{m/s}$ on the two sides so that the endpoints have velocities of $(\pm 1, 0, 0)\,\text{m/s}$, allowing the rod to start rotating.
We measure the evolution of the $x$-, $y$-, and $z$-component total angular momentum over 5 seconds. As shown in \autoref{fig:unit-test-conservation-of-angular-momentum}, these components remain nearly constant throughout the simulation (the maximum deviations of the x and y components from $0$ are $4.6 \cdot 10^{-7} \,\text{kg} \cdot \text{m/s}$ and $2.5 \cdot 10^{-6} \,\text{kg} \cdot \text{m/s}$, and the maximum deviation of the z component from $0.01 \,\text{kg} \cdot \text{m/s}$ is $6 \cdot 10^{-5} \,\text{kg} \cdot \text{m/s}$, resulting in an $L_\infty$-error rate of $\frac{6 \cdot 10^{-5}\,\text{kg} \cdot \text{m/s}}{0.01 \,\text{kg} \cdot \text{m/s}} = 6 \cdot 10^{-3}$). These results confirm that our method effectively conserves total angular momentum during the simulation.

\subsection{Comparisons} \label{sec:comparisons}

In this subsection, we present several test cases to evaluate the performance improvements and behavioral differences between our CK-MPM and the traditional quadratic B-spline MPM.

\subsubsection{Efficiency}  
As discussed in \autoref{sec:implementation}, our method incorporates an adapted version of the G2P2G algorithm \cite{wang2020massively}. Since G2P2G represents the most computationally intensive operation in each substep (around 80\% of the total computation time), we begin by comparing the speed achieved in G2P2G computations using our CK-MPM approach.

\begin{table}
\caption{Average G2P2G kernel time per substep (in milliseconds) over 100 frames (48 frames per second) for all examples, simulated using MLS-MPM in both \citet{wang2020massively} and our implementation. All simulations run with a grid resolution of \((256, 256, 256)\).}
\label{table:g2p2g-speedup}
\begin{tabular}{llllllllll}
\toprule
Example & \citet{wang2020massively} & Ours \\
\midrule 
Two Dragons Falling & 0.7 & 0.64 \\
Fluid Dam Break (4 Million) & 3.7 &  2.9  \\
Fluid Dam Break (8 Million) & 7.2 &  5.7  \\
Sand Armadillos & 0.74 & 0.66 \\
\bottomrule
\end{tabular}
\end{table}

\begin{table*}
\caption{Breakdown of average CUDA kernel execution times per substep over the first 100 frames (in milliseconds). Here, \textit{Copy Grid Block} duplicates grid block data for the subsequent timestep; \textit{Update Partition} updates the sparse grid structure following particle advection; \textit{Update Buffer} refreshes particle-related data; \textit{Activate Blocks} registers sparsity information for grid blocks containing particles.}
\label{table:time-breakdown}
\begin{tabular}{lcccccc}
\toprule
Example & G2P2G & Grid Update & Copy Grid Block & Update Partition & Update Buffer & Activate Blocks \\
\midrule 
Two Dragons Falling            & 0.64 (77.5\%) & 0.017 (2.1\%) & 0.026 (3.1\%) & 0.073 (8.8\%) & 0.026 (3.1\%) & 0.044 (5.4\%) \\ 
Fluid Dam Break (4 Million)    & 2.90 (87.5\%) & 0.055 (1.7\%) & 0.095 (2.9\%) & 0.160 (4.8\%) & 0.060 (1.8\%) & 0.044 (1.3\%) \\ 
Fluid Dam Break (8 Million)    & 5.70 (89.3\%) & 0.092 (1.4\%) & 0.160 (2.5\%) & 0.270 (4.2\%) & 0.110 (1.7\%) & 0.049 (0.9\%) \\ 
Sand Armadillos                & 0.66 (72.8\%) & 0.027 (3.0\%) & 0.042 (4.6\%) & 0.098 (10.8\%) & 0.035 (3.9\%) & 0.044 (4.9\%) \\ 
\bottomrule
\end{tabular}
\end{table*}

We compare four of our test cases against the implementation by \citet{wang2020massively}. For both methods, we measure and calculate the average computation time of the G2P2G kernel. The experiments are carefully designed to ensure comparable behavior across the first 100 frames, enabling a consistent and fair performance evaluation.

\paragraph{Two Dragons Falling}  
We initialize two dragon models with zero initial velocity, discretized into a total of 775,196 particles. The dragons are simulated using a Fixed Corotated hyperelasticity model with Young’s modulus $E = 6 \times 10^5\,\text{Pa}$ and Poisson’s ratio $\nu = 0.4$. The gravity is set to $-4\,\text{m/s}^2$.

\paragraph{Fluid Dam Break}  
% \begin{figure}[ht]
% 		\centering
% 		\includegraphics[width=0.5\columnwidth, trim=0 0 0 0, clip]{figures/dam_break_21_clipped.jpg}
% 		\caption{\textbf{Fluid Dam Break.}}
%         \label{fig:dam-break}
% \end{figure}
We simulate the same test case with both 4,175,808 and 8,994,048 particles that are initialized with a uniform distribution in a cuboid to simulate a single dam break. The fluid is modeled using the equation of state formulation \cite{monaghan1994simulating,tampubolon2017multi}, or the $J$-based fluid model, with parameters: $B = 10 Pa$, $\gamma = 7.15$, and viscosity $\mu = 0.1$. Gravity is set to $-9.8\,\text{m/s}^2$. 
% As shown in \autoref{fig:dam-break}, the simulation captures the realistic flow dynamics as the fluid rushes outward after the dam break.

\begin{figure*}
    \begin{subfigure}[b]{0.33\linewidth}
           \includegraphics[width=\linewidth,trim={5cm 2cm 5cm 0},clip]{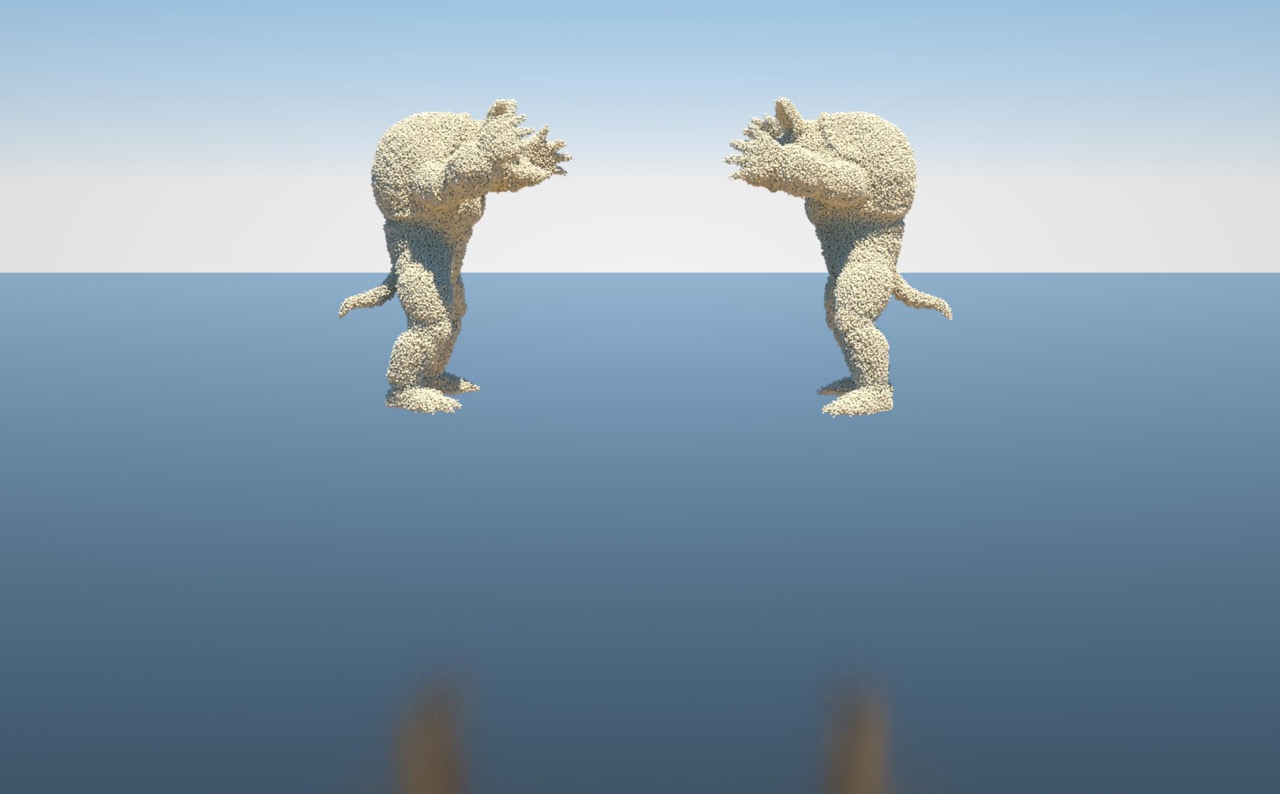}
    \end{subfigure}%
    \hfill%
    \begin{subfigure}[b]{0.33\linewidth}
           \includegraphics[width=\linewidth,trim={5cm 2cm 5cm 0},clip]{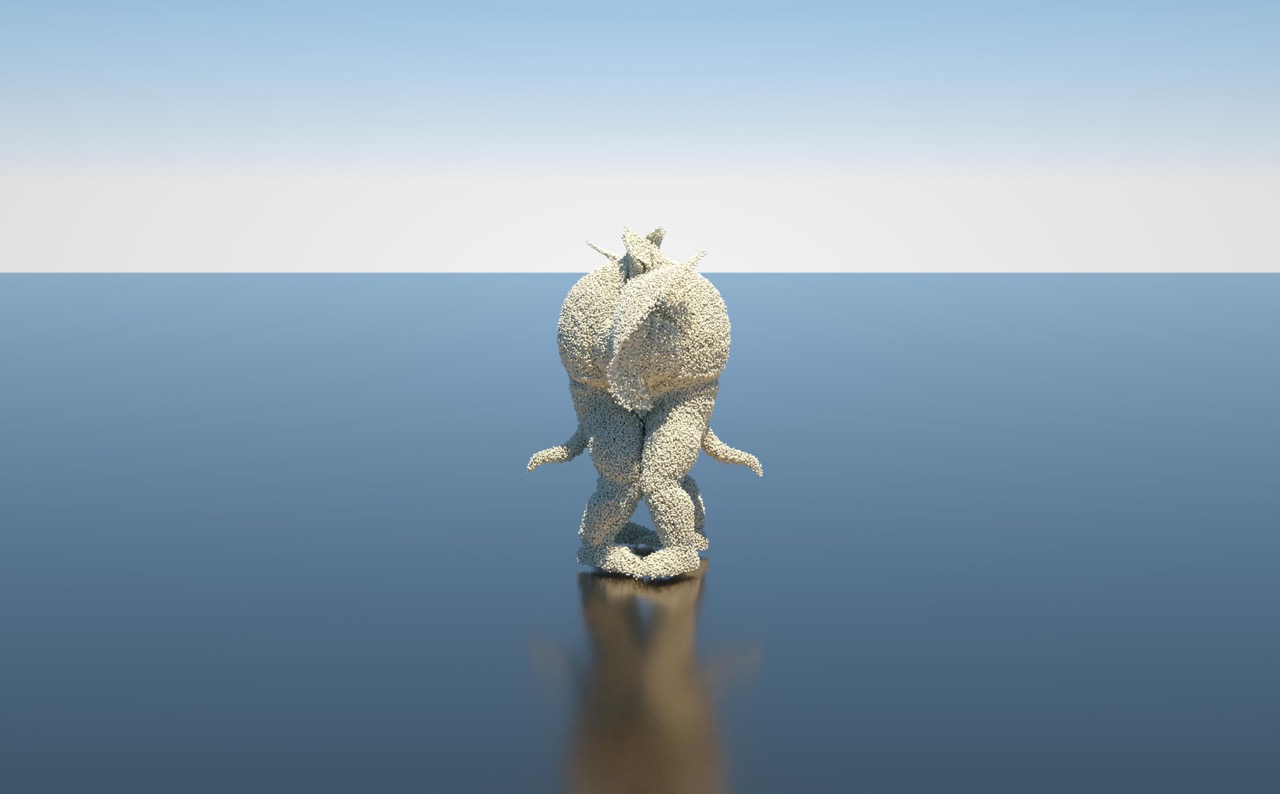}
    \end{subfigure}%
    \hfill%
    \begin{subfigure}[b]{0.33\linewidth}
        \includegraphics[width=\linewidth,trim={5cm 2cm 5cm 0},clip]{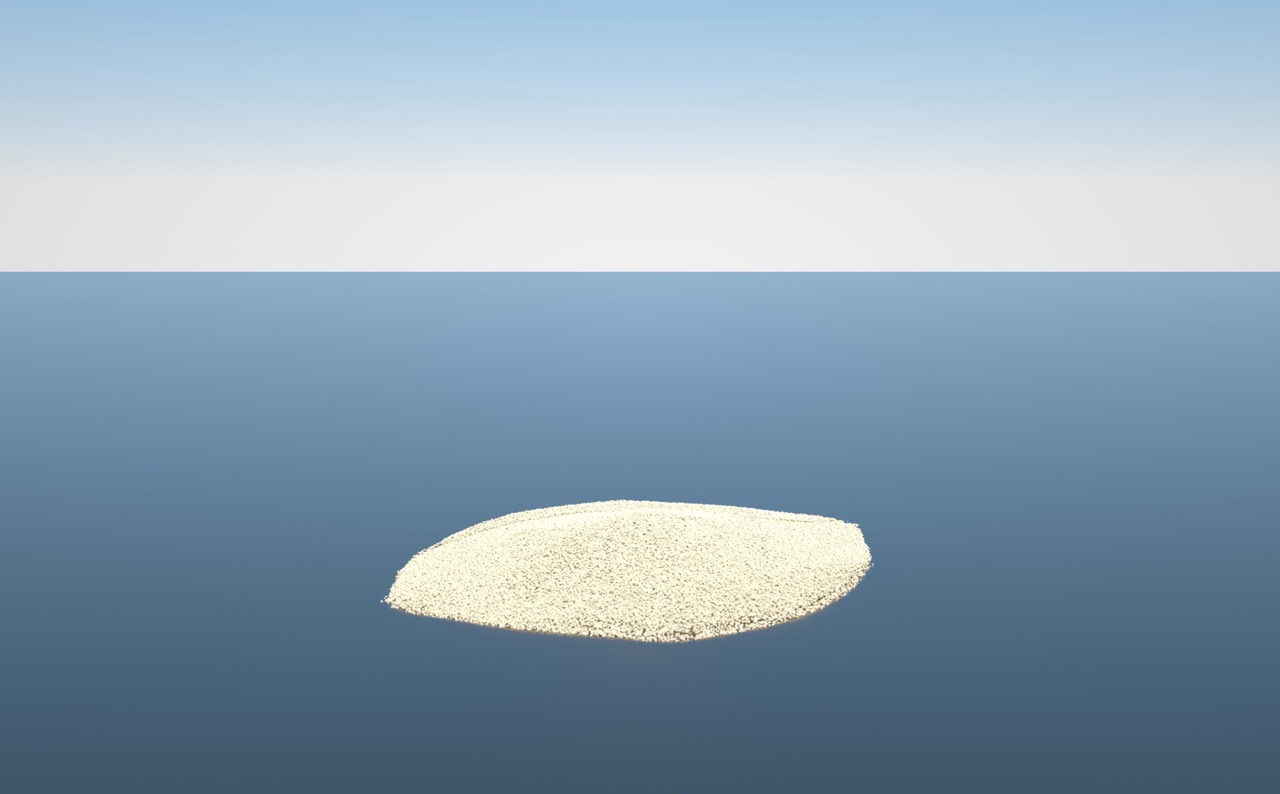}
    \end{subfigure}
    \caption{\textbf{Sand Armadillos.}}
    \label{fig:sand-armadillos}
\end{figure*}

\paragraph{Sand Armadillos}  
Two armadillo models are initialized with opposing initial velocities of $(0, 0, -0.5)$ and $(0, 0, 0.5)\,\text{m/s}$, respectively, and discretized into 511,902 particles. The simulation uses the Drucker-Prager elastoplasticity model with parameters $E = 10^4\,\text{Pa}$, $\nu = 0.4$, and friction angle of $30$\textdegree. Gravity is set to $-2\,\text{m/s}^2$. As shown in \autoref{fig:sand-armadillos}, the simulation captures the collapse and dispersion of the armadillos with fine granularity, demonstrating realistic sand behavior.
\\

In \autoref{table:g2p2g-speedup}, we observe that our compact kernel can achieve a comparable speed in G2P2G to the original quadratic kernel and demonstrated a slight speedup (around 10\%) in each G2P2G substep.
As discussed in \autoref{sec:implementation}, we have also provided a Taichi implementation of compact kernel and quadratic kernel MPM. We compare the performance of the standard P2G, Grid update, and G2P pipeline using PIC scheme of MPM. As shown in \autoref{table:taichi-speedup}, we observe a $1.5\times$ speedup on average for the standard pipeline of MPM. 
The speedup observed with Taichi and PIC aligns more closely with theoretical expectations, as our kernel reduces the number of nodes associated with each particle during the P2G step by 40\%. In contrast, the less pronounced speedup with CUDA and MLS is likely attributable to the matrix inversion required in the MLS formulation and the lack of extensive GPU-specific optimizations in our implementation. Additionally, \autoref{table:time-breakdown} provides a breakdown of kernel timings for each example, confirming that G2P2G remains the dominant cost across all configurations.

\begin{table}
\caption{Examples simulated with Taichi and PIC scheme. The total runtimes in seconds are compared between compact kernel and quadratic kernel. All examples are simulated for 100 frames with 48 frames per second and a grid resolution of $(256, 256, 256)$.}
\label{table:taichi-speedup}
\begin{tabular}{llllllllll}
\toprule
Example & Compact Kernel & Quadratic Kernel & Speedup\\
\midrule 
Jelly Falling & 733.4s & 1088.7s & 1.48$\times$\\
Fruit Falling & 165.8s &  242.8s  & 1.46$\times$\\
\bottomrule
\end{tabular}
\end{table}

\begin{figure}
    \includegraphics[width=0.75\linewidth,clip]{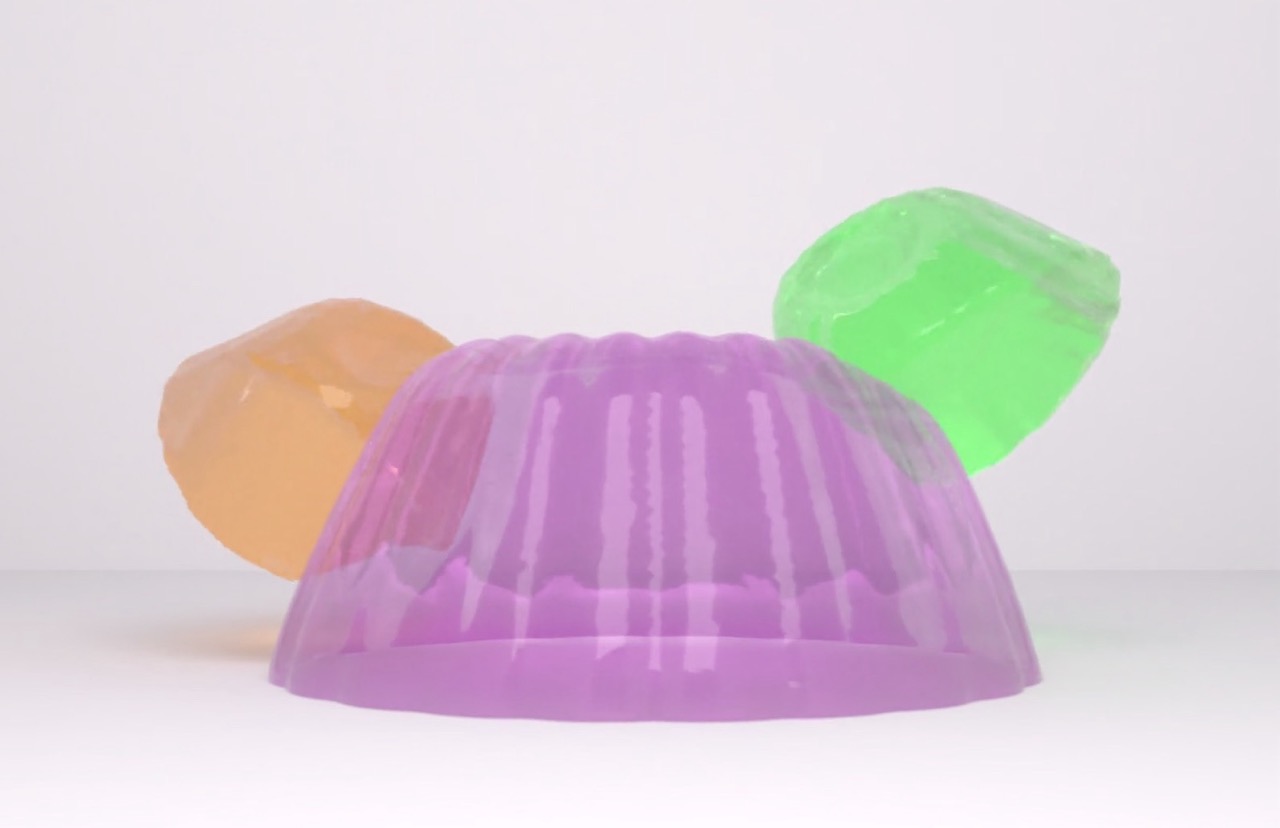}
        \caption{\textbf{Jelly Falling}.}
    \label{fig:jelly}
\end{figure}

We also compare the memory footprints of MPM implementations using quadratic kernels (as in \cite{wang2020massively}) and our compact kernel approach. Particle representations remain identical in both cases, but our method requires maintaining an additional staggered grid, effectively doubling the memory usage for grid storage. To quantify this difference, we report detailed memory usage across several test cases in \autoref{tab:mem-breakdown}.
These results highlight a trade-off between memory usage and improved computational efficiency and accuracy of our method. In large-scale simulations where memory is a limiting factor, it would be valuable to explore specialized implementations that mitigate memory overhead. For example, decoupling the Particle-to-Grid (P2G) and Grid-to-Particle (G2P) processes and performing transfers for each grid independently could help reduce memory requirements.
\begin{table}[h]
  \centering
  \caption{Comparison of grid memory usage across selected examples, assuming a maximum of 64 particles per cell. ‘Q’ denotes the use of the quadratic kernel, while ‘C’ represents the compact kernel.}
  \label{tab:mem-breakdown}
  \begin{tabular}{lcc}
    \toprule
    \textbf{Example}  &
    \textbf{\# Grid Blocks} &
    \textbf{Memory Usage}\\
    \midrule
    Two Dragons Falling (Q) & 6000 & 11.7\ MB  \\
    Two Dragons Falling (C) & 6000 & 23.4\ MB  \\
    Fluid Dam Break (Q) & 50000 & 97.7\ MB\\
    Fluid Dam Break (C) & 50000 & 195.3\ MB\\
    % --- add additional rows as needed ---
    \bottomrule
  \end{tabular}
\end{table}

\subsubsection{Behavior Analysis}  
We compare the behavioral differences between MPM simulations using our compact kernel and the standard quadratic kernel.

\paragraph{Fracture Behavior}  
Due to the smaller kernel radius, MPM simulations using our compact kernel are more prone to generating fractures upon collision. To illustrate this behavioral difference, we present two test cases.

The first test case, \textit{Pumpkin Smash}, involves two pumpkin models, with one resting on the ground and the other falling from above. The pumpkins are simulated using the Non-Associated Cam Clay (NACC) \cite{wolper2019cd} model with parameters $E = 2000 Pa$, $\nu = 0.39$, $\alpha_0 = -0.04$, $\beta = 2$, $\xi = 3$, and $M = 2.36$. The initial drop height of the falling pumpkin is set to $\frac{100}{128}m$, with a grid resolution of $(256, 256, 256)$, $\Delta x = \frac{1}{128}m$, and gravity $g = -2\,\text{m/s}^2$. 

\begin{figure}
    \begin{subfigure}[b]{0.49\linewidth}
        \includegraphics[width=\linewidth,trim={0 5cm 0 10cm},clip]{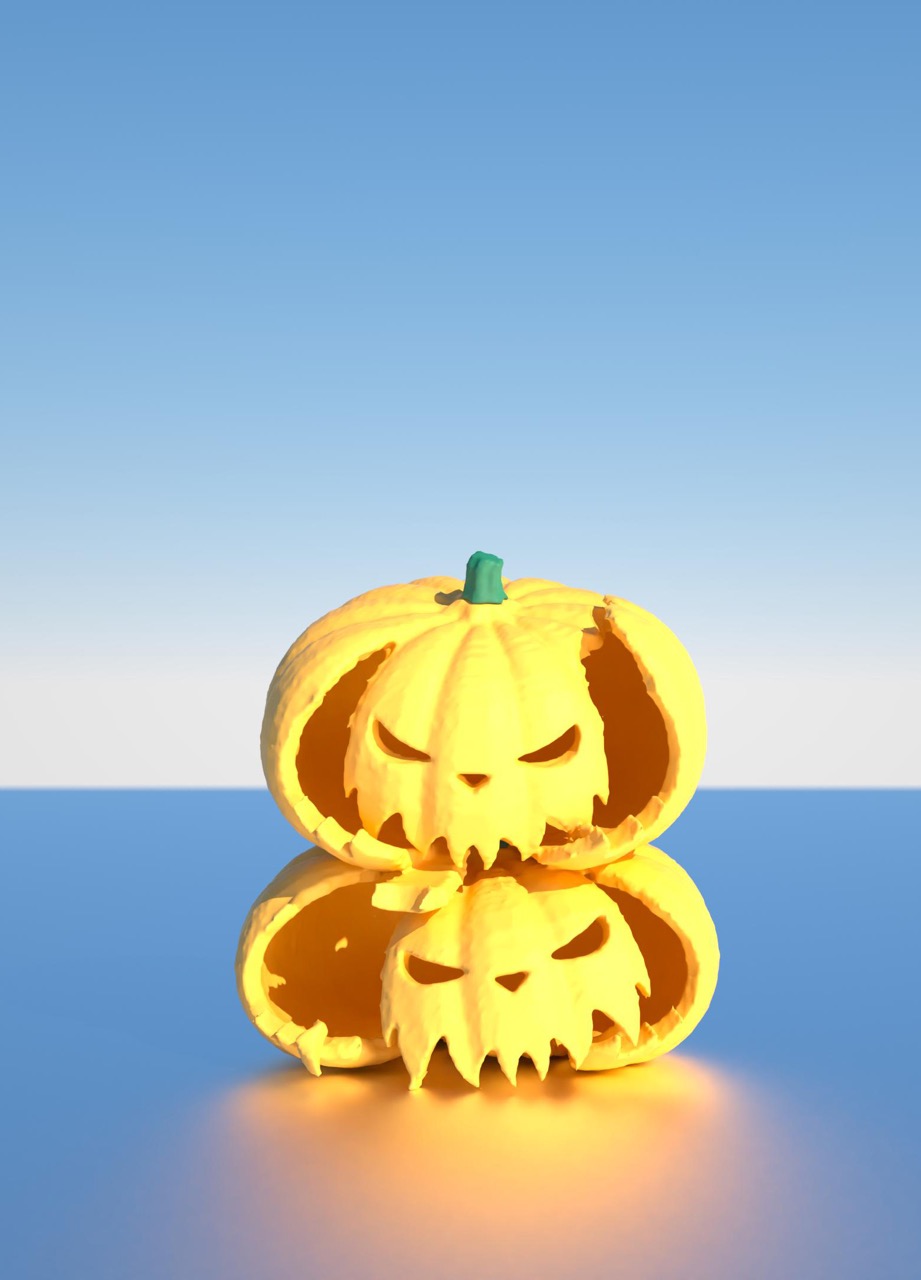}
        \caption{Compact Kernel}
    \end{subfigure}%
    \begin{subfigure}[b]{0.49\linewidth}
    \centering
        \includegraphics[width=\linewidth,trim={0 5cm 0 10cm},clip]{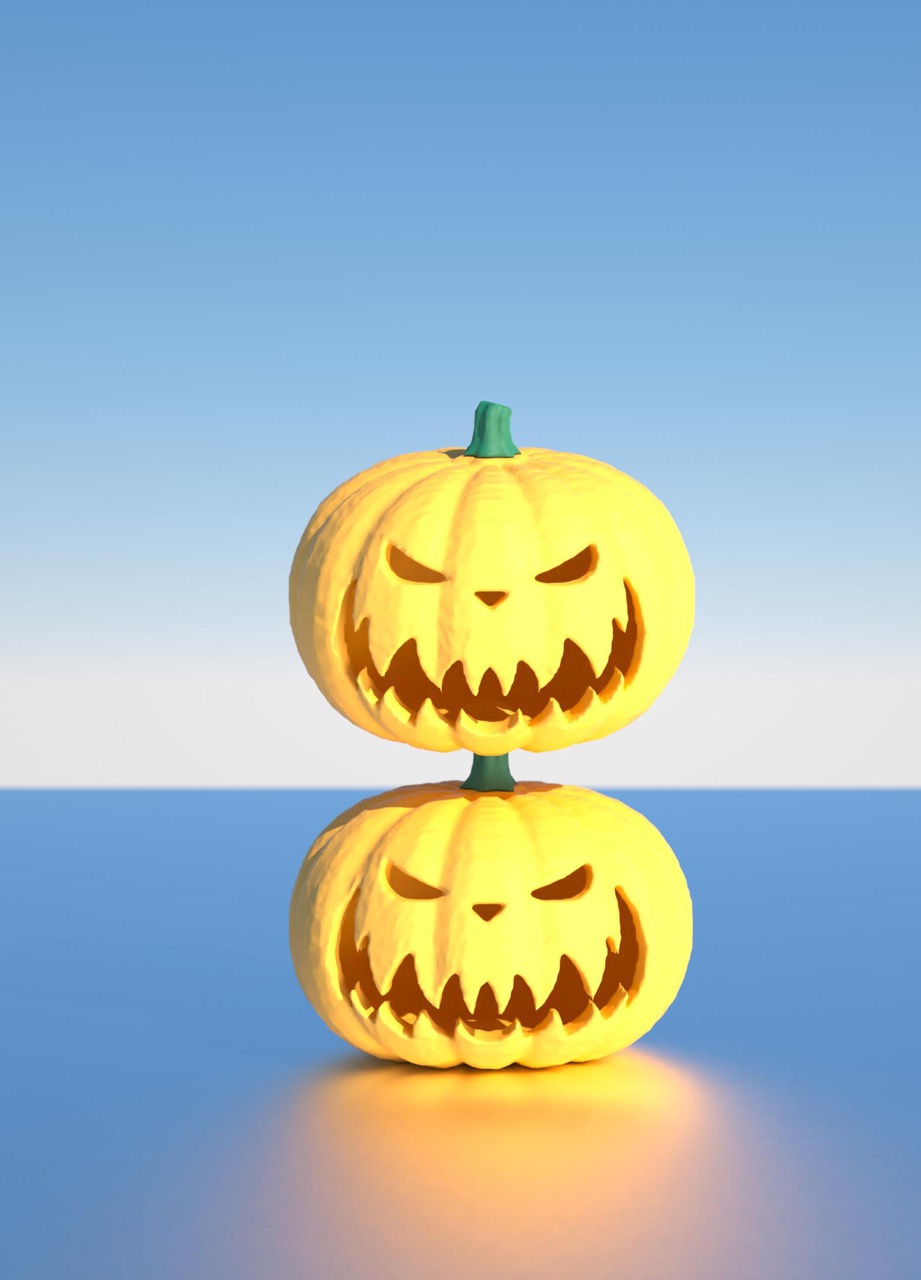}
        \caption{Quadratic Kernel}
    \end{subfigure}
    \caption{\textbf{Pumpkin Smash.}}
    \label{fig:pumpkin}
\end{figure}
As shown in \autoref{fig:pumpkin}, the pumpkins simulated with the compact kernel produce significant fractures, while those simulated with the quadratic kernel exhibit elastic behavior, bouncing back after a brief period of compression.

\begin{figure}
    \begin{subfigure}[b]{0.25\textwidth}
        \centering
        \includegraphics[height=4cm]{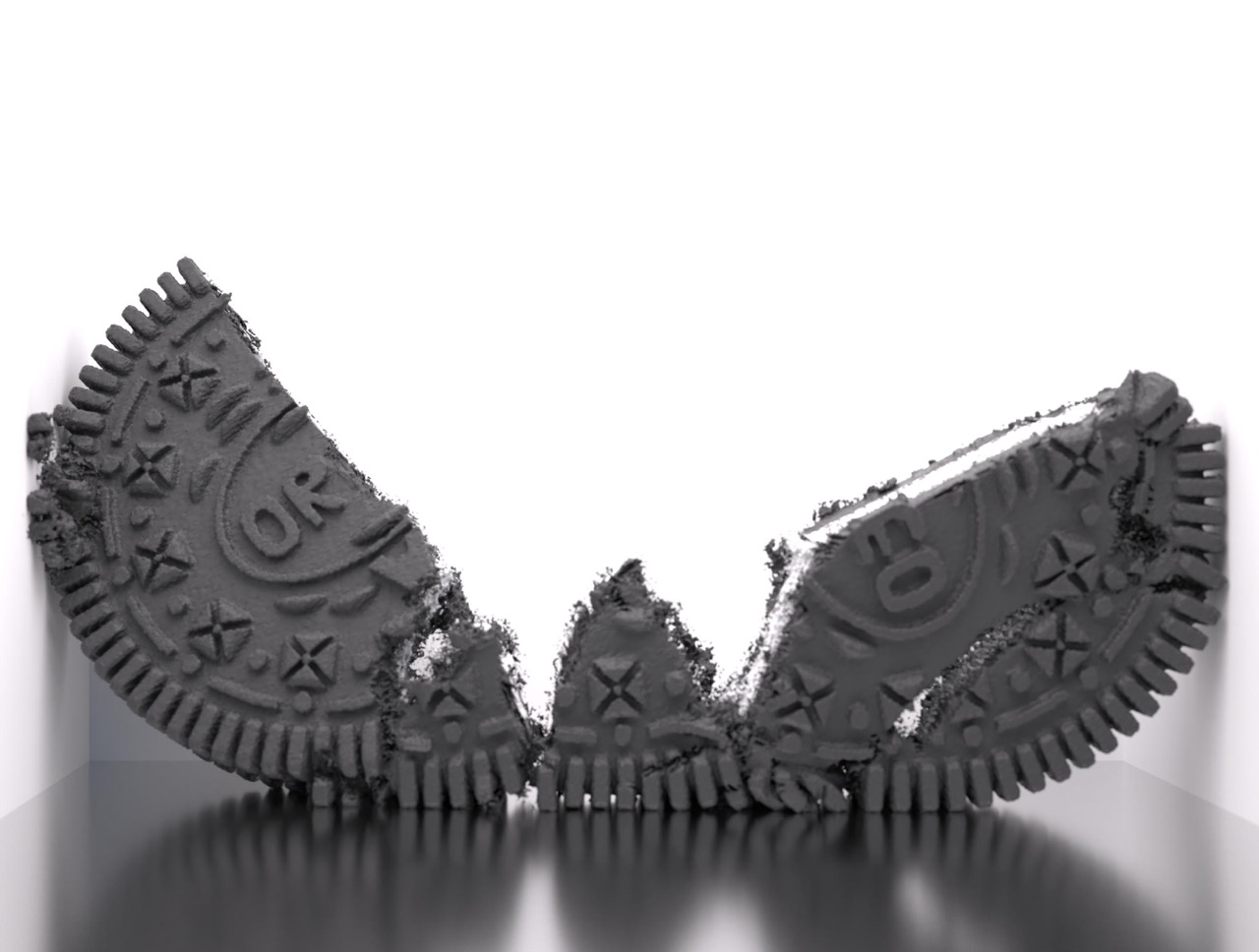}
        \caption{Compact Kernel}
    \end{subfigure}%
    \begin{subfigure}[b]{0.25\textwidth}
        \centering
        \includegraphics[height=4cm]{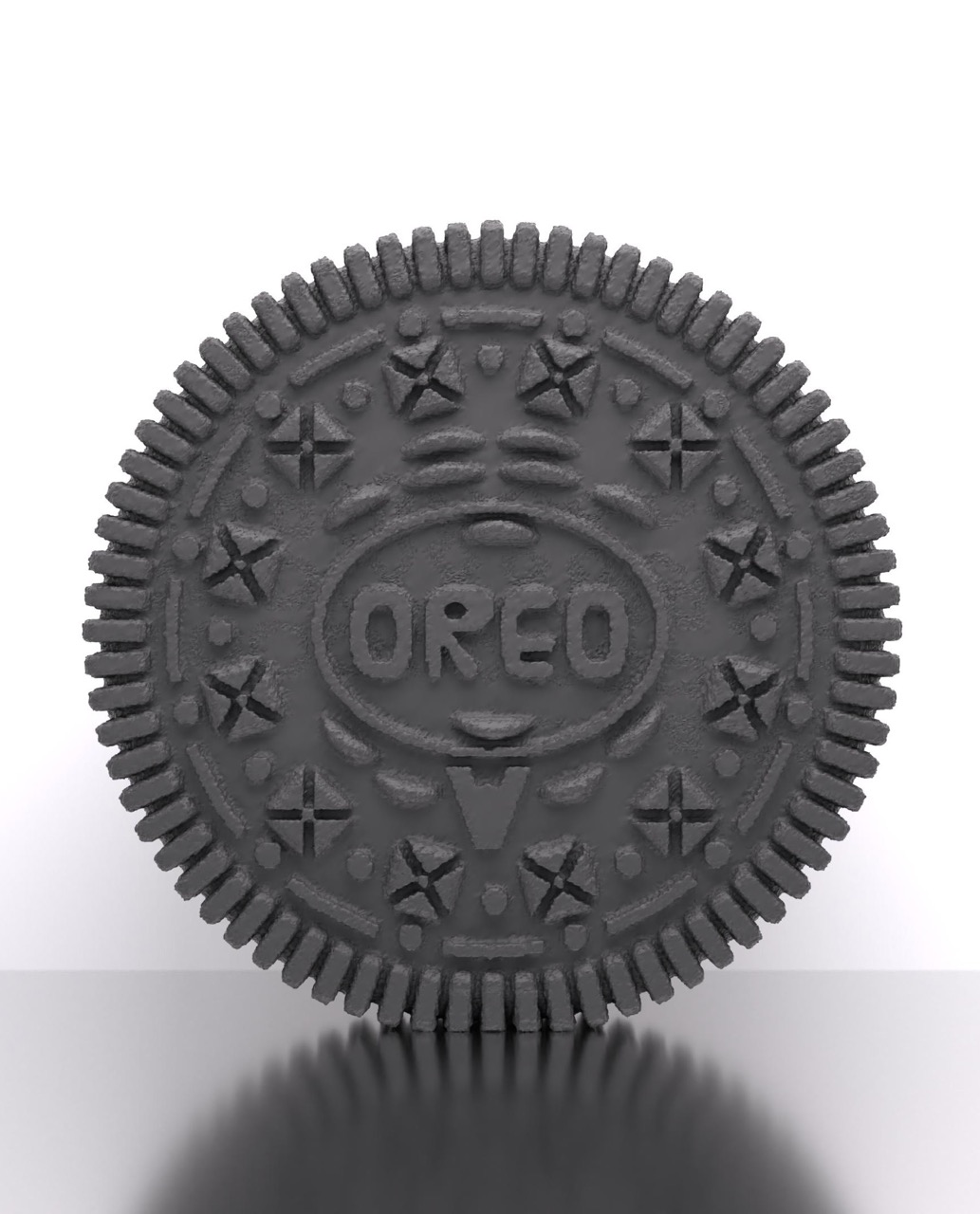}
        \caption{Quadratic Kernel}
    \end{subfigure}
    \caption{\textbf{Oreo Drop.} }
    \label{fig:oreo}
\end{figure}

In the second test case, \textit{Oreo Drop}, we simulate an Oreo-like structure consisting of filing and chocolate wafers falling to the ground. Both parts are modeled using the NACC model with parameters $E = 2 \times 10^4 Pa$, $\nu = 0.4$, $\alpha_0 = -0.01$, $\beta = 0.1$, $\xi = 0.8$, and $M = 2.36$. The initial drop height of the Oreo is set to $\frac{16}{256}$m, and the simulation is conducted with gravity $g = -9.8\,\text{m/s}^2$. As shown in \autoref{fig:oreo}, the Oreo simulated with the compact kernel fractures and falls apart, while the one simulated with the quadratic kernel retains its original shape. Additionally, the Oreo demonstrates more brittle fractures compared to the pumpkin due to its complex geometry and structural subtleties.

These results highlight the different fracture behaviors induced by the compact kernel, particularly in scenarios involving collisions and complex geometries. Although purely numerical, our method more easily produces fractures without relying on a phase-field model, which explicitly tracks fracture surfaces and softens materials to facilitate crack generation, as demonstrated in \cite{wolper2019cd,wolper2020anisompm}.

\begin{table*}[t]
\caption{Simulation statistics of our stress tests.}
\label{table:stress-test}
\begin{tabular}{llllllllll}
\toprule
Example &  Average sec/frame & Frame $\Delta t$ (s) & Max step $\Delta t$ (s) & Particle count & $\Delta x$ (m) & Grid resolution \\
\midrule
    Fluid Flush with Two Loongs        &           16.662        & $\frac{1}{48}$          &  $4.6 \times 10^{-5}$   &  84,404,827              & $\frac{1}{512}$   &  $(1024, 512, 256)$         \\
      Bullet Impact on Tungsten         &        4.847           & $\frac{1}{10^4}$           &  $9.03 \times 10^{-8}$  &   8,655,462              &  $\frac{1}{1024}$  &  (1024, 1024, 1024)\\
       Sand Castle Crashing       &    6.96               &  $\frac{1}{480}$           &  $9.26 \times 10^{-6}$  &  45,958,733              & $\frac{1}{512}$    &  $(2048, 1024, 1024)$          \\
    Fire Hydrant Pumping & 26.462 & $\frac{1}{240}$ & $5.34 \times 10^{-6}$ & 7333580 & $\frac{1}{512}$ & (512, 768, 512)\\
\bottomrule
\end{tabular}
\end{table*}

\paragraph{Fracture Avoidance}
Although the small kernel radius may lead to more frequent fracture behavior, we show that increasing the particle sampling density per cell could mitigate the unwanted fracture. 
\begin{figure}[H]
    \begin{subfigure}[b]{\linewidth}
        \includegraphics[width=\linewidth]{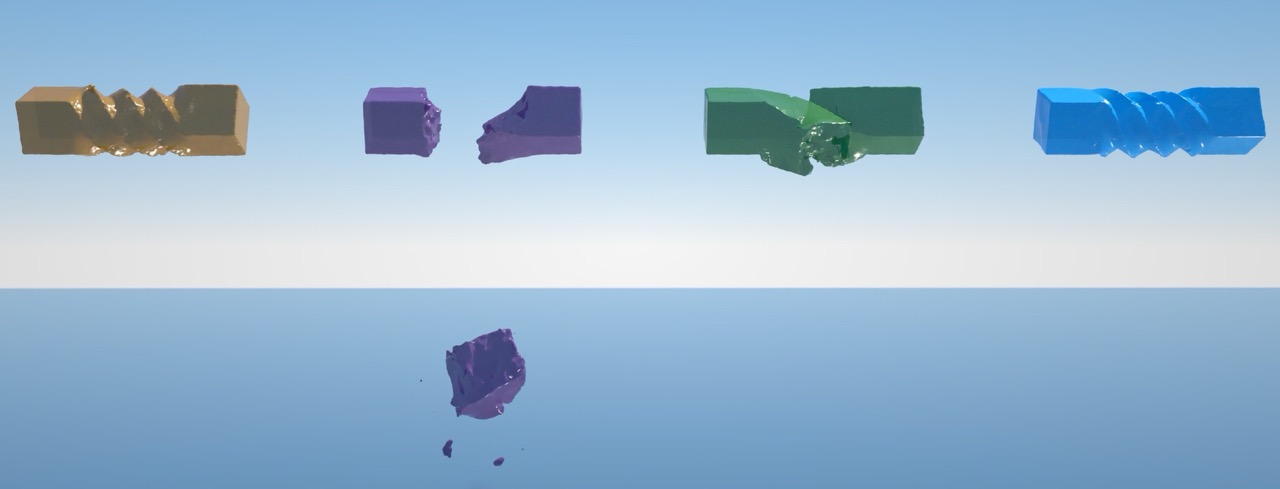}
    \end{subfigure}%
    \caption{\textbf{Twisting Elastic Bar.} The yellow bar is simulated with quadratic kernel with 8 particles per cell. Others are simulated using compact kernel with 8 (purple), 16 (green), and 27 (blue) particles per cell.}
    \label{fig:twisting-bar}
\end{figure}

To examine the performance of our method when simulating elastic objects where fracture is unwanted, we twist an elastic bar simulated using Fixed Corotated hyperelasticity model with Young's modulus $E = 100\,\text{Pa}$, $\nu = 0.4$, and density of $2\,\text{kg/m}^3$. We fix two ends of the elastic bar and rotate them in opposite directions. With a framerate of 48, we observe \autoref{fig:twisting-bar}, taken at frame 135, that the blue elastic bar simulated with 27 particles per cell remain connected as the yellow elastic bar simulated using quadartic kernel. We can also observe that the green bar simulated with 16 particles per cell demonstrated less fracture behavior than the purple bar simulated with 8 particles per cell. Therefore, it is possible to avoid unwanted fracture behavior by increasing the sampling density.

\paragraph{Contact Behavior}
The small kernel radius of our compact kernel enables a more precise contact behavior comparing to quadratic kernel. To demonstrate this difference, we present the test case with a ball sliding against the wall of a hollow cylinder. 
\begin{figure}[H]
    \includegraphics[width=0.8\linewidth]{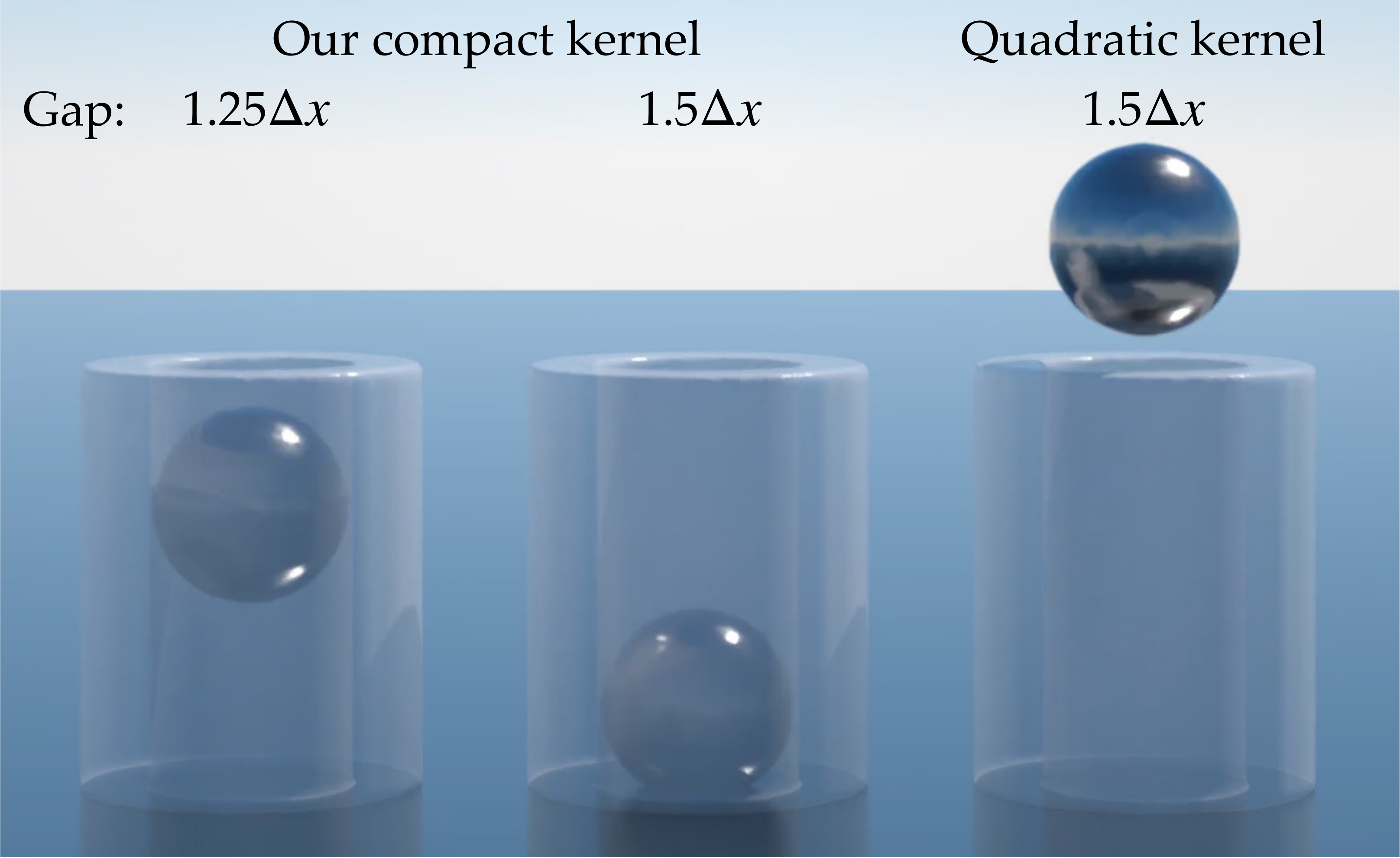}
    \caption{\textbf{Contact Behavior Test.} }
    \label{fig:contact-behavior}
\end{figure}
Both the ball and the cylinder are simulated using the Fixed Corotated hyperelasticity model with Young's modulus $E = 10^6 Pa$ and $\nu = 0.4$. We initialize the grid $\Delta x = \frac{1}{256}m$, and the hollow cylinder have inner diameter of $\frac{8}{256}m$. We simulate cases with varying diameters of the ball to test the difference in contact behavior between compact kernel and quadratic kernel. For compact kernel, we simulate with diameters of $\frac{5.5}{256}m$ and $\frac{5}{256}m$, i.e. the distance between the surface of the ball and the inner surface of the cylinder are $1.25 \Delta x$ and $1.5 \Delta x$ respectively. For quadratic kernel, we also simulate with a diameter of $\frac{5}{256}m$. We expect a precise contact resolution would allow the ball to fall and bounce freely without interfered by the velocity of the cylinder.

We observe in \autoref{fig:contact-behavior} that the two cases simulated with our compact kernel allow the ball to fall to the bottom. While the case with surface distance of $1.25 \Delta x$ slows down during the falling, the case with $1.5 \Delta x$ demonstrates a contact-free falling behavior and bounces up. In comparison, we note that, with a $1.5\Delta x$ margin, the ball simulated using quadratic kernel is stuck at the top of the cylinder.  

\paragraph{Numerical Diffusion}

\begin{figure*}
    \includegraphics[width=0.85\linewidth]{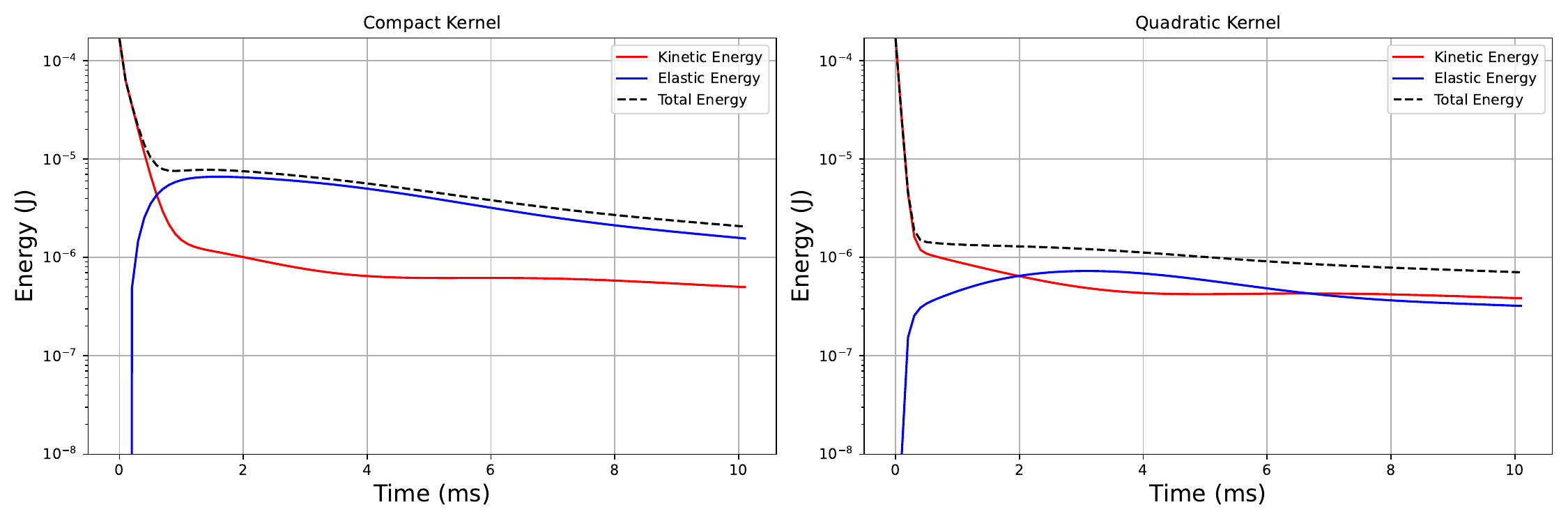}
    \caption{\textbf{Numerical Diffusion.} Kinetic, elastic, and total energy plotted on a logarithmic scale for simulations using the compact and quadratic kernels. }
    \label{fig:numerical-behavior}
\end{figure*}

Finally, we evaluate the difference in numerical diffusion between our compact kernel and the quadratic kernel. The scene is initialized with a rectangular cuboid of size \(\frac{20}{256} \times \frac{10}{256} \times \frac{140}{256}\) along the \(x\)-, \(y\)-, and \(z\)-axes, respectively. Particles are assigned a high-frequency initial velocity in the \(y\)-direction as \(0.2\sin(500 z_p)\), where \(z_p\) is the particle's \(z\)-position. The simulation is run at 10,000 frames per second for 100 frames with a time step of \(\Delta t = 2 \times 10^{-5}\).
We compare the energy decay over time for simulations using the compact and quadratic kernels. As shown in \autoref{fig:numerical-behavior}, the cuboid simulated with the compact kernel exhibits noticeably slower energy loss, indicating reduced numerical diffusion. We attribute this improved energy preservation to the compact kernel’s ability to better capture high-frequency features with minimal smoothing.

%We present a simple test case of a jelly cube falling on the ground to test the numerical diffusion of kernels. 

%\begin{figure}
 %       \includegraphics[width=0.6\linewidth]{figures/numerical_diffusion.png}
  %  \caption{\textbf{Numerical Behavior Test.}}
   % \label{fig:numerical-behavior}
%\end{figure}
%The cube is simulated with Fixed Corotated hyperelasticity model with Young's modulus $E = 10^5 Pa$ and $\nu = 0.4$. We setup the grid with $\Delta x = \frac{1}{256}m$, and we initialize the cube with side length of $\frac{20}{256}m$, density of $10^3\,\text{kg}/m^3$, and center at $(\frac{1}{2}m,\frac{25}{128}m,\frac{1}{2}m)$. The floor is initialize with sticky boundary conidition. We compare both the vibration mode and vibration duration of the cubes simulated with different kernels. From figure \autoref{fig:numerical-behavior}, we observe that the cube simulated with compact kernel demonstrated a more intensive strech and vibration comparing to the cube simulated with quadratic kernel. Moreover, the cube simulated with compact kernel demonstrated less numerical diffusion in contact with the sticky floor boundary as it continues to show a high-frequency vibration pattern throughout the 4 seconds of the simulation.   

\subsection{Stress tests} \label{sec:stress-tests}
To evaluate the stability and robustness of our compact-kernel MPM under extreme conditions, we conducted a series of stress tests. A summary of the results is presented in \autoref{table:stress-test}. The stress tests are categorized into two primary types: (1) large-scale simulations involving high-resolution grids and a large number of particles, and (2) simulations with extreme material parameter settings, designed to push the limits of stability and performance.
All simulations were performed on an Intel Xeon w7-3455 CPU and a single NVIDIA RTX 6000 Ada GPU, using CUDA 12.4 and CUDA driver version 550.54.14. 
% For the C++ backend, we employed gcc/g++ 12.3 as the compiler. 
Below, we provide a concise summary of each test case, highlighting the challenges posed and the performance of our method.

\begin{figure}
    \begin{subfigure}[b]{\linewidth}
        \centering
        \includegraphics[width=\linewidth]{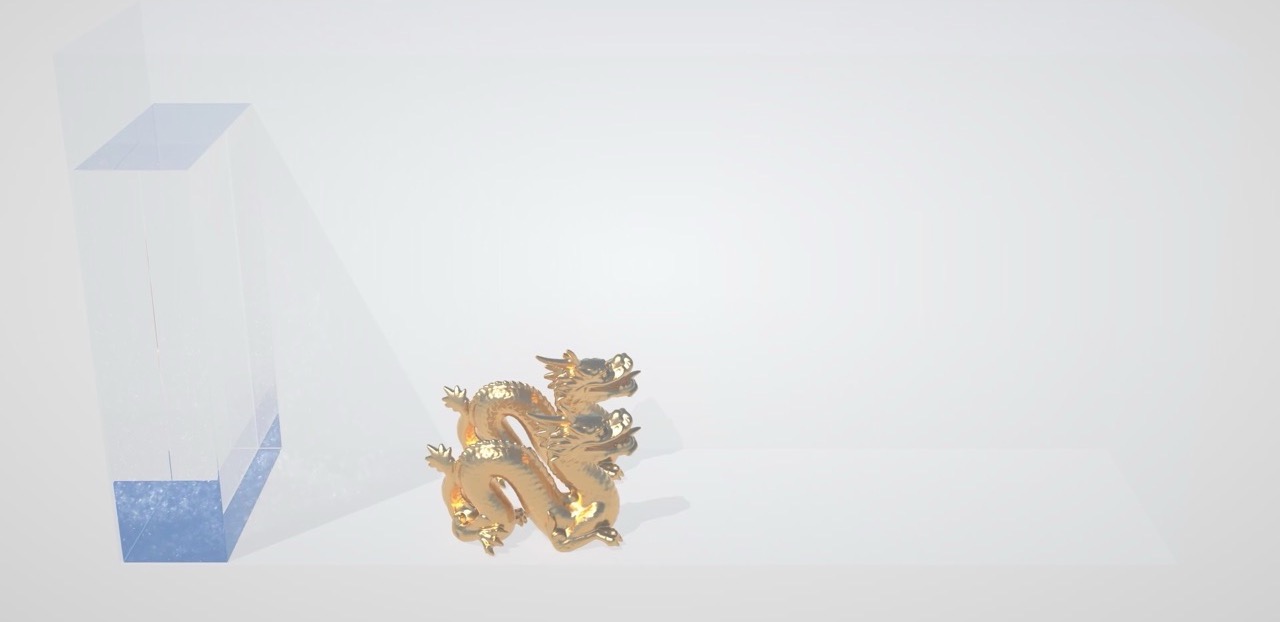}
    \end{subfigure}
    
    \begin{subfigure}[b]{\linewidth}
    \centering
        \includegraphics[width=\linewidth]{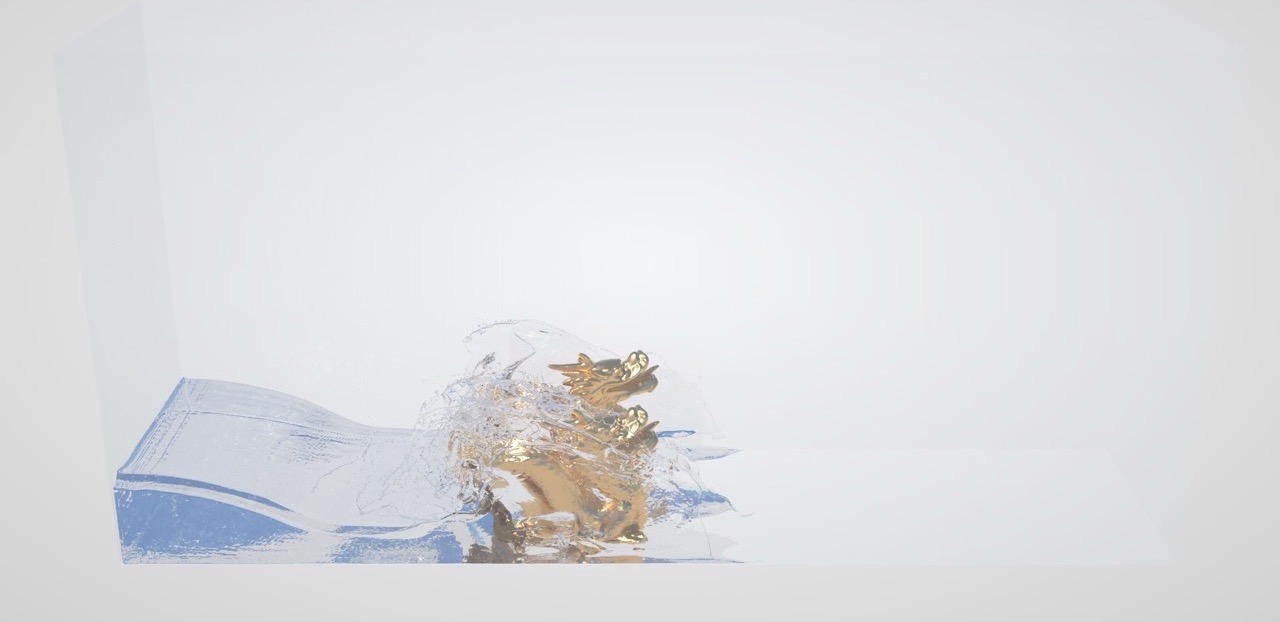}
    \end{subfigure}
    
    \begin{subfigure}[b]{\linewidth}
        \centering
        \includegraphics[width=\linewidth]{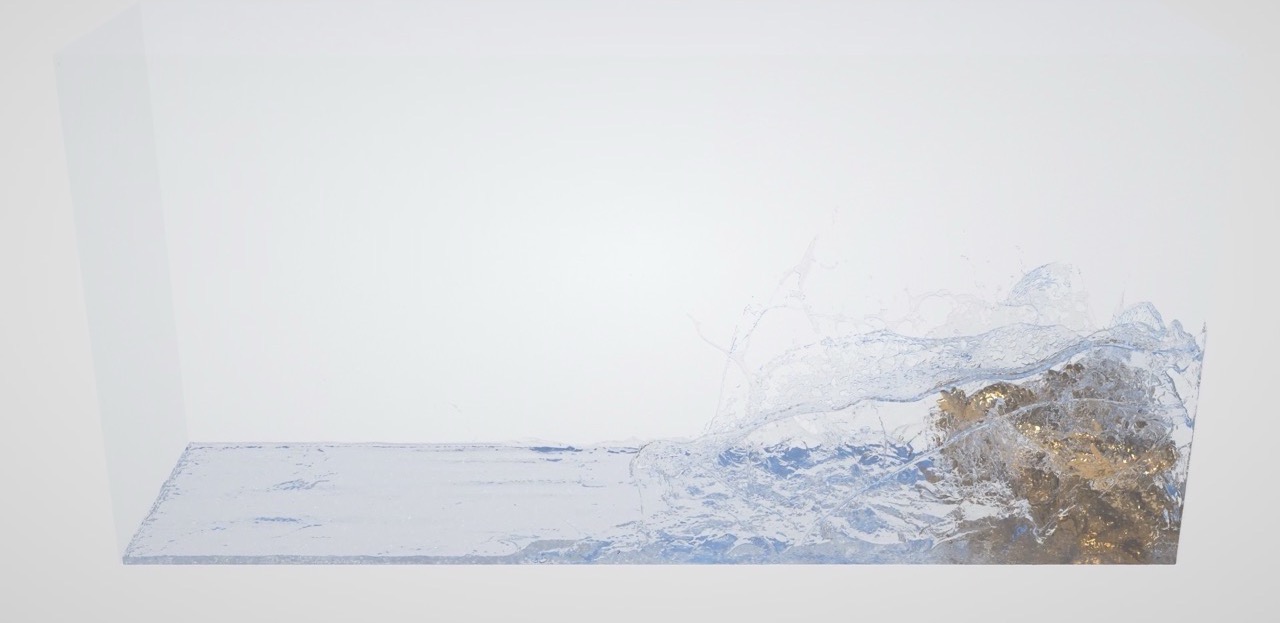}
    \end{subfigure}%
    \caption{\textbf{Fluid Flush with Two Loongs.}}
    \label{fig:fluid-with-two-longs}
\end{figure}
\paragraph{Fluid Flush with Two Loongs (Chinese dragon)}  
We simulate a large-scale scene with grid resolutions of $(1024, 512, 512)$ and a grid spacing of $\Delta x = \frac{1}{512} m$. The simulation features a cuboid of water, uniformly sampled with 75,264,000 particles, and two loongs placed near the center of the scene, discretized with 8,219,227 particles. The fluid is modeled using the $J$-based fluid model with $B = 10 Pa$, $\gamma = 7.15$, and viscosity $\mu = 0.1$. The loongs are simulated using the Fixed Corotated hyperelasticity model with Young’s modulus $E = 10^6\,\text{Pa}$ and Poisson’s ratio $\nu = 0.3$. As shown in \autoref{fig:fluid-with-two-longs}, the simulation demonstrates detailed interactions between water and the loongs, with clear splashing and turbulence effects.

\paragraph{Sand Castle Crashing}  
This test involves a sandcastle discretized with 45,925,181 particles on a grid with resolution $(2048, 1024, 1024)$ and a grid spacing of $\Delta x = \frac{1}{512}\,\text{m}$. The sand is modeled using the Non-Associative Cam Clay (NACC) model \cite{wolper2019cd} with parameters $E = 10^4\,\text{Pa}$, $\nu = 0.4$, $\alpha_0 = -0.006$, $\beta = 0.3$, $\xi = 0.5$, and $M = 1.85$. Additionally, a cannonball, discretized with 33,552 particles, is initialized with an initial velocity of $(10, 0, 0)\,\text{m/s}$ and simulated using the Fixed Corotated model with $E = 10^7\,\text{Pa}$ and $\nu = 0.2$. We set gravity to $g = 9.8\,\text{m/s}^2$. \autoref{fig:teaser} captures the dramatic destruction of the sandcastle as the cannonball collides, preserving intricate details of the collapsing structure.

\begin{figure}
    \begin{subfigure}[b]{0.49\linewidth}
        \centering
        \includegraphics[width=\linewidth]{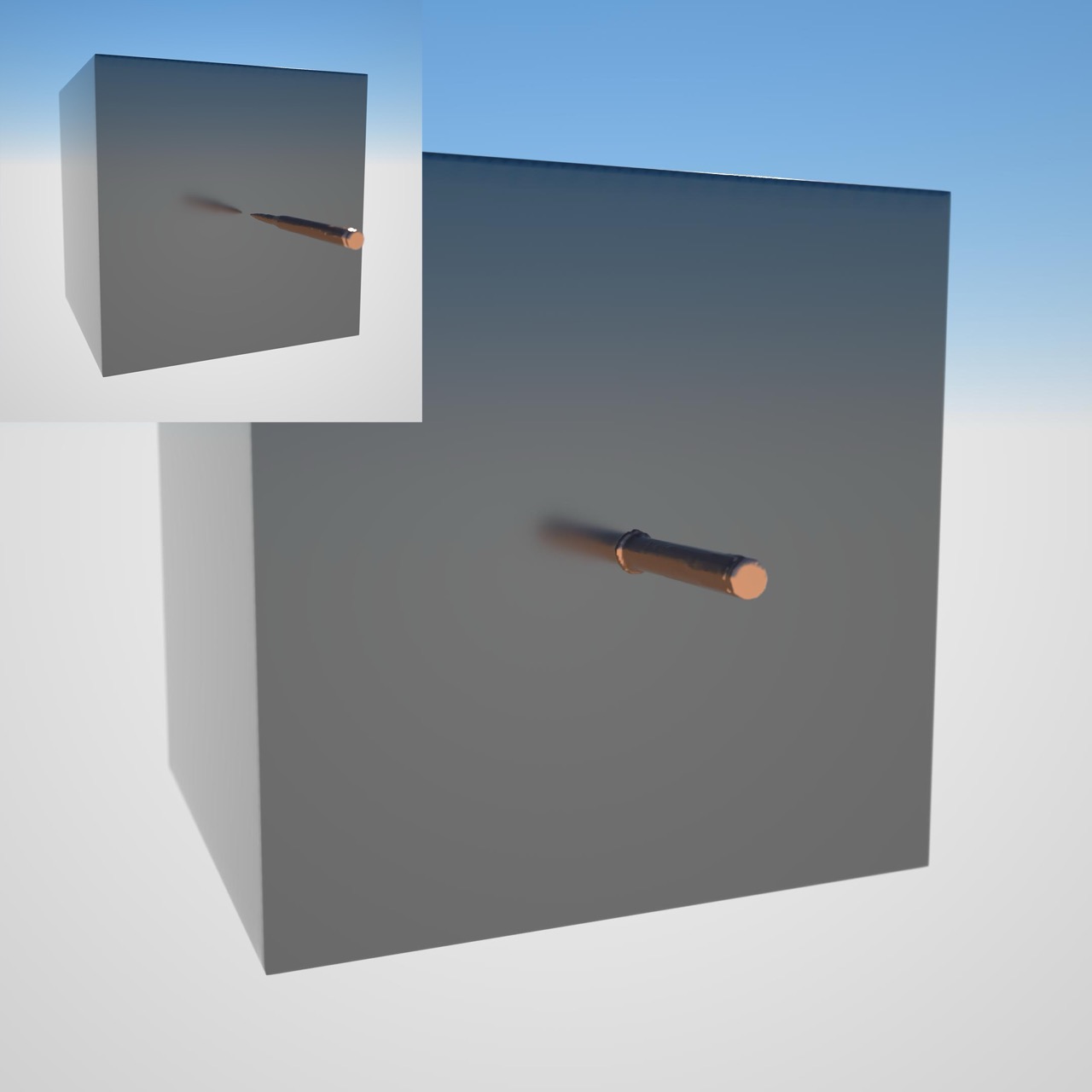}
        \caption{Before / During Impact}
    \end{subfigure}
        \begin{subfigure}[b]{0.49\linewidth}
        \centering
        \includegraphics[width=\linewidth]{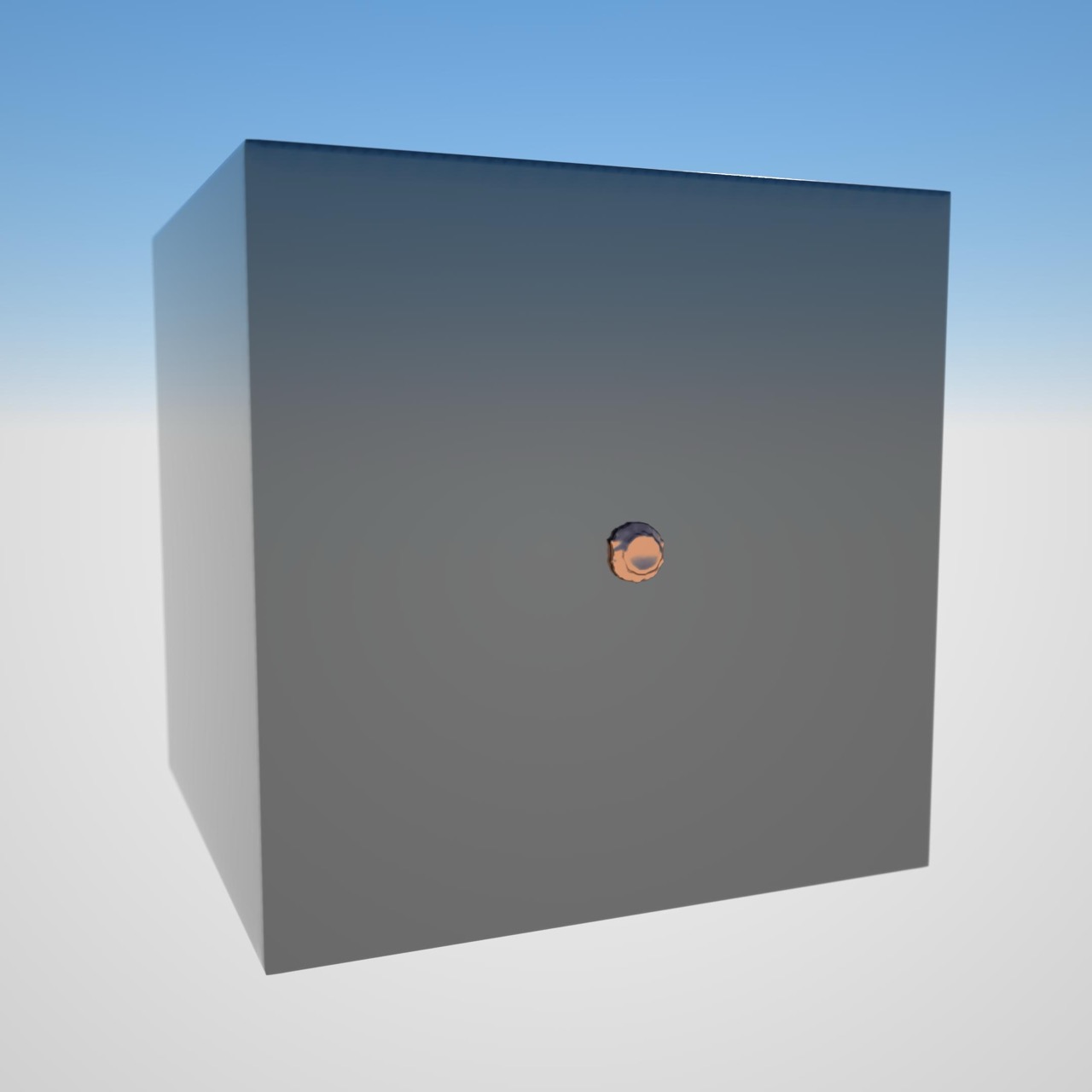}
        \caption{After Impact}
    \end{subfigure}
    \caption{\textbf{Bullet Impact on Tungsten.}}
    \label{fig:bullet}
\end{figure}

\begin{figure}
    \begin{subfigure}[b]{0.51\linewidth}
        \centering
        \includegraphics[width=\linewidth]{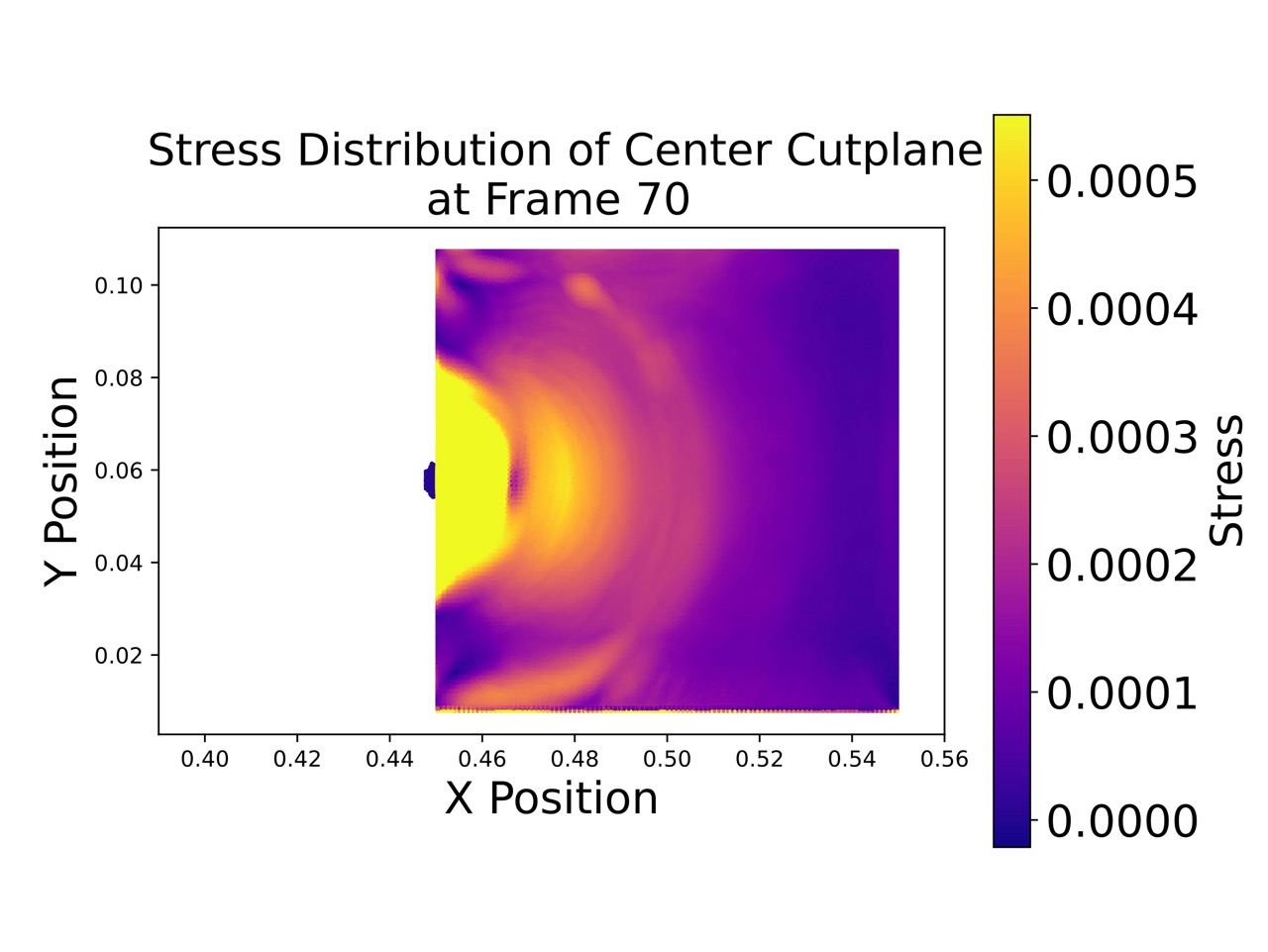}
    \end{subfigure}
        \begin{subfigure}[b]{0.48\linewidth}
        \centering
        \includegraphics[width=\linewidth]{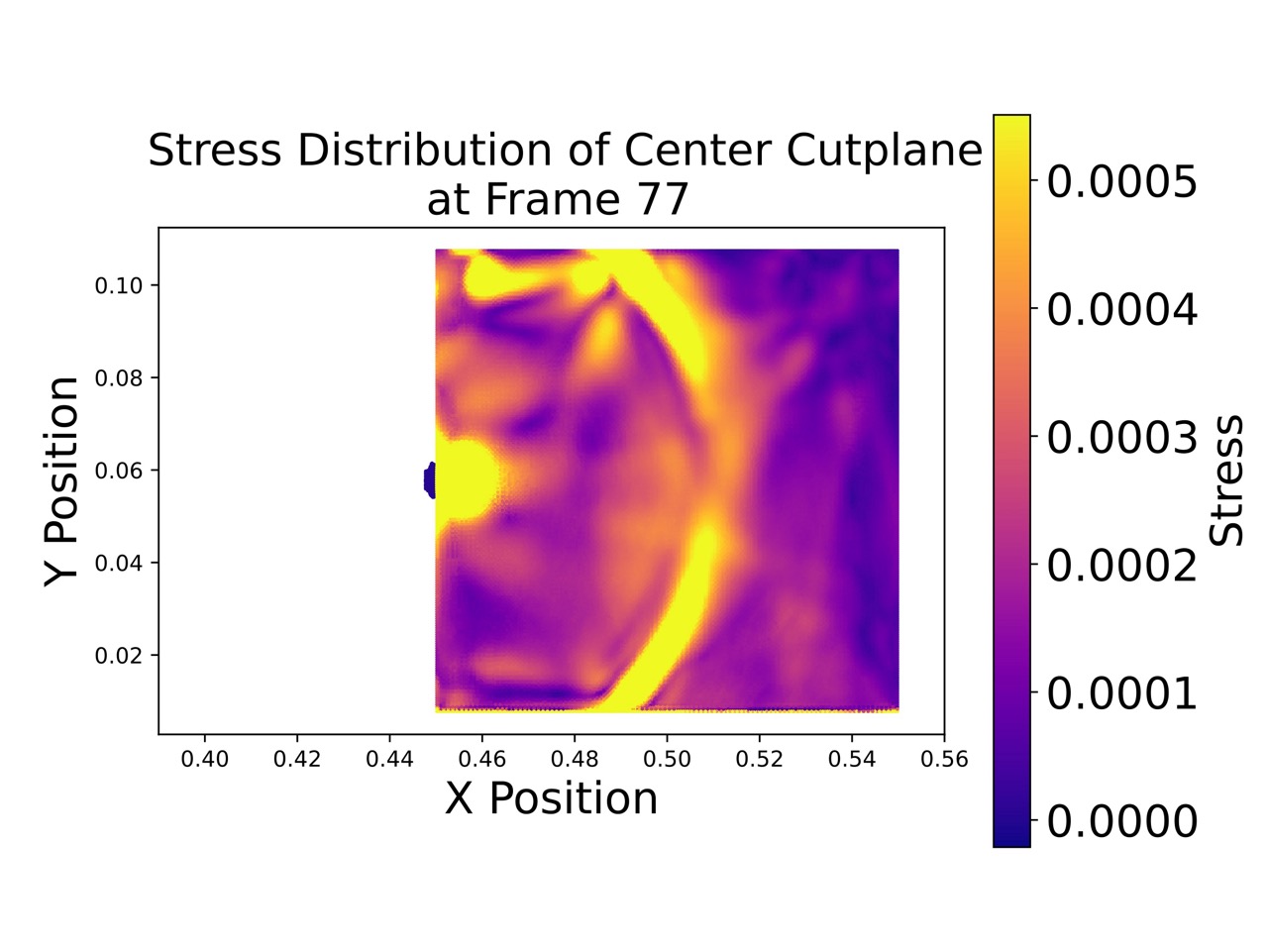}
    \end{subfigure}
    \caption{\textbf{Bullet Impact on Tungsten.} Cross-sectional view of stress distribution.}
    \label{fig:bullet-cross-section}
\end{figure}

\paragraph{Bullet Impact on Tungsten}  
This test explores the stability of our method under extreme impact conditions. The setup includes a lead bullet with an initial velocity of $(300, 0, 0)\,\text{m/s}$ striking a tungsten cube with a side length of $0.1\,\text{m}$. The lead bullet is simulated with real-world material parameters: gravity $g = 9.8\,\text{m/s}^2$, Young’s modulus $E = 1.5 \times 10^{10}\,\text{Pa}$, and Poisson’s ratio $\nu = 0.435$. The tungsten cube is modeled using the Fixed Corotated hyperelasticity model with $E = 4.5 \times 10^{11}\,\text{Pa}$ and $\nu = 0.27$. As illustrated in \autoref{fig:bullet}, the sequence shows the bullet's progression before, during, and after impact, with visible deformation and subtle vibration modes on the tungsten cube. In \autoref{fig:bullet-cross-section}, we show cross-sectional views of the stress distribution near the center cut-plane, where we observed two major wave propagations across the cube.
\begin{figure*}
    \begin{subfigure}[b]{0.33\linewidth}
        \centering
        \includegraphics[width=\linewidth]{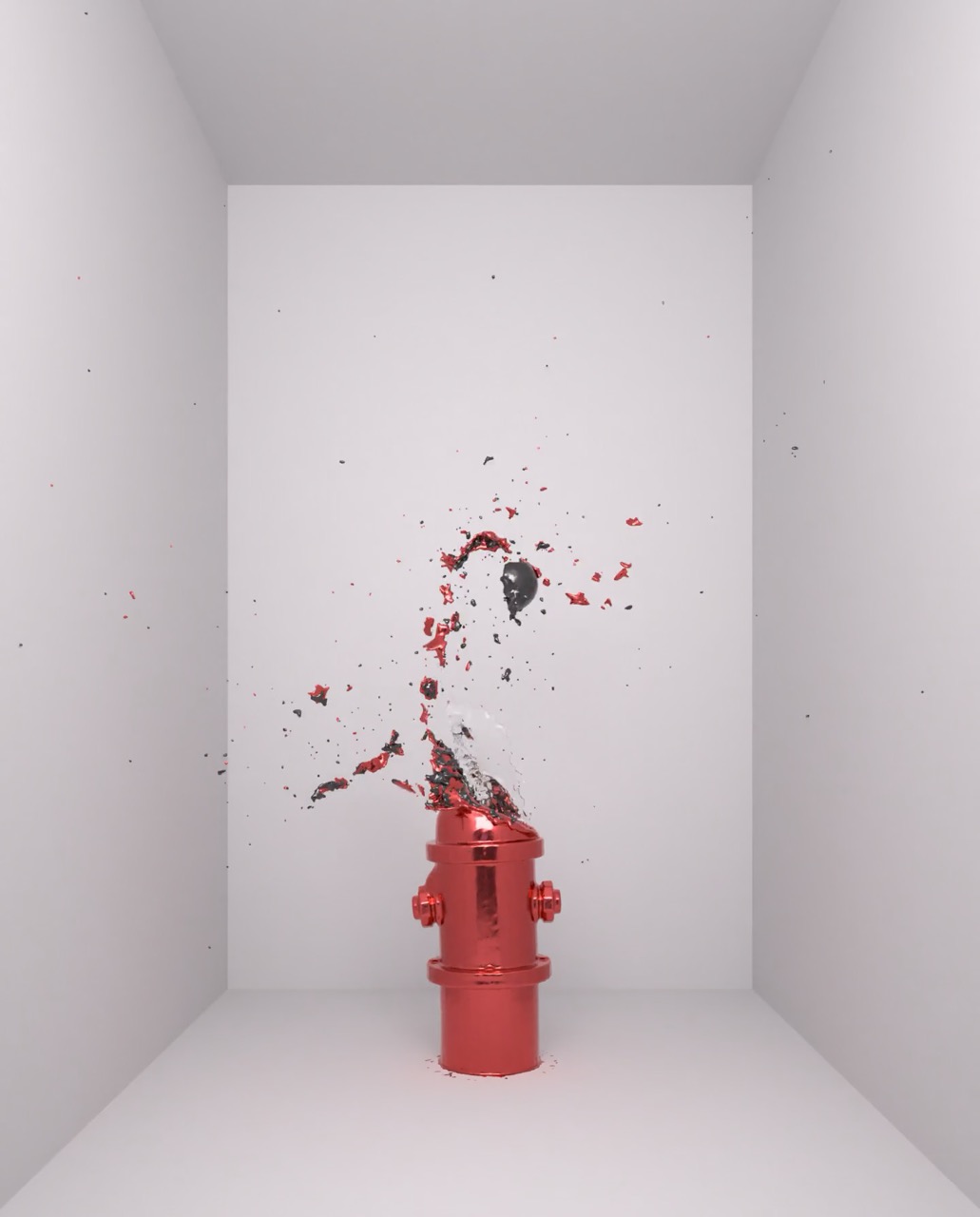}
        \caption{Initial Impact}
    \end{subfigure}
    \begin{subfigure}[b]{0.33\linewidth}
        \centering
        \includegraphics[width=\linewidth]{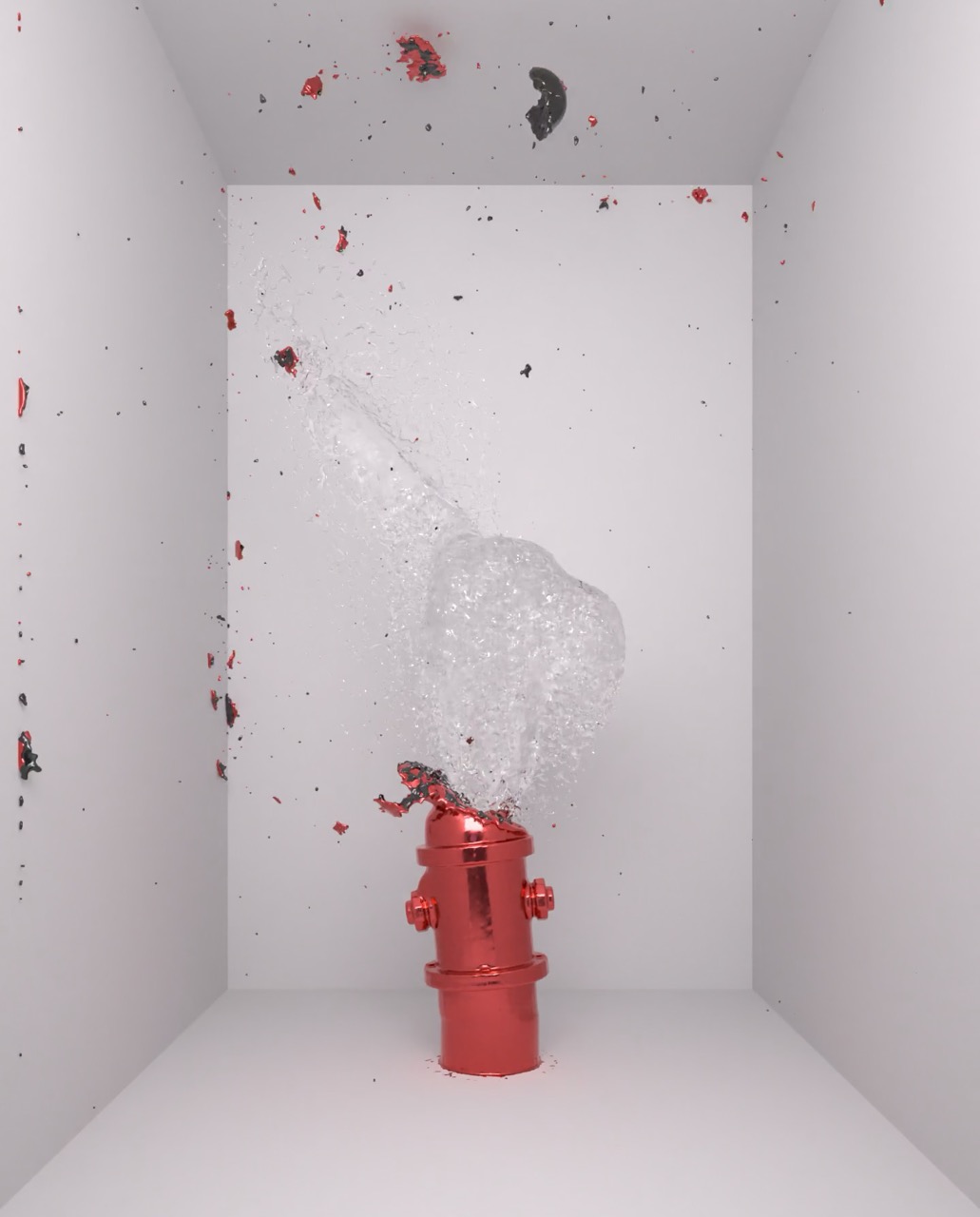}
        \caption{Fluid Burst}
    \end{subfigure}
        \begin{subfigure}[b]{0.33\linewidth}
        \centering
        \includegraphics[width=\linewidth]{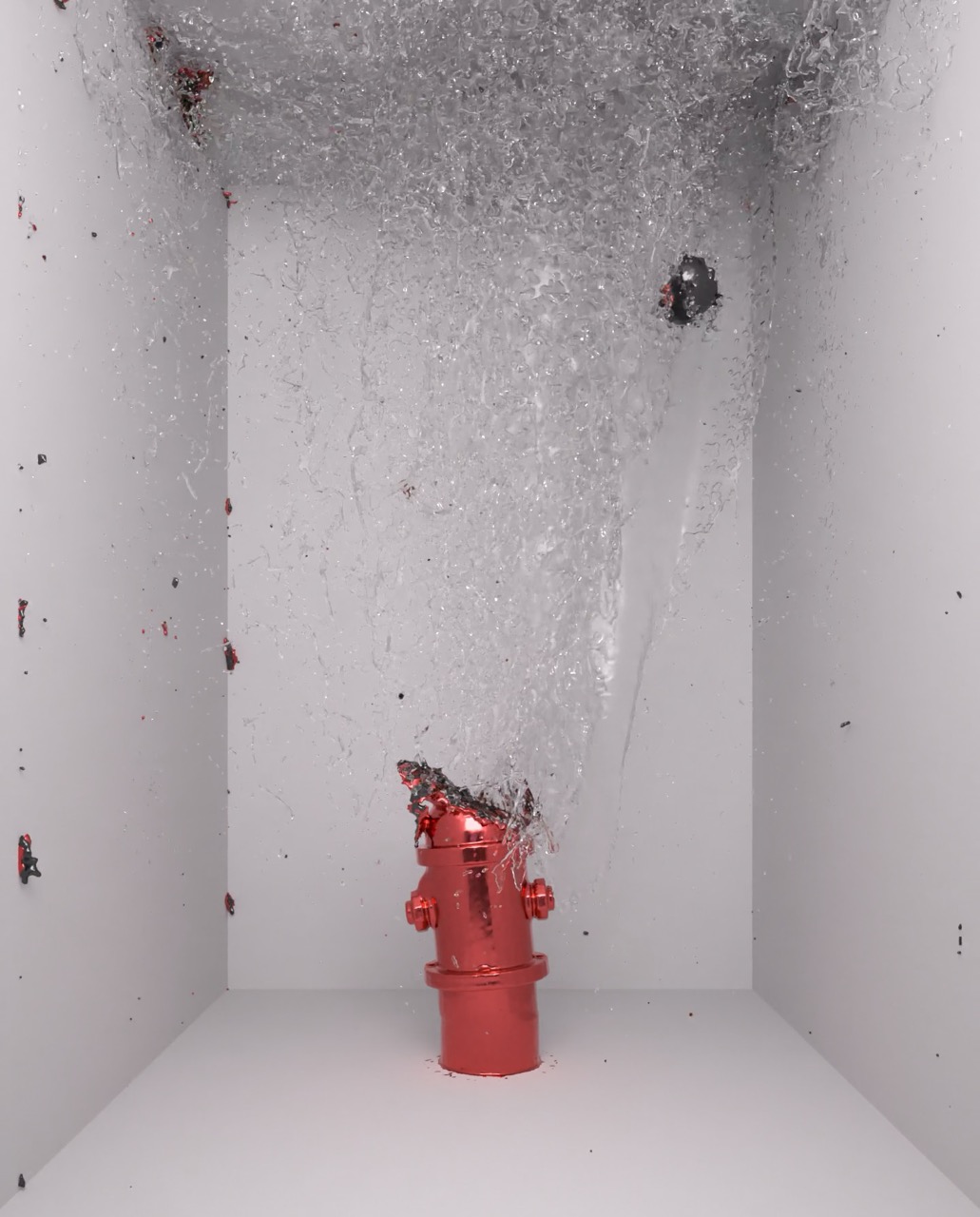}
        \caption{Final Outflow}
    \end{subfigure}
    \caption{\textbf{Fire Hydrant Pumping.}}
    \label{fig:fire-hydrant}
\end{figure*}
\paragraph{Fire Hydrant} 
This test explores the capability of our method handling complex scenes with multiple materials. The setup involves a ball traveling at $(100, 0, 0)\,\text{m/s}$ crashing a fire hydrant located at $(0.5, 0, 0.5)\,\text{m}$. We simulate the ball using 161,717 particles with Fixed Corotated hyperelasticity model with Young's modulus $E = 10^8\,\text{Pa}$, $\nu = 0.4$, and density of $10^4\,\text{kg/m}^3$. The fire hydrant is discretized into 3,999,705 particles, and we simulate it using von Mises model with Young's modulus $E = 10^9\,\text{Pa}$, $\nu = 0.4$, yield stress $3\times 10^6$. and density of $10^5\,\text{kg/m}^3$. Within the fire hydrant, it contains 3,172,158 fluid particles with $B = 10Pa$, $\gamma = 7.15$, viscosity $\mu = 0.1$, density of $10^3\,\text{kg/m}^3$, and initial $J = 0.2$ to represent a compressed form. The grid has resolution of $(512, 768, 512)$ and grid spacing of $\frac{1}{512}m$. As illustrated in \autoref{fig:fire-hydrant}, we observe metallic fractures of the fire hydrant after the initial impact. Then, the compressed fluid burst out of the fire hydrant. Eventually, all of the fluid has been pumped out of the fire hydrant. 


\section{Conclusion}

In this work, we introduced CK-MPM, a compact-kernel material point method, featuring a novel $C^2$-continuous compact kernel integrated into a staggered dual-grid framework. Our approach ensures that each particle is exclusively associated with the grid nodes of the cell it occupies per grid, enabling reductions in particle-grid-transfer costs and numerical diffusion compared to quadratic B-spline MPM. 
CK-MPM is fully compatible with existing PIC, Affine PIC, and MLS-MPM schemes, preserving critical physical properties such as linear and angular momentum. Through extensive testing, we demonstrated the efficacy of our method across a wide range of large-scale simulations, including scenarios involving extreme stiffness, high-speed impacts, and other challenging setups.
By combining compact support, high-order continuity, and compatibility with established MPM schemes, CK-MPM represents a pioneering investigation in pushing the boundaries of MPM simulations. We believe our approach lays a strong foundation for further advancements in physics-based simulation, empowering applications in engineering, computer graphics, and beyond.

% While our compact kernel only associates each particle with 2 grid nodes per dimension on each grid, this additional grid would require more GPU memory for storage. Moreover, it requires a large amount of shared memory per CUDA block in the fused G2P2G kernel, restricting the number of concurrent CUDA blocks. One possible future work may consider reducing the memory overhead caused by the auxiliary grid. 
\paragraph{Limitations \& Future Work}
While the staggered grid framework in our method adds complexity to the implementation, it opens avenues for meaningful future research. For instance, further optimizing memory layouts on the GPU, multi-GPU, or multinode computing environments could unlock additional speedups, enabling more efficient simulations.
Another intriguing direction lies in the trade-off between efficiency and accuracy in the CUDA trigonometric functions. The intrinsic CUDA functions $\mathrm{\_\_sinf}$ and $\mathrm{\_\_cosf}$, though faster, lack the precision required to accurately preserve momentum. To ensure accuracy, we currently rely on the slower $\sin$ and $\cos$ functions. Future advancements in CUDA's implementation of faster and more precise trigonometric functions would directly enhance the performance of our method.
Our compact kernel, with its small support size, has a tendency to facilitate material fractures. This can be advantageous for fracture simulations but presents challenges for scenarios requiring fracture-free large deformations. In such cases, increasing the particle sampling density per cell could mitigate unwanted fractures.
Finally, we envision future work that further leverages the mathematical properties of our kernel. One promising direction involves integrating our kernel with the PolyPIC \cite{fu2017polynomial} method to achieve higher-order accuracy for particle-grid transfer. Additionally, the unique structure of our kernel enables independent solutions to the systems arising in implicit MPM time integration, potentially offering significant speedup.

% Finally, our method preserves the differentiability of MPM simulations, making it highly suitable for solving inverse problems, such as those explored in \citet{zhang2025physdreamer,li2021soft}. We envision that the combination of high efficiency and differentiability could significantly advance applications in 3D reconstruction, data-driven design, etc.

%%
%% The acknowledgments section is defined using the "acks" environment
%% (and NOT an unnumbered section). This ensures the proper
%% identification of the section in the article metadata, and the
%% consistent spelling of the heading.
\begin{acks}
This work was supported in part by the Junior Faculty Startup Fund of Carnegie Mellon University. We thank the reviewers for their detailed and insightful feedback, and are especially grateful to Kemeng Huang and Taku Komura for generously providing part of the computing resources used in our experiments, and to Muyuan Ma for creating the illustrative figures of the dual-grid scheme.
\end{acks}

%%
%% The next two lines define the bibliography style to be used, and
%% the bibliography file.
\bibliographystyle{ACM-Reference-Format}
\bibliography{main}

%%% -*-BibTeX-*-
%%% Do NOT edit. File created by BibTeX with style
%%% ACM-Reference-Format-Journals [18-Jan-2012].

\begin{thebibliography}{50}

%%% ====================================================================
%%% NOTE TO THE USER: you can override these defaults by providing
%%% customized versions of any of these macros before the \bibliography
%%% command.  Each of them MUST provide its own final punctuation,
%%% except for \shownote{}, \showDOI{}, and \showURL{}.  The latter two
%%% do not use final punctuation, in order to avoid confusing it with
%%% the Web address.
%%%
%%% To suppress output of a particular field, define its macro to expand
%%% to an empty string, or better, \unskip, like this:
%%%
%%% \newcommand{\showDOI}[1]{\unskip}   % LaTeX syntax
%%%
%%% \def \showDOI #1{\unskip}           % plain TeX syntax
%%%
%%% ====================================================================

\ifx \showCODEN    \undefined \def \showCODEN     #1{\unskip}     \fi
\ifx \showDOI      \undefined \def \showDOI       #1{#1}\fi
\ifx \showISBNx    \undefined \def \showISBNx     #1{\unskip}     \fi
\ifx \showISBNxiii \undefined \def \showISBNxiii  #1{\unskip}     \fi
\ifx \showISSN     \undefined \def \showISSN      #1{\unskip}     \fi
\ifx \showLCCN     \undefined \def \showLCCN      #1{\unskip}     \fi
\ifx \shownote     \undefined \def \shownote      #1{#1}          \fi
\ifx \showarticletitle \undefined \def \showarticletitle #1{#1}   \fi
\ifx \showURL      \undefined \def \showURL       {\relax}        \fi
% The following commands are used for tagged output and should be
% invisible to TeX
\providecommand\bibfield[2]{#2}
\providecommand\bibinfo[2]{#2}
\providecommand\natexlab[1]{#1}
\providecommand\showeprint[2][]{arXiv:#2}

\bibitem[Bardenhagen et~al\mbox{.}(2004)]%
        {bardenhagen2004generalized}
\bibfield{author}{\bibinfo{person}{Scott~G Bardenhagen},
  \bibinfo{person}{Edward~M Kober}, {et~al\mbox{.}}}
  \bibinfo{year}{2004}\natexlab{}.
\newblock \showarticletitle{The generalized interpolation material point
  method}.
\newblock \bibinfo{journal}{\emph{Computer Modeling in Engineering and
  Sciences}} \bibinfo{volume}{5}, \bibinfo{number}{6} (\bibinfo{year}{2004}),
  \bibinfo{pages}{477--496}.
\newblock


\bibitem[Brackbill et~al\mbox{.}(1988)]%
        {brackbill1988flip}
\bibfield{author}{\bibinfo{person}{Jeremiah~U Brackbill},
  \bibinfo{person}{Douglas~B Kothe}, {and} \bibinfo{person}{Hans~M Ruppel}.}
  \bibinfo{year}{1988}\natexlab{}.
\newblock \showarticletitle{FLIP: a low-dissipation, particle-in-cell method
  for fluid flow}.
\newblock \bibinfo{journal}{\emph{Computer Physics Communications}}
  \bibinfo{volume}{48}, \bibinfo{number}{1} (\bibinfo{year}{1988}),
  \bibinfo{pages}{25--38}.
\newblock


\bibitem[Bridson(2015)]%
        {bridson2015fluid}
\bibfield{author}{\bibinfo{person}{Robert Bridson}.}
  \bibinfo{year}{2015}\natexlab{}.
\newblock \bibinfo{booktitle}{\emph{Fluid simulation for computer graphics}}.
\newblock \bibinfo{publisher}{AK Peters/CRC Press}.
\newblock


\bibitem[Cao et~al\mbox{.}(2022)]%
        {cao2022efficient}
\bibfield{author}{\bibinfo{person}{Yadi Cao}, \bibinfo{person}{Yunuo Chen},
  \bibinfo{person}{Minchen Li}, \bibinfo{person}{Yin Yang},
  \bibinfo{person}{Xinxin Zhang}, \bibinfo{person}{Mridul Aanjaneya}, {and}
  \bibinfo{person}{Chenfanfu Jiang}.} \bibinfo{year}{2022}\natexlab{}.
\newblock \showarticletitle{An efficient b-spline lagrangian/eulerian method
  for compressible flow, shock waves, and fracturing solids}.
\newblock \bibinfo{journal}{\emph{ACM Transactions on Graphics (TOG)}}
  \bibinfo{volume}{41}, \bibinfo{number}{5} (\bibinfo{year}{2022}),
  \bibinfo{pages}{1--13}.
\newblock


\bibitem[Chen et~al\mbox{.}(2021)]%
        {chen2021hybrid}
\bibfield{author}{\bibinfo{person}{Peter~Yichen Chen}, \bibinfo{person}{Maytee
  Chantharayukhonthorn}, \bibinfo{person}{Yonghao Yue}, \bibinfo{person}{Eitan
  Grinspun}, {and} \bibinfo{person}{Ken Kamrin}.}
  \bibinfo{year}{2021}\natexlab{}.
\newblock \showarticletitle{Hybrid discrete-continuum modeling of shear
  localization in granular media}.
\newblock \bibinfo{journal}{\emph{Journal of the Mechanics and Physics of
  Solids}}  \bibinfo{volume}{153} (\bibinfo{year}{2021}),
  \bibinfo{pages}{104404}.
\newblock


\bibitem[De~Vaucorbeil et~al\mbox{.}(2020)]%
        {de2020material}
\bibfield{author}{\bibinfo{person}{Alban De~Vaucorbeil},
  \bibinfo{person}{Vinh~Phu Nguyen}, \bibinfo{person}{Sina Sinaie}, {and}
  \bibinfo{person}{Jian~Ying Wu}.} \bibinfo{year}{2020}\natexlab{}.
\newblock \showarticletitle{Material point method after 25 years: theory,
  implementation, and applications}.
\newblock \bibinfo{journal}{\emph{Advances in applied mechanics}}
  \bibinfo{volume}{53} (\bibinfo{year}{2020}), \bibinfo{pages}{185--398}.
\newblock


\bibitem[Ding et~al\mbox{.}(2019)]%
        {ding2019thermomechanical}
\bibfield{author}{\bibinfo{person}{Mengyuan Ding}, \bibinfo{person}{Xuchen
  Han}, \bibinfo{person}{Stephanie Wang}, \bibinfo{person}{Theodore~F Gast},
  {and} \bibinfo{person}{Joseph~M Teran}.} \bibinfo{year}{2019}\natexlab{}.
\newblock \showarticletitle{A thermomechanical material point method for baking
  and cooking}.
\newblock \bibinfo{journal}{\emph{ACM Transactions on Graphics (TOG)}}
  \bibinfo{volume}{38}, \bibinfo{number}{6} (\bibinfo{year}{2019}),
  \bibinfo{pages}{1--14}.
\newblock


\bibitem[Fang et~al\mbox{.}(2019)]%
        {fang2019silly}
\bibfield{author}{\bibinfo{person}{Yu Fang}, \bibinfo{person}{Minchen Li},
  \bibinfo{person}{Ming Gao}, {and} \bibinfo{person}{Chenfanfu Jiang}.}
  \bibinfo{year}{2019}\natexlab{}.
\newblock \showarticletitle{Silly rubber: an implicit material point method for
  simulating non-equilibrated viscoelastic and elastoplastic solids}.
\newblock \bibinfo{journal}{\emph{ACM Transactions on Graphics (TOG)}}
  \bibinfo{volume}{38}, \bibinfo{number}{4} (\bibinfo{year}{2019}),
  \bibinfo{pages}{1--13}.
\newblock


\bibitem[Fang et~al\mbox{.}(2020)]%
        {fang2020iq}
\bibfield{author}{\bibinfo{person}{Yu Fang}, \bibinfo{person}{Ziyin Qu},
  \bibinfo{person}{Minchen Li}, \bibinfo{person}{Xinxin Zhang},
  \bibinfo{person}{Yixin Zhu}, \bibinfo{person}{Mridul Aanjaneya}, {and}
  \bibinfo{person}{Chenfanfu Jiang}.} \bibinfo{year}{2020}\natexlab{}.
\newblock \showarticletitle{IQ-MPM: an interface quadrature material point
  method for non-sticky strongly two-way coupled nonlinear solids and fluids}.
\newblock \bibinfo{journal}{\emph{ACM Transactions on Graphics (TOG)}}
  \bibinfo{volume}{39}, \bibinfo{number}{4} (\bibinfo{year}{2020}),
  \bibinfo{pages}{51--1}.
\newblock


\bibitem[Fei et~al\mbox{.}(2018)]%
        {fei2018multi}
\bibfield{author}{\bibinfo{person}{Yun Fei}, \bibinfo{person}{Christopher
  Batty}, \bibinfo{person}{Eitan Grinspun}, {and} \bibinfo{person}{Changxi
  Zheng}.} \bibinfo{year}{2018}\natexlab{}.
\newblock \showarticletitle{A multi-scale model for simulating liquid-fabric
  interactions}.
\newblock \bibinfo{journal}{\emph{ACM Transactions on Graphics (TOG)}}
  \bibinfo{volume}{37}, \bibinfo{number}{4} (\bibinfo{year}{2018}),
  \bibinfo{pages}{1--16}.
\newblock


\bibitem[Fei et~al\mbox{.}(2019)]%
        {fei2019multi}
\bibfield{author}{\bibinfo{person}{Yun Fei}, \bibinfo{person}{Christopher
  Batty}, \bibinfo{person}{Eitan Grinspun}, {and} \bibinfo{person}{Changxi
  Zheng}.} \bibinfo{year}{2019}\natexlab{}.
\newblock \showarticletitle{A multi-scale model for coupling strands with
  shear-dependent liquid}.
\newblock \bibinfo{journal}{\emph{ACM Transactions on Graphics (TOG)}}
  \bibinfo{volume}{38}, \bibinfo{number}{6} (\bibinfo{year}{2019}),
  \bibinfo{pages}{1--20}.
\newblock


\bibitem[Fei et~al\mbox{.}(2021a)]%
        {fei2021revisiting}
\bibfield{author}{\bibinfo{person}{Yun Fei}, \bibinfo{person}{Qi Guo},
  \bibinfo{person}{Rundong Wu}, \bibinfo{person}{Li Huang}, {and}
  \bibinfo{person}{Ming Gao}.} \bibinfo{year}{2021}\natexlab{a}.
\newblock \showarticletitle{Revisiting integration in the material point
  method: a scheme for easier separation and less dissipation}.
\newblock \bibinfo{journal}{\emph{ACM Transactions on Graphics (TOG)}}
  \bibinfo{volume}{40}, \bibinfo{number}{4} (\bibinfo{year}{2021}),
  \bibinfo{pages}{1--16}.
\newblock


\bibitem[Fei et~al\mbox{.}(2021b)]%
        {fei2021principles}
\bibfield{author}{\bibinfo{person}{Yun Fei}, \bibinfo{person}{Yuhan Huang},
  {and} \bibinfo{person}{Ming Gao}.} \bibinfo{year}{2021}\natexlab{b}.
\newblock \showarticletitle{Principles towards real-time simulation of material
  point method on modern GPUs}.
\newblock \bibinfo{journal}{\emph{arXiv preprint arXiv:2111.00699}}
  (\bibinfo{year}{2021}).
\newblock


\bibitem[Fei et~al\mbox{.}(2017)]%
        {fei2017multi}
\bibfield{author}{\bibinfo{person}{Yun Fei}, \bibinfo{person}{Henrique~Teles
  Maia}, \bibinfo{person}{Christopher Batty}, \bibinfo{person}{Changxi Zheng},
  {and} \bibinfo{person}{Eitan Grinspun}.} \bibinfo{year}{2017}\natexlab{}.
\newblock \showarticletitle{A multi-scale model for simulating liquid-hair
  interactions}.
\newblock \bibinfo{journal}{\emph{ACM Transactions on Graphics (TOG)}}
  \bibinfo{volume}{36}, \bibinfo{number}{4} (\bibinfo{year}{2017}),
  \bibinfo{pages}{1--17}.
\newblock


\bibitem[Fu et~al\mbox{.}(2017)]%
        {fu2017polynomial}
\bibfield{author}{\bibinfo{person}{Chuyuan Fu}, \bibinfo{person}{Qi Guo},
  \bibinfo{person}{Theodore Gast}, \bibinfo{person}{Chenfanfu Jiang}, {and}
  \bibinfo{person}{Joseph Teran}.} \bibinfo{year}{2017}\natexlab{}.
\newblock \showarticletitle{A polynomial particle-in-cell method}.
\newblock \bibinfo{journal}{\emph{ACM Transactions on Graphics (TOG)}}
  \bibinfo{volume}{36}, \bibinfo{number}{6} (\bibinfo{year}{2017}),
  \bibinfo{pages}{1--12}.
\newblock


\bibitem[Gao et~al\mbox{.}(2017)]%
        {gao2017adaptive}
\bibfield{author}{\bibinfo{person}{Ming Gao}, \bibinfo{person}{Andre~Pradhana
  Tampubolon}, \bibinfo{person}{Chenfanfu Jiang}, {and}
  \bibinfo{person}{Eftychios Sifakis}.} \bibinfo{year}{2017}\natexlab{}.
\newblock \showarticletitle{An adaptive generalized interpolation material
  point method for simulating elastoplastic materials}.
\newblock \bibinfo{journal}{\emph{ACM Transactions on Graphics (TOG)}}
  \bibinfo{volume}{36}, \bibinfo{number}{6} (\bibinfo{year}{2017}),
  \bibinfo{pages}{1--12}.
\newblock


\bibitem[Gao et~al\mbox{.}(2018)]%
        {gao2018gpu}
\bibfield{author}{\bibinfo{person}{Ming Gao}, \bibinfo{person}{Xinlei Wang},
  \bibinfo{person}{Kui Wu}, \bibinfo{person}{Andre Pradhana},
  \bibinfo{person}{Eftychios Sifakis}, \bibinfo{person}{Cem Yuksel}, {and}
  \bibinfo{person}{Chenfanfu Jiang}.} \bibinfo{year}{2018}\natexlab{}.
\newblock \showarticletitle{GPU optimization of material point methods}.
\newblock \bibinfo{journal}{\emph{ACM Transactions on Graphics (TOG)}}
  \bibinfo{volume}{37}, \bibinfo{number}{6} (\bibinfo{year}{2018}),
  \bibinfo{pages}{1--12}.
\newblock


\bibitem[Gaume et~al\mbox{.}(2018)]%
        {gaume2018dynamic}
\bibfield{author}{\bibinfo{person}{Johan Gaume}, \bibinfo{person}{T Gast},
  \bibinfo{person}{Joseph Teran}, \bibinfo{person}{Alec van Herwijnen}, {and}
  \bibinfo{person}{Chenfanfu Jiang}.} \bibinfo{year}{2018}\natexlab{}.
\newblock \showarticletitle{Dynamic anticrack propagation in snow}.
\newblock \bibinfo{journal}{\emph{Nature communications}} \bibinfo{volume}{9},
  \bibinfo{number}{1} (\bibinfo{year}{2018}), \bibinfo{pages}{3047}.
\newblock


\bibitem[Guo et~al\mbox{.}(2018)]%
        {guo2018material}
\bibfield{author}{\bibinfo{person}{Qi Guo}, \bibinfo{person}{Xuchen Han},
  \bibinfo{person}{Chuyuan Fu}, \bibinfo{person}{Theodore Gast},
  \bibinfo{person}{Rasmus Tamstorf}, {and} \bibinfo{person}{Joseph Teran}.}
  \bibinfo{year}{2018}\natexlab{}.
\newblock \showarticletitle{A material point method for thin shells with
  frictional contact}.
\newblock \bibinfo{journal}{\emph{ACM Transactions on Graphics (TOG)}}
  \bibinfo{volume}{37}, \bibinfo{number}{4} (\bibinfo{year}{2018}),
  \bibinfo{pages}{1--15}.
\newblock


\bibitem[Han et~al\mbox{.}(2019)]%
        {han2019hybrid}
\bibfield{author}{\bibinfo{person}{Xuchen Han}, \bibinfo{person}{Theodore~F
  Gast}, \bibinfo{person}{Qi Guo}, \bibinfo{person}{Stephanie Wang},
  \bibinfo{person}{Chenfanfu Jiang}, {and} \bibinfo{person}{Joseph Teran}.}
  \bibinfo{year}{2019}\natexlab{}.
\newblock \showarticletitle{A hybrid material point method for frictional
  contact with diverse materials}.
\newblock \bibinfo{journal}{\emph{Proceedings of the ACM on Computer Graphics
  and Interactive Techniques}} \bibinfo{volume}{2}, \bibinfo{number}{2}
  (\bibinfo{year}{2019}), \bibinfo{pages}{1--24}.
\newblock


\bibitem[Harlow(1962)]%
        {harlow1962particle}
\bibfield{author}{\bibinfo{person}{Francis~H Harlow}.}
  \bibinfo{year}{1962}\natexlab{}.
\newblock \bibinfo{booktitle}{\emph{The particle-in-cell method for numerical
  solution of problems in fluid dynamics}}.
\newblock \bibinfo{type}{{T}echnical {R}eport}. \bibinfo{institution}{Los
  Alamos National Lab.(LANL), Los Alamos, NM (United States)}.
\newblock


\bibitem[Hu et~al\mbox{.}(2018)]%
        {hu2018moving}
\bibfield{author}{\bibinfo{person}{Yuanming Hu}, \bibinfo{person}{Yu Fang},
  \bibinfo{person}{Ziheng Ge}, \bibinfo{person}{Ziyin Qu},
  \bibinfo{person}{Yixin Zhu}, \bibinfo{person}{Andre Pradhana}, {and}
  \bibinfo{person}{Chenfanfu Jiang}.} \bibinfo{year}{2018}\natexlab{}.
\newblock \showarticletitle{A moving least squares material point method with
  displacement discontinuity and two-way rigid body coupling}.
\newblock \bibinfo{journal}{\emph{ACM Transactions on Graphics (TOG)}}
  \bibinfo{volume}{37}, \bibinfo{number}{4} (\bibinfo{year}{2018}),
  \bibinfo{pages}{1--14}.
\newblock


\bibitem[Hu et~al\mbox{.}(2019)]%
        {hu2019taichi}
\bibfield{author}{\bibinfo{person}{Yuanming Hu}, \bibinfo{person}{Tzu-Mao Li},
  \bibinfo{person}{Luke Anderson}, \bibinfo{person}{Jonathan Ragan-Kelley},
  {and} \bibinfo{person}{Fr{\'e}do Durand}.} \bibinfo{year}{2019}\natexlab{}.
\newblock \showarticletitle{Taichi: a language for high-performance computation
  on spatially sparse data structures}.
\newblock \bibinfo{journal}{\emph{ACM Transactions on Graphics (TOG)}}
  \bibinfo{volume}{38}, \bibinfo{number}{6} (\bibinfo{year}{2019}),
  \bibinfo{pages}{1--16}.
\newblock


\bibitem[Jiang et~al\mbox{.}(2017a)]%
        {jiang2017anisotropic}
\bibfield{author}{\bibinfo{person}{Chenfanfu Jiang}, \bibinfo{person}{Theodore
  Gast}, {and} \bibinfo{person}{Joseph Teran}.}
  \bibinfo{year}{2017}\natexlab{a}.
\newblock \showarticletitle{Anisotropic elastoplasticity for cloth, knit and
  hair frictional contact}.
\newblock \bibinfo{journal}{\emph{ACM Transactions on Graphics (TOG)}}
  \bibinfo{volume}{36}, \bibinfo{number}{4} (\bibinfo{year}{2017}),
  \bibinfo{pages}{1--14}.
\newblock


\bibitem[Jiang et~al\mbox{.}(2015)]%
        {jiang2015affine}
\bibfield{author}{\bibinfo{person}{Chenfanfu Jiang}, \bibinfo{person}{Craig
  Schroeder}, \bibinfo{person}{Andrew Selle}, \bibinfo{person}{Joseph Teran},
  {and} \bibinfo{person}{Alexey Stomakhin}.} \bibinfo{year}{2015}\natexlab{}.
\newblock \showarticletitle{The affine particle-in-cell method}.
\newblock \bibinfo{journal}{\emph{ACM Transactions on Graphics (TOG)}}
  \bibinfo{volume}{34}, \bibinfo{number}{4} (\bibinfo{year}{2015}),
  \bibinfo{pages}{1--10}.
\newblock


\bibitem[Jiang et~al\mbox{.}(2017b)]%
        {jiang2017angular}
\bibfield{author}{\bibinfo{person}{Chenfanfu Jiang}, \bibinfo{person}{Craig
  Schroeder}, {and} \bibinfo{person}{Joseph Teran}.}
  \bibinfo{year}{2017}\natexlab{b}.
\newblock \showarticletitle{An angular momentum conserving
  affine-particle-in-cell method}.
\newblock \bibinfo{journal}{\emph{J. Comput. Phys.}}  \bibinfo{volume}{338}
  (\bibinfo{year}{2017}), \bibinfo{pages}{137--164}.
\newblock


\bibitem[Jiang et~al\mbox{.}(2016)]%
        {jiang2016material}
\bibfield{author}{\bibinfo{person}{Chenfanfu Jiang}, \bibinfo{person}{Craig
  Schroeder}, \bibinfo{person}{Joseph Teran}, \bibinfo{person}{Alexey
  Stomakhin}, {and} \bibinfo{person}{Andrew Selle}.}
  \bibinfo{year}{2016}\natexlab{}.
\newblock \showarticletitle{The material point method for simulating continuum
  materials}.
\newblock In \bibinfo{booktitle}{\emph{Acm siggraph 2016 courses}}.
  \bibinfo{pages}{1--52}.
\newblock


\bibitem[Kala et~al\mbox{.}(2024)]%
        {kala2024thermomechanical}
\bibfield{author}{\bibinfo{person}{Victoria Kala}, \bibinfo{person}{Jingyu
  Chen}, \bibinfo{person}{David Hyde}, \bibinfo{person}{Alexey Stomakhin},
  {and} \bibinfo{person}{Joseph Teran}.} \bibinfo{year}{2024}\natexlab{}.
\newblock \showarticletitle{A Thermomechanical Hybrid Incompressible Material
  Point Method}.
\newblock \bibinfo{journal}{\emph{arXiv preprint arXiv:2408.07276}}
  (\bibinfo{year}{2024}).
\newblock


\bibitem[Kl{\'a}r et~al\mbox{.}(2016)]%
        {klar2016drucker}
\bibfield{author}{\bibinfo{person}{Gergely Kl{\'a}r}, \bibinfo{person}{Theodore
  Gast}, \bibinfo{person}{Andre Pradhana}, \bibinfo{person}{Chuyuan Fu},
  \bibinfo{person}{Craig Schroeder}, \bibinfo{person}{Chenfanfu Jiang}, {and}
  \bibinfo{person}{Joseph Teran}.} \bibinfo{year}{2016}\natexlab{}.
\newblock \showarticletitle{Drucker-prager elastoplasticity for sand
  animation}.
\newblock \bibinfo{journal}{\emph{ACM Transactions on Graphics (TOG)}}
  \bibinfo{volume}{35}, \bibinfo{number}{4} (\bibinfo{year}{2016}),
  \bibinfo{pages}{1--12}.
\newblock


\bibitem[Liang et~al\mbox{.}(2019)]%
        {liang2019efficient}
\bibfield{author}{\bibinfo{person}{Yong Liang}, \bibinfo{person}{Xiong Zhang},
  {and} \bibinfo{person}{Yan Liu}.} \bibinfo{year}{2019}\natexlab{}.
\newblock \showarticletitle{An efficient staggered grid material point method}.
\newblock \bibinfo{journal}{\emph{Computer Methods in Applied Mechanics and
  Engineering}}  \bibinfo{volume}{352} (\bibinfo{year}{2019}),
  \bibinfo{pages}{85--109}.
\newblock


\bibitem[Monaghan(1994)]%
        {monaghan1994simulating}
\bibfield{author}{\bibinfo{person}{Joe~J Monaghan}.}
  \bibinfo{year}{1994}\natexlab{}.
\newblock \showarticletitle{Simulating free surface flows with SPH}.
\newblock \bibinfo{journal}{\emph{Journal of computational physics}}
  \bibinfo{volume}{110}, \bibinfo{number}{2} (\bibinfo{year}{1994}),
  \bibinfo{pages}{399--406}.
\newblock


\bibitem[Moutsanidis et~al\mbox{.}(2020)]%
        {moutsanidis2020iga}
\bibfield{author}{\bibinfo{person}{Georgios Moutsanidis},
  \bibinfo{person}{Christopher~C Long}, {and} \bibinfo{person}{Yuri Bazilevs}.}
  \bibinfo{year}{2020}\natexlab{}.
\newblock \showarticletitle{IGA-MPM: the isogeometric material point method}.
\newblock \bibinfo{journal}{\emph{Computer Methods in Applied Mechanics and
  Engineering}}  \bibinfo{volume}{372} (\bibinfo{year}{2020}),
  \bibinfo{pages}{113346}.
\newblock


\bibitem[Qiu et~al\mbox{.}(2023)]%
        {qiu2023sparse}
\bibfield{author}{\bibinfo{person}{Yuxing Qiu}, \bibinfo{person}{Samuel~Temple
  Reeve}, \bibinfo{person}{Minchen Li}, \bibinfo{person}{Yin Yang},
  \bibinfo{person}{Stuart~Ryan Slattery}, {and} \bibinfo{person}{Chenfanfu
  Jiang}.} \bibinfo{year}{2023}\natexlab{}.
\newblock \showarticletitle{A sparse distributed gigascale resolution material
  point method}.
\newblock \bibinfo{journal}{\emph{ACM Transactions on Graphics}}
  \bibinfo{volume}{42}, \bibinfo{number}{2} (\bibinfo{year}{2023}),
  \bibinfo{pages}{1--21}.
\newblock


\bibitem[Sadeghirad et~al\mbox{.}(2011)]%
        {sadeghirad2011convected}
\bibfield{author}{\bibinfo{person}{Alireza Sadeghirad},
  \bibinfo{person}{Rebecca~M Brannon}, {and} \bibinfo{person}{Jeff Burghardt}.}
  \bibinfo{year}{2011}\natexlab{}.
\newblock \showarticletitle{A convected particle domain interpolation technique
  to extend applicability of the material point method for problems involving
  massive deformations}.
\newblock \bibinfo{journal}{\emph{International Journal for numerical methods
  in Engineering}} \bibinfo{volume}{86}, \bibinfo{number}{12}
  (\bibinfo{year}{2011}), \bibinfo{pages}{1435--1456}.
\newblock


\bibitem[Steffen et~al\mbox{.}(2008)]%
        {steffen2008analysis}
\bibfield{author}{\bibinfo{person}{Michael Steffen}, \bibinfo{person}{Robert~M
  Kirby}, {and} \bibinfo{person}{Martin Berzins}.}
  \bibinfo{year}{2008}\natexlab{}.
\newblock \showarticletitle{Analysis and reduction of quadrature errors in the
  material point method (MPM)}.
\newblock \bibinfo{journal}{\emph{International journal for numerical methods
  in engineering}} \bibinfo{volume}{76}, \bibinfo{number}{6}
  (\bibinfo{year}{2008}), \bibinfo{pages}{922--948}.
\newblock


\bibitem[Stomakhin et~al\mbox{.}(2012)]%
        {stomakhin2012energetically}
\bibfield{author}{\bibinfo{person}{Alexey Stomakhin}, \bibinfo{person}{Russell
  Howes}, \bibinfo{person}{Craig~A Schroeder}, {and} \bibinfo{person}{Joseph~M
  Teran}.} \bibinfo{year}{2012}\natexlab{}.
\newblock \showarticletitle{Energetically Consistent Invertible Elasticity.}.
  In \bibinfo{booktitle}{\emph{Symposium on Computer Animation}},
  Vol.~\bibinfo{volume}{1}.
\newblock


\bibitem[Stomakhin et~al\mbox{.}(2013)]%
        {stomakhin2013material}
\bibfield{author}{\bibinfo{person}{Alexey Stomakhin}, \bibinfo{person}{Craig
  Schroeder}, \bibinfo{person}{Lawrence Chai}, \bibinfo{person}{Joseph Teran},
  {and} \bibinfo{person}{Andrew Selle}.} \bibinfo{year}{2013}\natexlab{}.
\newblock \showarticletitle{A material point method for snow simulation}.
\newblock \bibinfo{journal}{\emph{ACM Transactions on Graphics (TOG)}}
  \bibinfo{volume}{32}, \bibinfo{number}{4} (\bibinfo{year}{2013}),
  \bibinfo{pages}{1--10}.
\newblock


\bibitem[Stomakhin et~al\mbox{.}(2014)]%
        {stomakhin2014augmented}
\bibfield{author}{\bibinfo{person}{Alexey Stomakhin}, \bibinfo{person}{Craig
  Schroeder}, \bibinfo{person}{Chenfanfu Jiang}, \bibinfo{person}{Lawrence
  Chai}, \bibinfo{person}{Joseph Teran}, {and} \bibinfo{person}{Andrew Selle}.}
  \bibinfo{year}{2014}\natexlab{}.
\newblock \showarticletitle{Augmented MPM for phase-change and varied
  materials}.
\newblock \bibinfo{journal}{\emph{ACM Transactions on Graphics (TOG)}}
  \bibinfo{volume}{33}, \bibinfo{number}{4} (\bibinfo{year}{2014}),
  \bibinfo{pages}{1--11}.
\newblock


\bibitem[Su et~al\mbox{.}(2021)]%
        {su2021unified}
\bibfield{author}{\bibinfo{person}{Haozhe Su}, \bibinfo{person}{Tao Xue},
  \bibinfo{person}{Chengguizi Han}, \bibinfo{person}{Chenfanfu Jiang}, {and}
  \bibinfo{person}{Mridul Aanjaneya}.} \bibinfo{year}{2021}\natexlab{}.
\newblock \showarticletitle{A unified second-order accurate in time MPM
  formulation for simulating viscoelastic liquids with phase change}.
\newblock \bibinfo{journal}{\emph{ACM Transactions on Graphics (TOG)}}
  \bibinfo{volume}{40}, \bibinfo{number}{4} (\bibinfo{year}{2021}),
  \bibinfo{pages}{1--18}.
\newblock


\bibitem[Sulsky et~al\mbox{.}(1995)]%
        {sulsky1995application}
\bibfield{author}{\bibinfo{person}{Deborah Sulsky}, \bibinfo{person}{Shi-Jian
  Zhou}, {and} \bibinfo{person}{Howard~L Schreyer}.}
  \bibinfo{year}{1995}\natexlab{}.
\newblock \showarticletitle{Application of a particle-in-cell method to solid
  mechanics}.
\newblock \bibinfo{journal}{\emph{Computer physics communications}}
  \bibinfo{volume}{87}, \bibinfo{number}{1-2} (\bibinfo{year}{1995}),
  \bibinfo{pages}{236--252}.
\newblock


\bibitem[Tampubolon et~al\mbox{.}(2017)]%
        {tampubolon2017multi}
\bibfield{author}{\bibinfo{person}{Andre~Pradhana Tampubolon},
  \bibinfo{person}{Theodore Gast}, \bibinfo{person}{Gergely Kl{\'a}r},
  \bibinfo{person}{Chuyuan Fu}, \bibinfo{person}{Joseph Teran},
  \bibinfo{person}{Chenfanfu Jiang}, {and} \bibinfo{person}{Ken Museth}.}
  \bibinfo{year}{2017}\natexlab{}.
\newblock \showarticletitle{Multi-species simulation of porous sand and water
  mixtures}.
\newblock \bibinfo{journal}{\emph{ACM Transactions on Graphics (TOG)}}
  \bibinfo{volume}{36}, \bibinfo{number}{4} (\bibinfo{year}{2017}),
  \bibinfo{pages}{1--11}.
\newblock


\bibitem[Wang et~al\mbox{.}(2020)]%
        {wang2020massively}
\bibfield{author}{\bibinfo{person}{Xinlei Wang}, \bibinfo{person}{Yuxing Qiu},
  \bibinfo{person}{Stuart~R Slattery}, \bibinfo{person}{Yu Fang},
  \bibinfo{person}{Minchen Li}, \bibinfo{person}{Song-Chun Zhu},
  \bibinfo{person}{Yixin Zhu}, \bibinfo{person}{Min Tang},
  \bibinfo{person}{Dinesh Manocha}, {and} \bibinfo{person}{Chenfanfu Jiang}.}
  \bibinfo{year}{2020}\natexlab{}.
\newblock \showarticletitle{A massively parallel and scalable multi-GPU
  material point method}.
\newblock \bibinfo{journal}{\emph{ACM Transactions on Graphics (TOG)}}
  \bibinfo{volume}{39}, \bibinfo{number}{4} (\bibinfo{year}{2020}),
  \bibinfo{pages}{30--1}.
\newblock


\bibitem[Wilson et~al\mbox{.}(2021)]%
        {wilson2021distillation}
\bibfield{author}{\bibinfo{person}{Peter Wilson}, \bibinfo{person}{Roland
  W{\"u}chner}, {and} \bibinfo{person}{Dilum Fernando}.}
  \bibinfo{year}{2021}\natexlab{}.
\newblock \showarticletitle{Distillation of the material point method cell
  crossing error leading to a novel quadrature-based C 0 remedy}.
\newblock \bibinfo{journal}{\emph{Internat. J. Numer. Methods Engrg.}}
  \bibinfo{volume}{122}, \bibinfo{number}{6} (\bibinfo{year}{2021}),
  \bibinfo{pages}{1513--1537}.
\newblock


\bibitem[Wolper et~al\mbox{.}(2020)]%
        {wolper2020anisompm}
\bibfield{author}{\bibinfo{person}{Joshuah Wolper}, \bibinfo{person}{Yunuo
  Chen}, \bibinfo{person}{Minchen Li}, \bibinfo{person}{Yu Fang},
  \bibinfo{person}{Ziyin Qu}, \bibinfo{person}{Jiecong Lu},
  \bibinfo{person}{Meggie Cheng}, {and} \bibinfo{person}{Chenfanfu Jiang}.}
  \bibinfo{year}{2020}\natexlab{}.
\newblock \showarticletitle{Anisompm: Animating anisotropic damage mechanics:
  Supplemental document}.
\newblock \bibinfo{journal}{\emph{ACM Trans. Graph}} \bibinfo{volume}{39},
  \bibinfo{number}{4} (\bibinfo{year}{2020}).
\newblock


\bibitem[Wolper et~al\mbox{.}(2019)]%
        {wolper2019cd}
\bibfield{author}{\bibinfo{person}{Joshuah Wolper}, \bibinfo{person}{Yu Fang},
  \bibinfo{person}{Minchen Li}, \bibinfo{person}{Jiecong Lu},
  \bibinfo{person}{Ming Gao}, {and} \bibinfo{person}{Chenfanfu Jiang}.}
  \bibinfo{year}{2019}\natexlab{}.
\newblock \showarticletitle{CD-MPM: continuum damage material point methods for
  dynamic fracture animation}.
\newblock \bibinfo{journal}{\emph{ACM Transactions on Graphics (TOG)}}
  \bibinfo{volume}{38}, \bibinfo{number}{4} (\bibinfo{year}{2019}),
  \bibinfo{pages}{1--15}.
\newblock


\bibitem[Yue et~al\mbox{.}(2015)]%
        {yue2015continuum}
\bibfield{author}{\bibinfo{person}{Yonghao Yue}, \bibinfo{person}{Breannan
  Smith}, \bibinfo{person}{Christopher Batty}, \bibinfo{person}{Changxi Zheng},
  {and} \bibinfo{person}{Eitan Grinspun}.} \bibinfo{year}{2015}\natexlab{}.
\newblock \showarticletitle{Continuum foam: A material point method for
  shear-dependent flows}.
\newblock \bibinfo{journal}{\emph{ACM Transactions on Graphics (TOG)}}
  \bibinfo{volume}{34}, \bibinfo{number}{5} (\bibinfo{year}{2015}),
  \bibinfo{pages}{1--20}.
\newblock


\bibitem[Yue et~al\mbox{.}(2018)]%
        {yue2018hybrid}
\bibfield{author}{\bibinfo{person}{Yonghao Yue}, \bibinfo{person}{Breannan
  Smith}, \bibinfo{person}{Peter~Yichen Chen}, \bibinfo{person}{Maytee
  Chantharayukhonthorn}, \bibinfo{person}{Ken Kamrin}, {and}
  \bibinfo{person}{Eitan Grinspun}.} \bibinfo{year}{2018}\natexlab{}.
\newblock \showarticletitle{Hybrid grains: Adaptive coupling of discrete and
  continuum simulations of granular media}.
\newblock \bibinfo{journal}{\emph{ACM Transactions on Graphics (TOG)}}
  \bibinfo{volume}{37}, \bibinfo{number}{6} (\bibinfo{year}{2018}),
  \bibinfo{pages}{1--19}.
\newblock


\bibitem[Zhang et~al\mbox{.}(2011)]%
        {zhang2011material}
\bibfield{author}{\bibinfo{person}{Duan~Z Zhang}, \bibinfo{person}{Xia Ma},
  {and} \bibinfo{person}{Paul~T Giguere}.} \bibinfo{year}{2011}\natexlab{}.
\newblock \showarticletitle{Material point method enhanced by modified gradient
  of shape function}.
\newblock \bibinfo{journal}{\emph{J. Comput. Phys.}} \bibinfo{volume}{230},
  \bibinfo{number}{16} (\bibinfo{year}{2011}), \bibinfo{pages}{6379--6398}.
\newblock


\bibitem[Zhao et~al\mbox{.}(2023)]%
        {zhao2023coupled}
\bibfield{author}{\bibinfo{person}{Yidong Zhao}, \bibinfo{person}{Jinhyun
  Choo}, \bibinfo{person}{Yupeng Jiang}, {and} \bibinfo{person}{Liuchi Li}.}
  \bibinfo{year}{2023}\natexlab{}.
\newblock \showarticletitle{Coupled material point and level set methods for
  simulating soils interacting with rigid objects with complex geometry}.
\newblock \bibinfo{journal}{\emph{Computers and Geotechnics}}
  \bibinfo{volume}{163} (\bibinfo{year}{2023}), \bibinfo{pages}{105708}.
\newblock


\bibitem[Zhao et~al\mbox{.}(2024)]%
        {zhao2024mapped}
\bibfield{author}{\bibinfo{person}{Yidong Zhao}, \bibinfo{person}{Minchen Li},
  \bibinfo{person}{Chenfanfu Jiang}, {and} \bibinfo{person}{Jinhyun Choo}.}
  \bibinfo{year}{2024}\natexlab{}.
\newblock \showarticletitle{Mapped material point method for large deformation
  problems with sharp gradients and its application to soil-structure
  interactions}.
\newblock \bibinfo{journal}{\emph{International Journal for Numerical and
  Analytical Methods in Geomechanics}} (\bibinfo{year}{2024}).
\newblock


\end{thebibliography}

        \newpage
        \clearpage
        

\section*{Supplementary material}

# CK-MPM: A Compact-Kernel Material Point Method Supplemental Document

MICHAEL LIU, Carnegie Mellon University, USA
XINLEI WANG, ZenusTech Inc., China
MINCHEN LI, Carnegie Mellon University, USA

## Contents

## 1  Proof of Linear Momentum Conservation with PIC

Traditional PIC pipeline preserves linear momentum in P2G, grid update, and G2P steps. We show a similar result in our new dual grid system.

THEOREM 1.1 (CONSERVATION OF LINEAR MOMENTUM WITH PIC). *The total linear momentum is preserved in P2G, grid update, and P2G steps in PIC scheme by defining the total linear momentum of the grid as:*

$$\mathbf{p}^{\alpha}_{\text{grid total},t_n} \coloneqq \frac{1}{2} \sum_{k \in \{-1,+1\}} \mathbf{p}^{\alpha}_{\mathcal{G}_k,t_n} = \frac{1}{2} \sum_{k \in \{\pm 1\}} \sum_i m_{i,\mathcal{G}_k,t_n} \mathbf{v}^{\alpha}_{i,\mathcal{G}_k,t_n} \tag{1}$$

PROOF. Let's assume that the current time step is $t_n$. In P2G step, the total linear momentum of particle is $\sum_p m_p \mathbf{v}^{\alpha}_{p,\mathcal{G}_0,t_n}$. With the particle-to-grid transfer for momentum, we note that:

$$\sum_p m_p \mathbf{v}^{\alpha}_{p,\mathcal{G}_0,t_n}$$
$$= \sum_p m_p \mathbf{v}^{\alpha}_{p,\mathcal{G}_0,t_n} \left( \frac{1}{2} \sum_{k \in \{\pm 1\}} \sum_i w_{i,p,\mathcal{G}_k,t_n} \right)$$
$$= \frac{1}{2} \sum_{k \in \{-1,+1\}} \sum_i \sum_p w_{i,p,\mathcal{G}_k,t_n} m_p \mathbf{v}^{\alpha}_{p,t_n}$$
$$= \frac{1}{2} \sum_{k \in \{\pm 1\}} \sum_i m_{i,\mathcal{G}_k,t_n} \mathbf{v}^{\alpha}_{i,\mathcal{G}_k,t_n}$$

---

Authors' Contact Information: Michael Liu, Carnegie Mellon University, USA, appledorem.g@gmail.com; Xinlei Wang, ZenusTech Inc., China, wxlwxl1993@zju.edu.cn; Minchen Li, Carnegie Mellon University, USA, minchernl@gmail.com.

For grid update, we have:

$$\frac{1}{2} \sum_{k \in \{\pm 1\}} \sum_i m_{i,\mathcal{G}_k,t_n} \tilde{\mathbf{v}}^\alpha_{i,\mathcal{G}_k,t^{n+1}}$$
$$= \frac{1}{2} \sum_{k \in \{\pm 1\}} \sum_i \left( m_{i,\mathcal{G}_k,t_n} \mathbf{v}^\alpha_{i,\mathcal{G}_k,t^n} + \Delta t \mathbf{f}^\alpha_{i,\mathcal{G}_k,t_n} \right)$$
$$= \frac{1}{2} \sum_{k \in \{\pm 1\}} \sum_i m_{i,\mathcal{G}_k,t_n} \mathbf{v}^\alpha_{i,\mathcal{G}_k,t_n} + \frac{\Delta t}{2} \sum_{k \in \{\pm 1\}} \sum_i \sum_p V_0 (\mathbf{P}_{p,\mathcal{G}_0,t_n})^\alpha_\beta (\mathbf{F}_{p,\mathcal{G}_0,t_n})^\beta_\nu (\nabla w_{i,p,\mathcal{G}_k,t_n})^\nu$$
$$= \frac{1}{2} \sum_{k \in \{\pm 1\}} \sum_i m_{i,\mathcal{G}_k,t_n} \mathbf{v}^\alpha_{i,\mathcal{G}_k,t_n} + \Delta t \sum_p V_0 (\mathbf{P}_{p,\mathcal{G}_0,t_n})^\alpha_\beta (\mathbf{F}_{p,\mathcal{G}_0,t_n})^\beta_\nu \left( \frac{1}{2} \sum_{k \in \{\pm 1\}} \sum_i (\nabla w_{i,p,\mathcal{G}_k,t_n})^\nu \right)$$
$$= \frac{1}{2} \sum_{k \in \{\pm 1\}} \sum_i m_{i,\mathcal{G}_k,t_n} \mathbf{v}^\alpha_{i,\mathcal{G}_k,t_n}$$

Finally, note:

$$\mathbf{v}^\alpha_{p,t^{n+1}} = \frac{1}{2 m_p} \sum_{k \in \{\pm 1\}} \sum_i m_{i,\mathcal{G}_k,t_n} w_{i,p,\mathcal{G}_k,t_n} \tilde{\mathbf{v}}^\alpha_{i,\mathcal{G}_k,t^{n+1}}$$

We thus have:

$$\frac{1}{2} \sum_{k \in \{\pm 1\}} \sum_i m_{i,\mathcal{G}_k,t_n} \tilde{\mathbf{v}}^\alpha_{i,\mathcal{G}_k,t^{n+1}}$$
$$= \frac{1}{2} \sum_{k \in \{\pm 1\}} \sum_i m_{i,\mathcal{G}_k,t_n} \tilde{\mathbf{v}}^\alpha_{i,\mathcal{G}_k,t^{n+1}} \left( \sum_p w_{i,p,\mathcal{G}_k,t_n} \right)$$
$$= \frac{1}{2} \sum_{k \in \{\pm 1\}} \sum_i \sum_p m_{i,\mathcal{G}_k,t_n} w_{i,p,\mathcal{G}_k,t_n} \tilde{\mathbf{v}}^\alpha_{i,\mathcal{G}_k,t^{n+1}}$$
$$= \sum_p \left( \frac{1}{2} \sum_{k \in \{\pm 1\}} \sum_i m_{i,\mathcal{G}_k,t_n} w_{i,p,\mathcal{G}_k,t_n} \tilde{\mathbf{v}}^\alpha_{i,\mathcal{G}_k,t^{n+1}} \right)$$
$$= \sum_p m_p \mathbf{v}^\alpha_{p,t^{n+1}}$$

□

## 2 Proof of Momentum Conservation with APIC

Theorem 2.1 (Conservation of linear momentum with APIC).

Proof. Note that there is no difference in the grid evolution and grid-to-particle transfer from the PIC scheme. Hence, it suffices to show that the particle-to-grid transfer conserves the total linear momentum.

Furthermore, we note that it suffices to show that the extra terms in the APIC particle-to-grid transfer, comparing to the PIC transfer, sum to zero across both grids, i.e.

$$\frac{1}{2} \sum_{k \in \{\pm 1\}} \sum_i \sum_p w_{i,p,\mathcal{G}_k,t_n} m_p (\mathbf{B}_{p,t_n})^\alpha_\nu ((\mathbf{D}_{p,t_n})^{-1})^\nu_\beta (\mathbf{x}^\beta_{i,\mathcal{G}_k,t_n} - \mathbf{x}^\beta_{p,\mathcal{G}_0,t_n}) = 0$$

It is trivial to show that the above holds with the 1-order accuracy property. □

Theorem 2.2 (Conservation of angular momentum with APIC). *The total angular momentum is preserved in P2G, grid update, and P2G steps in APIC scheme by defining the total angular momentum of the grid as:*

$$\mathbf{L}^{\alpha}_{\text{grid total},t_n} := \frac{1}{2} \sum_{k \in \{\pm 1\}} \mathbf{L}^{\alpha}_{\mathcal{G}_k,t_n} = \frac{1}{2} \sum_{k \in \{\pm 1\}} \sum_{i} (\mathbf{x}_{i,\mathcal{G}_k,t_n} \times m_{i,\mathcal{G}_k,t_n} \mathbf{v}_{i,\mathcal{G}_k,t_n})^{\alpha} \quad (2)$$

Proof. In this part of the proof, we mainly follow the proof presented in the APIC paper. We first demonstrate that the P2G process will conserve angular momentum. We first note that we may write the cross product of two vector **p**, **q**:

$$(\mathbf{u} \times \mathbf{v})^{\alpha} = \delta^{\alpha\beta} \varepsilon_{\beta\gamma\eta} \mathbf{p}^{\gamma} \mathbf{q}^{\eta}$$

Moreover, note that the APIC transfer, we have

$$m_{i,\mathcal{G}_k,t_n} \mathbf{v}^{\eta}_{i,\mathcal{G}_k,t_n} = \sum_p w_{i,p,\mathcal{G}_k,t_n} m_p \left( \mathbf{v}^{\eta}_{p,\mathcal{G}_0,t_n} + (\mathbf{B}_{p,\mathcal{G}_0,t_n})^{\eta}_{\nu} ((\mathbf{D}_{p,\mathcal{G}_0,t_n})^{-1})^{\nu}_{\mu} (\mathbf{x}^{\mu}_{i,\mathcal{G}_k,t_n} - \mathbf{x}^{\mu}_{p,\mathcal{G}_0,t_n})) \right)$$

Therefore, the total angular momentum is:

$$\frac{1}{2} \sum_{k \in \{\pm 1\}} \sum_{i} (\mathbf{x}_{i,\mathcal{G}_k,t_n} \times m_{i,\mathcal{G}_k,t_n} \mathbf{v}_{i,\mathcal{G}_k,t_n})^{\alpha}$$

$$= \frac{1}{2} \sum_{k \in \{\pm 1\}} \sum_{i} \sum_{p} w_{i,p,\mathcal{G}_k,t_n} m_p \delta^{\alpha\beta} \varepsilon_{\beta\gamma\eta} \mathbf{x}^{\gamma}_{i,\mathcal{G}_k,t_n} \left( \mathbf{v}^{\eta}_{p,\mathcal{G}_0,t_n} + (\mathbf{B}_{p,\mathcal{G}_0,t_n})^{\eta}_{\nu} ((\mathbf{D}_{p,\mathcal{G}_0,t_n})^{-1})^{\nu}_{\mu} (\mathbf{x}^{\mu}_{i,\mathcal{G}_k,t_n} - \mathbf{x}^{\mu}_{p,\mathcal{G}_0,t_n}) \right)$$

We then simplify part of above as in APIC:

$$\frac{1}{2} \sum_{k \in \{\pm 1\}} \sum_{i} \sum_{p} w_{i,p,\mathcal{G}_k,t_n} m_p \delta^{\alpha\beta} \varepsilon_{\beta\gamma\eta} \mathbf{x}^{\gamma}_{i,\mathcal{G}_k,t_n} (\mathbf{B}_{p,\mathcal{G}_0,t_n})^{\eta}_{\nu} ((\mathbf{D}_{p,\mathcal{G}_0,t_n})^{-1})^{\nu}_{\mu} (\mathbf{x}^{\mu}_{i,\mathcal{G}_k,t_n} - \mathbf{x}^{\mu}_{p,\mathcal{G}_0,t_n})$$

$$= \frac{1}{2} \sum_{p} m_p \delta^{\alpha\beta} \varepsilon_{\beta\gamma\eta} (\mathbf{B}_{p,\mathcal{G}_0,t_n})^{\eta}_{\nu} ((\mathbf{D}_{p,\mathcal{G}_0,t_n})^{-1})^{\nu}_{\mu} \sum_{k \in \{\pm 1\}} \sum_{i} w_{i,p,\mathcal{G}_k,t_n} \left( (\mathbf{x}^{\gamma}_{i,\mathcal{G}_k,t_n} - \mathbf{x}^{\gamma}_{p,\mathcal{G}_0,t_n})(\mathbf{x}^{\mu}_{i,\mathcal{G}_k,t_n} - \mathbf{x}^{\mu}_{p,\mathcal{G}_0,t_n}) + \mathbf{x}^{\gamma}_{p,\mathcal{G}_0,t_n} (\mathbf{x}^{\mu}_{i,\mathcal{G}_k,t_n} - \mathbf{x}^{\mu}_{p,\mathcal{G}_0,t_n}) \right)$$

$$= \frac{1}{2} \sum_{p} m_p \delta^{\alpha\beta} \varepsilon_{\beta\gamma\eta} (\mathbf{B}_{p,\mathcal{G}_0,t_n})^{\eta}_{\nu} ((\mathbf{D}_{p,\mathcal{G}_0,t_n})^{-1})^{\nu}_{\mu} \sum_{k \in \{\pm 1\}} \sum_{i} w_{i,p,\mathcal{G}_k,t_n} \left( (\mathbf{x}^{\gamma}_{i,\mathcal{G}_k,t_n} - \mathbf{x}^{\gamma}_{p,\mathcal{G}_0,t_n})(\mathbf{x}^{\mu}_{i,\mathcal{G}_k,t_n} - \mathbf{x}^{\mu}_{p,\mathcal{G}_0,t_n}) \right)$$

$$= \sum_{p} m_p \delta^{\alpha\beta} \varepsilon_{\beta}{}^{\gamma}{}_{\eta} (\mathbf{B}_{p,\mathcal{G}_0,t_n})^{\eta}_{\nu} ((\mathbf{D}_{p,\mathcal{G}_0,t_n})^{-1})^{\nu}_{\mu} \left( \frac{1}{2} \sum_{k \in \{\pm 1\}} \sum_{i} w_{i,p,\mathcal{G}_k,t_n} (((\mathbf{x}_{i,\mathcal{G}_k,t_n})^T - (\mathbf{x}_{p,\mathcal{G}_0,t_n})^T)_{\gamma} (\mathbf{x}^{\mu}_{i,\mathcal{G}_k,t_n} - \mathbf{x}^{\mu}_{p,\mathcal{G}_0,t_n})) \right)$$

$$= \sum_{p} m_p \delta^{\alpha\beta} \varepsilon_{\beta}{}^{\gamma}{}_{\eta} (\mathbf{B}_{p,\mathcal{G}_0,t_n})^{\eta}_{\nu} ((\mathbf{D}_{p,\mathcal{G}_0,t_n})^{-1})^{\nu}_{\mu} (\mathbf{D}_{p,\mathcal{G}_0,t_n})^{\mu}_{\gamma}$$

$$= \sum_{p} m_p \delta^{\alpha\beta} \varepsilon_{\beta}{}^{\gamma}{}_{\eta} (\mathbf{B}_{p,\mathcal{G}_0,t_n})^{\eta}_{\gamma}$$

Hence, we note that the total angular momentum can be reduced to:

$$\frac{1}{2} \sum_{k \in \{\pm 1\}} \sum_{i} \sum_{p} (w_{i,p,\mathcal{G}_k,t_n} \mathbf{x}_{i,\mathcal{G}_k,t_n} \times m_p \mathbf{v}_{p,\mathcal{G}_0,t_n})^{\alpha} + \sum_{p} m_p \delta^{\alpha\beta} \varepsilon_{\beta}{}^{\gamma}{}_{\eta} (\mathbf{B}_{p,\mathcal{G}_0,t_n})^{\eta}_{\gamma}$$

$$= \sum_{p} (\mathbf{x}_{p,\mathcal{G}_0,t_n} \times m_p \mathbf{v}_{p,\mathcal{G}_0,t_n})^{\alpha} + \sum_{p} m_p \delta^{\alpha\beta} \varepsilon_{\beta}{}^{\gamma}{}_{\eta} (\mathbf{B}_{p,\mathcal{G}_0,t_n})^{\eta}_{\gamma} \quad (3)$$

Following APIC formulation, we also define Equation 3 to be the total angular momentum on particles. This concludes the proof for the conservation of angular momentum in the particle-to-grid step.

For the grid evolution step, we first note that $\mathbf{F}_{p,\mathcal{G}_0,t_n}$ and $\mathbf{P}_{p,\mathcal{G}_0,t_n}$ shares the same singular space for isotropic materials, i.e. if the singular value decomposition is in form of

$$\mathbf{F}_{p,\mathcal{G}_0,t_n} = \mathbf{U}_{p,\mathcal{G}_0,t_n} \Sigma_{p,\mathcal{G}_0,t_n} \mathbf{V}_{p,\mathcal{G}_0,t_n} \tag{4}$$

then we would have $\mathbf{P}_{p,\mathcal{G}_0,t_n} = \mathbf{U}_{p,\mathcal{G}_0,t_n} \hat{\Sigma}_{p,\mathcal{G}_0,t_n} \mathbf{V}_{p,\mathcal{G}_0,t_n}$. This implies that:

$$(\mathbf{P}_{p,\mathcal{G}_0,t_n})^\alpha_\beta = (\mathbf{F}_{p,\mathcal{G}_0,t_n})^\alpha_\nu (\mathbf{A}_{p,\mathcal{G}_0,t_n})^\nu_\beta \tag{5}$$

where $(\mathbf{A}_{p,\mathcal{G}_0,t_n})^\nu_\beta = ((\mathbf{V}_{p,\mathcal{G}_0,t_n})^T)^\nu_\mu (\hat{\Sigma}_{p,\mathcal{G}_0,t_n})^\mu_\eta (\mathbf{V}_{p,\mathcal{G}_0,t_n})^\eta_\beta$ is a symmetric matrix.

Then, we denote

$$(\mathbf{G}_{p,\mathcal{G}_0,t_n})^\alpha_\beta = \frac{1}{2} \sum_{k\in\{\pm 1\}} \sum_i \tilde{\mathbf{x}}^\alpha_{i,k,t^{n+1}} ((\nabla w_{i,p,k,t_n})^T)_\beta \tag{6}$$

We see that:

$$(\mathbf{F}_{p,\mathcal{G}_0,t^{n+1}})^\alpha_\beta$$
$$= (\delta^\alpha_\nu + \frac{\Delta t}{2} \sum_{k\in\{\pm 1\}} \sum_i \tilde{\mathbf{v}}^\alpha_{i,\mathcal{G}_k,t^{n+1}} ((\nabla w_{i,p,\mathcal{G}_k,t_n})^T)_\nu)(\mathbf{F}_{p,\mathcal{G}_0,t_n})^\nu_\beta$$
$$= (\delta^\alpha_\nu + \frac{1}{2} \sum_{k\in\{\pm 1\}} \sum_i (\tilde{\mathbf{x}}^\alpha_{i,\mathcal{G}_k,t^{n+1}} - \mathbf{x}^\alpha_{i,\mathcal{G}_k,t^n})((\nabla w_{i,p,\mathcal{G}_k,t_n})^T)_\nu)(\mathbf{F}_{p,\mathcal{G}_0,t_n})^\nu_\beta$$
$$= (\delta^\alpha_\nu + (\mathbf{G}_{p,\mathcal{G}_0,t^{n+1}})^\alpha_\nu - \delta^\alpha_\nu)(\mathbf{F}_{p,\mathcal{G}_0,t_n})^\nu_\beta$$
$$= (\mathbf{G}_{p,\mathcal{G}_0,t^{n+1}})^\alpha_\nu (\mathbf{F}_{p,\mathcal{G}_0,t_n})^\nu_\beta$$

Hence, with Equation 5, we observe that:

$$\frac{1}{2} \sum_{k\in\{\pm 1\}} \sum_i (\mathbf{x}_{i,\mathcal{G}_k,t_n} \times \mathbf{f}_{i,\mathcal{G}_k,t_n})^\alpha$$
$$= \frac{1}{2} \sum_{k\in\{\pm 1\}} \sum_i \delta^{\alpha\beta} \varepsilon_{\beta\gamma\mu} \mathbf{x}^\gamma_{i,\mathcal{G}_k,t_n} \mathbf{f}^\mu_{i,\mathcal{G}_k,t_n}$$
$$= \frac{1}{2} \sum_{k\in\{\pm 1\}} \sum_i \delta^{\alpha\beta} \varepsilon_{\beta\gamma\mu} \mathbf{x}^\gamma_{i,\mathcal{G}_k,t_n} \left( \sum_p V_0 (\mathbf{P}_{p,\mathcal{G}_0,t_n})^\mu_\eta ((\mathbf{F}_{p,\mathcal{G}_0,t_n})^T)^\eta_\nu (\nabla w_{i,p,\mathcal{G}_k,t_n})^\nu \right)$$
$$= V_0 \sum_p \delta^{\alpha\beta} \varepsilon_{\beta\gamma\mu} (\mathbf{P}_{p,\mathcal{G}_0,t_n})^\mu_\eta ((\mathbf{F}_{p,\mathcal{G}_0,t_n})^T)^\eta_\nu \delta^{\nu\xi} \left( \frac{1}{2} \sum_{k\in\{\pm 1\}} \sum_i \mathbf{x}^\gamma_{i,\mathcal{G}_k,t_n} ((\nabla w_{i,p,\mathcal{G}_k,t_n})^T)_\xi \right)$$
$$= V_0 \sum_p \delta^{\alpha\beta} \varepsilon_{\beta\gamma\mu} (\mathbf{P}_{p,\mathcal{G}_0,t_n})^\mu_\eta ((\mathbf{F}_{p,\mathcal{G}_0,t_n})^T)^\eta_\nu \delta^{\nu\gamma}$$
$$= V_0 \sum_p \delta^{\alpha\beta} \varepsilon_\beta{}^\nu{}_\mu (\mathbf{F}_{p,\mathcal{G}_0,t_n})^\mu_\xi (\mathbf{A}_{p,\mathcal{G}_0,t_n})^\xi_\eta ((\mathbf{F}_{p,\mathcal{G}_0,t_n})^T)^\eta_\nu$$
$$= 0$$

where the last equation holds since $\mathbf{FAF}^T$ (with subscripts $p$, $\mathcal{G}_0$, $t_n$) is symmetric. From above, it is then true that:

$$\frac{1}{2} \sum_{k\in\{\pm 1\}} \sum_i (\mathbf{x}_{i,\mathcal{G}_k,t_n} \times m_{i,\mathcal{G}_k,t_n} (\tilde{\mathbf{v}}_{i,\mathcal{G}_k,t^{n+1}} - \mathbf{v}_{i,\mathcal{G}_k,t_n})) = 0$$

Therefore, we may conclude with:

$$\frac{1}{2} \sum_{k\in\{\pm 1\}} \sum_i \left( (\tilde{\mathbf{x}}_{i,\mathcal{G}_k,t^{n+1}} \times m_{i,\mathcal{G}_k,t_n} \tilde{\mathbf{v}}_{i,\mathcal{G}_k,t^{n+1}}) - (\mathbf{x}_{i,\mathcal{G}_k,t^n} \times m_{i,\mathcal{G}_k,t_n} \mathbf{v}_{i,\mathcal{G}_k,t_n}) \right)$$

$$= \frac{1}{2} \sum_{k\in\{\pm 1\}} \sum_i \left( (\tilde{\mathbf{x}}_{i,\mathcal{G}_k,t^{n+1}} \times m_{i,\mathcal{G}_k,t_n} \tilde{\mathbf{v}}_{i,\mathcal{G}_k,t^{n+1}}) - (\mathbf{x}_{i,\mathcal{G}_k,t^n} \times m_{i,\mathcal{G}_k,t_n} \mathbf{v}_{i,\mathcal{G}_k,t_n}) \right)$$

$$- \frac{1}{2} \sum_{k\in\{\pm 1\}} \sum_i (\mathbf{x}_{i,\mathcal{G}_k,t_n} \times m_{i,\mathcal{G}_k,t_n} (\tilde{\mathbf{v}}_{i,\mathcal{G}_k,t^{n+1}} - \mathbf{v}_{i,\mathcal{G}_k,t_n}))$$

$$= \frac{1}{2} \sum_{k\in\{\pm 1\}} \sum_i ((\tilde{\mathbf{x}}_{i,\mathcal{G}_k,t^{n+1}} - \mathbf{x}_{i,\mathcal{G}_k,t_n}) \times m_{i,\mathcal{G}_k,t_n} \tilde{\mathbf{v}}_{i,\mathcal{G}_k,t^{n+1}})$$

$$= \frac{\Delta t}{2} \sum_{k\in\{\pm 1\}} \sum_i (\tilde{\mathbf{v}}_{i,\mathcal{G}_k,t^{n+1}} \times m_{i,\mathcal{G}_k,t_n} \tilde{\mathbf{v}}_{i,\mathcal{G}_k,t^{n+1}})$$

$$= 0$$

Hence, we see that the grid update step also conserves angular momentum.

Finally, for the grid-to-particle transfer step, we note that:

$$\sum_p (\mathbf{x}_{p,\mathcal{G}_0,t^{n+1}} \times m_p \mathbf{v}_{p,\mathcal{G}_0,t^{n+1}})^\alpha + \sum_p m_p \delta^{\alpha\beta} \varepsilon_\beta{}^\nu{}_\mu (\mathbf{B}_{p,\mathcal{G}_0,t^{n+1}})^\mu_\nu$$

$$= \sum_p (\mathbf{x}_{p,\mathcal{G}_0,t^{n+1}} \times m_p \mathbf{v}_{p,\mathcal{G}_0,t^{n+1}})^\alpha + \frac{1}{2} \sum_p \sum_{k\in\{\pm 1\}} \sum_i m_p \delta^{\alpha\beta} \varepsilon_\beta{}^\nu{}_\mu w_{i,p,\mathcal{G}_k,t_n} \tilde{\mathbf{v}}^\mu_{i,\mathcal{G}_k,t^{n+1}} ((\mathbf{x}_{i,\mathcal{G}_k,t_n} - \mathbf{x}_{p,\mathcal{G}_0,t_n})^T)_\nu$$

$$= \frac{1}{2} \sum_p m_p \sum_{k\in\{\pm 1\}} \sum_i \left( w_{i,p,\mathcal{G}_k,t_n} \delta^{\alpha\eta} \varepsilon_{\eta\gamma\xi} \mathbf{x}^\gamma_{p,\mathcal{G}_0,t^{n+1}} \tilde{\mathbf{v}}^\xi_{i,\mathcal{G}_k,t^{n+1}} + w_{i,p,\mathcal{G}_k,t_n} \delta^{\alpha\beta} \varepsilon_\beta{}^\nu{}_\mu \tilde{\mathbf{v}}^\mu_{i,\mathcal{G}_k,t^{n+1}} ((\mathbf{x}_{i,\mathcal{G}_k,t_n} - \mathbf{x}_{p,\mathcal{G}_0,t_n})^T)_\nu \right)$$

$$= \frac{1}{2} \sum_p m_p \sum_{k\in\{\pm 1\}} \sum_i \left( w_{i,p,\mathcal{G}_k,t_n} \delta^{\alpha\eta} \varepsilon_{\eta\gamma\xi} \tilde{\mathbf{v}}^\xi_{i,\mathcal{G}_k,t^{n+1}} (\mathbf{x}^\gamma_{p,\mathcal{G}_0,t^{n+1}} + (\mathbf{x}_{i,\mathcal{G}_k,t_n} - \mathbf{x}_{p,\mathcal{G}_0,t_n})^\gamma) \right)$$

$$= \sum_p \delta^{\alpha\eta} \varepsilon_{\eta\gamma\xi} \Delta t \mathbf{v}^\gamma_{p,\mathcal{G}_0,t^{n+1}} m_p (\frac{1}{2} \sum_{k\in\{\pm 1\}} \sum_i w_{i,p,\mathcal{G}_k,t_n} \tilde{\mathbf{v}}^\xi_{i,\mathcal{G}_k,t^{n+1}}) + \frac{1}{2} \sum_{k\in\{\pm 1\}} \sum_i \delta^{\alpha\beta} \varepsilon_{\beta\nu\mu} \mathbf{x}^\nu_{i,\mathcal{G}_k,t_n} \tilde{\mathbf{v}}^\mu_{i,\mathcal{G}_k,t^{n+1}} \sum_p m_p w_{i,p,\mathcal{G}_k,t_n}$$

$$= \frac{1}{2} \sum_{k\in\{\pm 1\}} \sum_i \left( m_{i,\mathcal{G}_k,t_n} \delta^{\alpha\beta} \varepsilon_{\beta\nu\mu} \mathbf{x}^\nu_{i,\mathcal{G}_k,t_n} \tilde{\mathbf{v}}^\mu_{i,\mathcal{G}_k,t^{n+1}} \right)$$

$$= \frac{1}{2} \sum_{k\in\{\pm 1\}} \sum_i (\mathbf{x}_{i,\mathcal{G}_k,t_n} \times m_{i,\mathcal{G}_k,t_n} \tilde{\mathbf{v}}_{i,\mathcal{G}_k,t^{n+1}})^\alpha$$

Therefore, it concludes the proof for the conservation of angular momentum in the G2P step.

□



\end{document}